\documentclass[journal ]{new-aiaa}
\usepackage[utf8]{inputenc}
\usepackage{textcomp}

\usepackage{graphicx}
\usepackage{amsmath}
\usepackage[version=4]{mhchem}
\usepackage{siunitx}
\usepackage{longtable,tabularx}
\usepackage{comment}

\usepackage{newtxtext}
\usepackage{newtxmath}
\usepackage{tikz}
\usepackage{pstricks}
\usepackage{natbib}
\usepackage{hyperref}
\usepackage{nameref}
\usepackage{soul}

\newcommand{\pd}[2]{\frac{\partial #1}{\partial #2}}
\newcommand{\be}{\begin{equation}}
\newcommand{\ee}{\end{equation}}
\newcommand{\norm}[1]{\left\lVert#1\right\rVert}

\title{Data-driven Balanced Truncation for Predictive Model Order Reduction of Aeroacoustic Response}

\author{Elnaz Rezaian \footnote{Postdoctoral Fellow, Department of Aerospace Engineering; rezaian@umich.edu, AIAA member} and Karthik Duraisamy \footnote{Professor, Department of Aerospace Engineering; kdur@umich.edu, AIAA Associate Fellow}}
\affil{University of Michigan, Ann Arbor, MI 48109, USA}

\begin{document}

\maketitle

\begin{abstract}
Rapid prediction of the aeroacoustic response is a key component in the design of aircraft and turbomachinery. In this work, we propose a technique for highly accelerated prediction of aeroacoustic response using a data-driven model reduction approach based on the eigensystem realization algorithm (ERA). Specifically, we create and compare ERA reduced-order models (ROMs) based on the training data generated by solving the linearized and nonlinear Euler equations, and use them to predict the aeroacoustic response of an airfoil in a purely predictive setting subject to different types of gust loading. Activating each input channel separately in the full-order model (FOM) to generate the Markov sequence for training makes it computationally challenging to use ERA in aeroacoustics applications. We address this bottleneck first by proposing a multi-fidelity gappy POD method to reduce the computation cost on the FOM and ROM levels by querying the high-resolution FOM only for the input channels identified by gappy POD. Second, we use tangential interpolation at the ROM level to reduce the size of the Hankel matrix. The proposed methods enable application of ERA for accurate online acoustic response prediction, and reduce the offline computation cost of ROMs.

\end{abstract}

\section{Introduction}
\label{sec:intro}
\lettrine{I}{nteractions} of disturbances in unsteady flows with other disturbances and with solid bodies generate aerodynamic noise in turbomachinery and aircraft. 
The most accurate approach towards prediction of the acoustic response is the direct numerical simulation (DNS) through the Lattice-Boltzmann method \cite{Moreau:2019, Lallier:2017, Sanjose:19} or by solving the Navier-Stokes equations \cite{Freund:2001}. A review of  numerical simulation techniques for capturing the acoustic noise generated as a result of unsteady interactions with different components is given by Moreau \cite{Moreau:2022}. Despite being accurate, such computations are typically not affordable in the context of design optimization.
To avoid the cost of high-resolution DNS to capture the low-energy acoustic waves, a convenient approach is to consider a moving mesh configuration as in the field velocity method (FVM), where the grid velocity is set to the gust velocity but in the opposite direction~\cite{sitaraman2006field,Wales:2015}. 
 Dieste \cite{Dieste2011} solved the linearized Euler equations for the scattered field and modeled noise sources by synthetic turbulence at the airfoil boundary. The idea in this approach is to separate the sound propagation and noise generation mechanisms \cite{Billson2003}. Therefore, sound generation is modeled by the Euler equations, and noise is injected through synthetic turbulence using approaches such as the Fourier mode method \cite{Clair:2013}, synthetic eddies \cite{Kim2015}, digital filters \cite{Gea2017}, and random particle-mesh method \cite{Ewert2008}. Despite the speed up compared to DNS, repeated computations based on the Euler equations also become expensive in the time scale of predictive control and design optimization.
Rapid computation of the acoustic noise has been achieved by analytical methods \cite{Amiet1975, Mish2006, Amiet1976, Roger2010, Evers2002} for simple geometries and specific operating conditions. However, these methods are not easily adaptable in aerodynamic design optimization that requires predictions for a variety of geometric and operational scenarios. 

Model reduction is a promising approach for rapid model evaluation in many-query problems such as  design optimization. Reduced-order models (ROMs) are  developed by a variety of methods that can be classified based on their reliance on the governing equations. In projection-based ROMs, a low-dimensional subspace (trial space) is constructed in most cases using the proper orthogonal decomposition (POD) and more recently by nonlinear manifold learning methods such as autoencoder neural networks \cite{holmes:96bk, Berkooz:1993, sirovich:87qam, Lee:2020}. The solution snapshots obtained by solving the high-fidelity model in the training regime are used to identify the trial subspace in a data-driven manner.
A test subspace is then constructed usually by solving an optimization problem in Petrov-Galerkin projection (e.g., the least-squares Petrov-Galerkin method). In Galerkin projection on the other hand, the test subspace is considered to be equal to the trial subspace assuming orthogonal projection. The governing equations with the states approximated by a linear combination of the trial bases are then projected onto the test subspace \cite{rowley:04pdnp,Ghattas:2021,rezaian:2021IJNME}. Therefore, projection-based ROMs require full knowledge of the governing equations, which is in some applications out of reach (e.g. in weather prediction). 

Another class of model reduction approaches developed in the past decade adopt a less intrusive approach by using system identification, inference algorithms and other machine learning methods to identify the low-dimensional system operators from high-fidelity simulation data assuming only the structure of the governing equations is known \cite{Peherstorfer2016, Ma:2011}. 
Projection-based ROMs contribute to the majority of model reduction applications including complex multi-physics and multi-scale engineering problems such as prediction of rocket combustor dynamics and turbulent flow past an aircraft \cite{Huang:2020, Grimberg2021}. Standard projection-based methods do not guarantee theoretical error bounds except under specific conditions (e.g. symmetry of the Jacobian matrix). Different approaches were taken to construct projection-based ROMs with symmetric Jacobian matrices to improve stability.These include resorting to an energy-based inner product \cite{barone:09jcp}, provoking entropy-stability \cite{kalashnikova:11AIAA, Chan:2020}, and construction of an optimal test subspace via the least-squares Petrov-Galerkin approach \cite{Carlberg:2011}.
Most of these methods can be adapted to highly nonlinear systems with the aid of hyper-reduction techniques \cite{Chan:2020, Carlberg:2013}.
For linearized systems, however, it is possible to project the system operators on a lower-dimensional space using a special set of modes that transform the equations into a new coordinate system in which the controllability and observability Gramians of the system are equal and diagonal (i.e., balanced). This method is known as balanced truncation and the resulting reduced-order model satisfies theoretical error bounds \cite{Antoulas:05bk}.

On the other hand, POD is a linear dimensionality reduction method that computes a hierarchy of modes in the sense of an energy-based norm. Therefore, the more energetic modes are preserved through the model reduction and those with lower energy content are discarded. This is in particular problematic in aeroacoustics applications, where low-energy acoustic waves carry important  information. Transformation of the system to balance the Gramians addresses this challenge, as in the new coordinate system,  modes that are preserved are equally highly controllable and observable. Therefore, components like the acoustic waves that are dynamically observable will be preserved despite they do not carry a significant amount of energy.

In  standard balanced truncation, ROM matrices are directly constructed from the FOM operators, which are not easily accessible in many engineering applications. Therefore, a data-driven variant of this method is developed using a system identification method called the eigensystem realization algorithm (ERA) \cite{Juang1985, Juang:94bk, Ma:2011}. ERA uses the impulse response of the system to construct the Hankel matrix and the balanced ROM matrices. Unlike balanced truncation based on empirical Gramians \cite{Willcox:02, rowley:05}, this approach does not require any adjoint system simulations, as it does not directly compute the Gramians or the balancing modes \cite{Ma:2011}. It is important to note that balancing the system Gramians by ERA, and accuracy guarantees thereof, is tied to the sampling properties (i.e., sampling time and frequency) when collecting the impulse response time series. Clearly, the magnitude and duration of the impulsive input for training the ERA ROM depends on the physics of the problem and the numerical simulation setup. When the impulse response is generated and sampled properly, the error bounds given by balanced truncation are also satisfied by the ERA ROM \cite{Antoulas:05bk, Ma:2011}. ERA ROMs have shown to be truly predictive outside of the training regime, both in the sense of future state prediction and in response to inputs that are not included in the training data \cite{rezaian:2022}.

\section{Objectives and Outline of Present Work}
In this work we develop a methodology based on ERA to generate data-driven ROMs using balancing transformation. As mentioned earlier, for the prediction of the acoustic response,  the importance of low-amplitude acoustic waves renders POD-based ROMs inaccurate. We demonstrate the predictive performance of ERA ROMs for an airfoil subject to periodic gust in a subsonic flow. We first train the ERA ROMs using the response of the 
nonlinear full-order model (FOM) to Gaussian input. The nonlinear FOM solves the nonlinear Euler equations with a cell-centered finite-volume approach. The periodic gust is modeled by perturbing the entire far-field boundary points, which translates into a multi-input system with many input channels. 

\begin{figure}
  \centering
  \begin{minipage}[a]{1.0\textwidth}
    \includegraphics[trim=4 -0.1 4 4, clip, width=\textwidth]{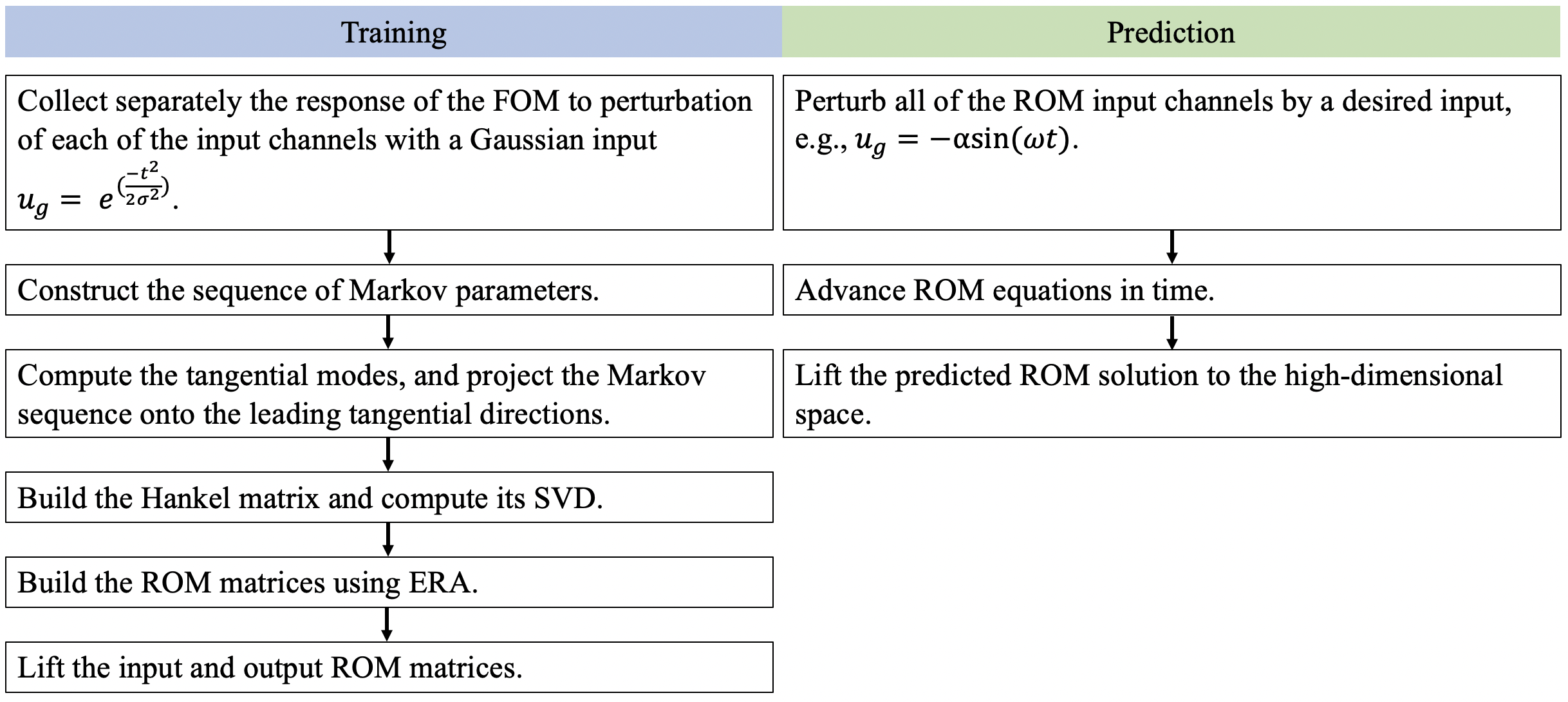}
  \end{minipage}
  \centering
  \caption{Training and prediction by the ERA ROM with tangential interpolation.} 
   \label{flowchart}
\end{figure}

To train the ERA ROMs, each of the input channels needs to be excited separately, and the collection of system responses to separately activated channels is used to build the Hankel matrix. Considering full-state output, therefore, we have a multi-input multi-output (MIMO) system with a total of 604 input and 40000 output channels. 
To reduce the size of the Hankel matrix, we use tangential interpolation to project the inputs and outputs of the system onto the leading tangential directions before computing the singular values and singular vectors of the Hankel matrix \cite{Kramer:2016, rezaian:2022}. The flow chart of the training and prediction process is shown in Figure~\ref{flowchart}. Note that the purpose of this work is to build purely predictive ROMs, in the sense that during the prediction, the system is perturbed by a different input from the training. 

Despite the computational savings at the ROM level by the tangential interpolation, running the FOM many times to separately activate each of the input channels, collecting a large number of time series to build the Hankel matrix and computing SVD of the training snapshots to compute the tangential directions is a computational bottleneck. To alleviate this limitation, we propose a multi-fidelity gappy POD approach. Flow field reconstruction based on sparse sensor measurements has been approached by the gappy POD method in the past decades \cite{Willcox:06}. Gappy POD was used with QR factorization for optimal sensor placement for maximum reconstruction accuracy \cite{Drmac:2016}. Here, we apply gappy POD for actuator placement. We use the pivot locations obtained by QR factorization of the POD modes to identify the input channels that have the most impact on the airfoil response reconstruction. To reduce the training cost of gappy POD and bypass the need to query a high-resolution FOM, we use the snapshots from a lower-fidelity FOM with a coarser mesh to train gappy POD. Instead of activating the entire far-field cells one by one, only the input channels identified by gappy POD are activated in the high-resolution FOM to generate the training data for the ERA ROMs.

The results show that the ERA ROMs with the tangential interpolation method accurately predict the acoustic response of the airfoil in a truly predictive setting when the perturbation input is not seen by the ROM during the training. In addition, two orders of magnitude reduction is achieved in the online computation cost compared to the FOM based on the nonlinear Euler equations. To further improve the computational savings, we create a second class of ROMs based on the training snapshots computed by the linearized Euler equations, which reduces the offline cost of constructing the ERA ROM matrices. The near real-time computations enabled by the ERA ROM make it a powerful substitute for the FOM in order to accelerate prediction of the acoustic response of the turbomachinery and aircraft components for predictive control and aerodynamic design optimization.

The paper is organized as follows. In section~\ref{sec:hfm} the fundamentals of the nonlinear and linearized high-fidelity solvers are explained. The theory behind balanced truncation and the eigensystem realization algorithm are reviewed in section~\ref{sec:rom}. The procedure for reducing the dimension of the Hankel matrix via the tangential interpolation method is outlined in section~\ref{tanint}. The proposed multi-fidelity gappy POD approach to identify the critical input channels is explained in section~\ref{gappy}. In section~\ref{sec:result} the numerical setup for simulation of the subsonic flow over the airfoil using the nonlinear and linearized Euler equations is described in addition to the details for construction of the ERA ROMs for prediction of the acoustic response of the airfoil. The results are also presented and discussed in this section. The final remarks and perspectives for future extension of this research are outlined in section~\ref{sec:conclusion}.

\section{Full-order Model}
\label{sec:hfm}
The high-fidelity solution is obtained by solving the Euler equations with a finite volume approach. Consider the nonlinear Euler equations in two dimensions,
\be
    \pd{\mathbf{q}}{t} + \pd{\mathbf{f}}{x} + \pd{\mathbf{g}}{y} = 0,
\ee
where $\mathbf{q}= \left[
\rho \ \ \rho u \ \ \rho v \ \ e \right]^T$ is the vector of conservative variables, $\rho$ is the density, $u$ and $v$ are the horizontal and vertical components of velocity, and $e$ is the total energy. Here, $\mathbf{f}$ and $\mathbf{g}$ are the flux vectors,
\be
\mathbf{f} = \begin{bmatrix}
    \rho u \\
    \rho u^2 + p \\
    \rho u v \\
    (e + p) u
\end{bmatrix},
\qquad
\mathbf{g} = \begin{bmatrix}
    \rho v \\
    \rho u v \\
    \rho v^2 + p \\
    (e + p) v
\end{bmatrix},
\ee
and $p$ is obtained by the perfect gas equation of state as,
$
p = \left(\gamma - 1 \right) \left[e - \frac{\rho}{2} (u^2 + v^2) \right],
$
where, $\gamma = 1.4$ is the ratio of specific heats. The integral form of the Euler equations is written as,
\be
\frac{d}{dt} \iint_{\Omega} \mathbf{q} dx dy + \int_{\partial \Omega} \mathbf{f} dy + \int_{\partial \Omega} \mathbf{g} dx = 0.
\ee
A cell-centered finite-volume approach is used to solve the Euler equations and fluxes are computed by the Roe scheme with the second-order reconstruction. The semi-discretized equations are integrated in time using the second-order Runge-Kutta (R-K) method. The chain rule is applied to implement the equations in curvilinear coordinates,
\be
\pd{\mathbf{q}}{t} + \pd{\xi}{x} \pd{\mathbf{f}}{\xi} + \pd{\eta}{x} \pd{\mathbf{f}}{\eta} + \pd{\xi}{y} \pd{\mathbf{g}}{\xi} + \pd{\eta}{y} \pd{\mathbf{g}}{\eta} = 0,
\ee
where, $\xi = \xi(x, y)$ and $\eta = \eta(x, y)$ are the generalized curvilinear coordinates.

We use the flow field decomposition $\mathbf{q}(\mathbf{x},t) = \Bar{\mathbf{q}}(\mathbf{x}) + \mathbf{q}'(\mathbf{x},t)$, and linearize Euler equations about the steady-state solution $\Bar{\mathbf{q}}(\mathbf{x})$ computed by solving the nonlinear Euler equations, to generate the linear time invariant (LTI) system of equations of the form,
\be
\label{linearODE}
\dot{\mathbf{q}}' = \mathbf{J} \mathbf{q}' + \mathbf{b},
\ee
where, $\mathbf{q}' \in \mathbb{R}^n$ is the perturbation solution, $\dot{\mathbf{q}}' = \frac{d \mathbf{q}'}{dt}$, $\mathbf{J}=\left[- \left(\pd{(\mathbf{f}+\mathbf{g})}{\mathbf{q}}\right)_C + \left(\pd{(\mathbf{f} + \mathbf{g})}{\mathbf{q}}\right)_L - \left(\pd{(\mathbf{f} + \mathbf{g})}{\mathbf{q}}\right)_R\right]_{\mathbf{q} = \Bar{\mathbf{q}}}$ is the Jacobian matrix, $\mathbf{b}$ is a vector containing boundary fluxes multiplied by the boundary perturbation variables. Here, $\mathbf{J} \in \mathbb{R}^{n \times n}$ and $\mathbf{b} \in \mathbb{R}^{n}$ are evaluated by the steady-state solution and subscripts C, L and R denote the current, left and right neighboring cells, respectively. Therefore, for each variable we have,
\be
\label{b_mat}
\begin{aligned}
& \mathbf{B}_{(1,j)} = \left[\left(\pd{\mathbf{f}}{\mathbf{q}}\right)_L\right]_{\mathbf{q}_{(1,j)} = \Bar{\mathbf{q}}_{(1,j)}} \mathbf{Q}'_{(1,j)}, \\
& \mathbf{B}_{(n_x,j)} = -\left[\left(\pd{\mathbf{f}}{\mathbf{q}}\right)_R\right]_{\mathbf{q}_{(n_x,j)} = \Bar{\mathbf{q}}_{(n_x,j)}} \mathbf{Q}'_{(n_x,j)}, \\
& \mathbf{B}_{(i,1)} = \left[\left(\pd{\mathbf{g}}{\mathbf{q}}\right)_L\right]_{\mathbf{q}_{(i,1)} = \Bar{\mathbf{q}}_{(i,1)}} \mathbf{Q}'_{(i,1)}, \\
& \mathbf{B}_{(i,n_y)} = -\left[\left(\pd{\mathbf{g}}{\mathbf{q}}\right)_R\right]_{\mathbf{q}_{(i,n_y)} = \Bar{\mathbf{q}}_{(i,n_y)}} \mathbf{Q}'_{(i,n_y)}, 
\end{aligned}
\ee
where, $i=1, \dots n_x$, and $j = 1,\dots, n_y$. The boundary input vector $\mathbf{b}$ is then obtained by reshaping $\mathbf{B} \in \mathbb{R}^{n_x \times n_y \times n_v}$, such that $\mathbf{B} \in \mathbb{R}^{n}$, where $n=n_x n_y n_v$, and $n_v$ is the number of variables. The only nonzero components of $\mathbf{B}$ are those shown in equation~\ref{b_mat}. Note that in equation~\ref{b_mat}, the flux derivative terms also contain the Roe flux dissipation term. More details on the linearized solver matrices are shown in \nameref{appA}.
Here, $\mathbf{Q}' \in \mathbb{R}^{n_x \times n_y \times n_v}$ contains the fluctuating components of the conservative variables. Note that to evaluate $\mathbf{Q}'$, the conservative variables are linearized about the steady-state solution.

Equation~\ref{linearODE} is integrated in time using the second-order R-K scheme, with $\mathbf{q}'(\mathbf{x},0) = 0$, and subject to far-field boundary conditions $u_{\infty}'=u_g(t)$, $v_{\infty}'=v_g(t)$, and $p'=\rho'=0$, where $u_g$ and $v_g$ are gust perturbations. Further details about the two high-fidelity solvers are given in section~\ref{sec:result}. 

\section{Reduced-order Model}
\label{sec:rom}
The eigensystem realization algorithm (ERA) is a data-driven system identification method developed based on the minimal realization theory \citep{Juang1985} that was later shown to be related to balanced truncation as a model reduction method \cite{Ma:2011}. 
The original balanced truncation  is an intrusive method that requires full access to the FOM operators. Therefore, making this connection enabled data-driven implementation of this powerful model reduction technique without the need to access the FOM operators and any adjoint system simulations. The latter also makes it possible to apply this method to experiments, where the adjoint system response is not available. In this section we introduce the basic form of the balanced truncation method followed by a description of the ERA method and the tangential interpolation approach to reduce the computational cost associated with the construction of ROM matrices in systems with multiple input and output channels. We also explain the proposed actuator selection framework to address the computational bottleneck for training ERA ROMs in systems with many input channels.

Two ROMs are constructed in this work: one based on the snapshots from the nonlinear Euler solver and another based on the snapshots computed by the linearized Euler equations. 

\subsection{Balanced Truncation}
Consider a linear time-invariant (LTI) system in state-space form,
\be
\label{FOMss}
\begin{aligned}
 \dot{\mathbf{x}}(t) = \mathbf{A} \mathbf{x}(t) + \mathbf{B} u(t)\ \ ; \ \
 \mathbf{y}(t) = \mathbf{C} \mathbf{x}(t) + \mathbf{D} u(t),
\end{aligned}
\ee
where, $\mathbf{x}(t) \in \mathbb{R}^{n}$ is the FOM state, $\dot{\mathbf{x}}(t)  = \frac{d \mathbf{x}(t)}{dt}$, $u(t) \in \mathbb{R}^{p}$ is the input, and $\mathbf{y}(t) \in \mathbb{R}^{q}$ is the system output ($q=n$ for full-state output). Without loss of generality, we remove the feedthrough term (i.e., $\mathbf{D} = \mathbf{0}$) in what follows. Here, $\mathbf{A} \in \mathbb{R}^{n \times n}$, $\mathbf{B} \in \mathbb{R}^{n \times p}$,  and $\mathbf{C} \in \mathbb{R}^{q \times n}$ are constant matrices. 
The idea is to find a transformation matrix that when applied to the LTI system, results in a change of coordinates that produces balanced (i.e., equal and diagonal) reachability and observability Gramians \cite{Antoulas:05bk, Willcox:02}. Reachability and observability Gramians are Hermitian matrices that are positive definite if and only if the system is reachable (i.e., $\mathcal{R}(\mathscr{P}) = n$) and observable (i.e., $\mathcal{R}(\mathscr{O}) = n$), respectively, where, $\mathcal{R}$ denotes matrix rank, $\mathscr{P}$ is the reachability matrix,
\be
\mathscr{P} = \begin{bmatrix}
    \mathbf{B} & \mathbf{A} \mathbf{B} & \dots & \mathbf{A}^{n-1} \mathbf{B} 
    \end{bmatrix}
\ee
and $\mathscr{O}$ is the observability matrix,
\be
\mathscr{O} = \begin{bmatrix}
        \mathbf{C}  \\ \mathbf{C} \mathbf{A} \\ \vdots \\ \mathbf{C}  \mathbf{A}^{n-1}
    \end{bmatrix}.
\ee

We define the reachability Gramian ($\mathcal{W}_p$) as a measure of the amount of energy required to drive the system from zero to a desired state \cite{Antoulas:05bk},
\be
\label{CWp}
\mathcal{W}_p = \int_0^{\infty} e^{\mathbf{A}t} \mathbf{B} \mathbf{B}^{*} e^{\mathbf{A}^* t} dt,
\ee
for continuous-time systems and $\mathcal{W}_p = \mathscr{P} \mathscr{P}^{*}$ for discrete-time systems, where, ($^*$) denotes complex conjugate transpose. This Gramian is obtained as the solution of a Lyaounov equation,
\be
\label{ctrbLyap}
\mathbf{A}  \mathcal{W}_p + \mathcal{W}_p \mathbf{A} ^{*} + \mathbf{B} \mathbf{B}^{*} = 0.
\ee
The dual of the reachability Gramian is the observability Gramian ($\mathcal{W}_o$) that measures the impact of the state on the system output, when $u(t) = 0$. For continuous-time systems we have,
\be
\label{CWo}
\mathcal{W}_o = \int_0^{\infty} e^{\mathbf{A}^* t} \mathbf{C}^* \mathbf{C} e^{\mathbf{A} t} dt,
\ee
which reduces to $\mathcal{W}_o = \mathscr{O}^{*} \mathscr{O}$ in discrete-time systems. Similarly, the observability Gramian is the solution of the following Lyapunov equation,
\be
\label{bsvLyap}
\mathbf{A}^{*} \mathcal{W}_o + \mathcal{W}_o \mathbf{A} + \mathbf{C}^{*} \mathbf{C} = 0.
\ee
Equations (\ref{CWp}) and (\ref{CWo}) are defined only for stable systems. Therefore, the standard balanced truncation based on analytical Gramians is not applicable to unstable systems. 

We seek a transformation matrix $\mathbf{T}$ that balances the reachability and observability Gramians as,
\be
\mathbf{T}^{-1} \mathcal{W}_c \mathbf{T}^{-*} = \mathbf{T}^{*} \mathcal{W}_o \mathbf{T} = \boldsymbol{\Sigma}, 
\ee
where, $\boldsymbol{\Sigma}$ is a diagonal matrix. Its diagonal entries are known as Hankel singular values, that are invariant under coordinate transformation and represent the input-output energy of the system. Subsequently, the balancing transformation $\mathbf{T}_r$ is computed by $\mathbf{T}_r = \mathbf{U} \mathbf{W}_r \boldsymbol{\Sigma}_r^{-1/2}$, where, subscript $r$ shows that the Hankel modes corresponding to the smaller singular values are truncated. $\mathbf{U}$ is the Cholesky factor of the reachability Gramian $\mathcal{W}_p = \mathbf{U} \mathbf{U}^*$, and $\mathbf{W}$ is obtained by SVD of the product $
\mathbf{U}^* \mathbf{L} = \mathbf{W} \boldsymbol{\Sigma} \mathbf{V}^*$, where $\mathbf{L}$ is computed by Cholesky factorization of the observability Gramian $\mathcal{W}_o = \mathbf{L} \mathbf{L}^*$. Column vectors of $\mathbf{T}_r$, and row vectors of the inverse transformation $\mathbf{T}_r^{-1} = \boldsymbol{\Sigma}_r^{-1/2} \mathbf{V}_r^{*} \mathbf{L}^{*}$ \citep{Antoulas:05bk} are known as direct and adjoint modes, respectively. Given the transformation matrix $\mathbf{T}$, balanced ROMs are obtained as,
\be
\begin{aligned}
\label{BROM}
\dot{\mathbf{x}}_r(t) = \mathbf{A}_r \mathbf{x}_r(t) + \mathbf{B}_r u(t)\ \ ; \ \
 \mathbf{y}(t) = \mathbf{C}_r \mathbf{x}_r(t),
\end{aligned}
\ee
where, 
\be
\label{BROMmatrices}
\begin{aligned}
& \mathbf{A}_r = \mathbf{T}_r^{-1} \mathbf{A} \mathbf{T}_r,\ \
& \mathbf{B}_r = \mathbf{T}_r^{-1} \mathbf{B},\ \
& \mathbf{C}_r = \mathbf{C} \mathbf{T}_r.
\end{aligned}
\ee
Balanced ROMs are known for their accuracy guarantees. Regardless of the method used to create ROMs, the error they generate is bounded below by,
\be
\norm{\mathbf{G} - \mathbf{G}_r}_{\infty} > \sigma_{r+1}.
\ee
However, prediction error in balanced ROMs is also bounded above as,
\be
\norm{\mathbf{G} - \mathbf{G}_r}_{\infty} < 2 \sum_{i=r+1}^n \sigma_i,
\ee
where, $\mathbf{G} = \mathbf{C} (s \mathbf{E} - \mathbf{A})^{-1} \mathbf{B}$ and $\mathbf{G}_r = \mathbf{C}_r (s \mathbf{E}_r - \mathbf{A}_r)^{-1} \mathbf{B}_r$ are the transfer functions of FOM and ROM, respectively \citep{Glover:1984, Hinrichsen:1990, Willcox:02}. Here, $\mathbf{E} = \mathbf{I} \in \mathbb{R}^{n \times n}$, $\mathbf{E}_r = \mathbf{I}_r \in \mathbb{R}^{r \times r}$, and $s \in \mathbb{C}$. It is important to note that BT is equivalent to the standard POD-Galerkin projection when the observability Gramian is used as the inner product \cite{rowley:05, Carlberg:15}.

\subsection{The Eigensystem Realization Algorithm}\label{era}
Computing system Gramians by solving the Lyapunov equations (\ref{ctrbLyap}) and (\ref{bsvLyap}) is expensive and requires access to the FOM operators. Using the method of snapshots \citep{sirovich:87qam}, it is possible to bypass direct computation of the Gramians and evaluate the Hankel matrix as $ \mathbf{H} = \mathscr{O}^{*} \mathscr{P}$. This is the idea behind approximate balanced truncation (a.k.a, balanced POD) \citep{Willcox:02, rowley:05}. Approximate balanced truncation however, requires access to the FOM operators and adjoint system simulations, that are not always available. ERA on the other hand, balances the system Gramians in a purely data-driven setting, while avoiding FOM operators, adjoint simulations and direct computation of the Gramians and the transformation matrices. The flexibility offered by this system identification approach therefore, has pushed the boundaries of balanced truncation in engineering applications.

For a discrete-time linear system,
\be
\label{dlti}
\begin{aligned}
 \mathbf{x}_{k+1} = \mathbf{A} \mathbf{x}_k +  \mathbf{B} u_k \ \ ; \ \
 \mathbf{y}_k = \mathbf{C} \mathbf{x}_k,
\end{aligned}
\ee
the Hankel matrix takes the following form,
\be
\begin{aligned}
\mathbf{H} & = \begin{bmatrix}
                \mathbf{y}_1 & \mathbf{y}_2 & \dots & \mathbf{y}_{m_p} \\
                \mathbf{y}_2 & \mathbf{y}_3 & \dots  & \mathbf{y}_{m_p + 1}  \\
                 \vdots & \dots & \ddots & \vdots \\
                 \mathbf{y}_{m_o} & \mathbf{y}_{m_o + 1}  & \dots & \mathbf{y}_{m_o + m_p - 1}   \\
\end{bmatrix}  
 = \begin{bmatrix}
                \mathbf{C}  \mathbf{B} & \mathbf{C} \mathbf{A}  \mathbf{B} & \dots & \mathbf{C} \mathbf{A}^{m_p - 1}  \mathbf{B} \\
                \mathbf{C} \mathbf{A}  \mathbf{B} & \mathbf{C} \mathbf{A}^2  \mathbf{B} & \dots & \mathbf{C} \mathbf{A}^{m_p}  \mathbf{B} \\
               \vdots & \dots & \ddots & \vdots \\
                \mathbf{C} \mathbf{A} ^{m_o - 1} \mathbf{B} & \mathbf{C} \mathbf{A}^{m_o}  \mathbf{B} & \dots & \mathbf{C} \mathbf{A}^{m_p + m_o - 2}  \mathbf{B} \\
\end{bmatrix},
\end{aligned}
\ee
where, the terms of the sequence,
\be
\label{markov}
\mathbf{y}_i = \mathbf{C} \mathbf{A}^i \mathbf{B}, \qquad i = 0, \dots, m_p + m_o - 2,
\ee
are known as Markov parameters \citep{Ma:2011}. For a linear system, Markov parameters are equivalent to the unit impulse response of the system \citep{Antoulas:05bk}. The SVD of the Hankel matrix gives $\mathbf{H} = \mathbf{U} \boldsymbol{\Sigma} \mathbf{V}^{*}$, where we retain only the singular values that capture most of the input-output energy. Advancing the sequence (\ref{markov}) one step in time yields the shifted Hankel matrix \citep{Ma:2011},
\be
\begin{aligned}
\mathbf{H}^{\prime} & = \begin{bmatrix}
                \mathbf{y}_2 & \mathbf{y}_3 & \dots & \mathbf{y}_{m_p + 1} \\
                \mathbf{y}_3 & \mathbf{y}_4 & \dots  & \mathbf{y}_{m_p + 2}  \\
                 \vdots & \dots & \ddots & \vdots \\
                 \mathbf{y}_{m_o + 1} & \mathbf{y}_{m_o + 2}  & \dots & \mathbf{y}_{m_o + m_p}   \\
\end{bmatrix} 
 = \begin{bmatrix}
                \mathbf{C}  \mathbf{A}  \mathbf{B} & \mathbf{C} \mathbf{A}^2  \mathbf{B} & \dots & \mathbf{C} \mathbf{A}^{m_p}  \mathbf{B} \\
                \mathbf{C} \mathbf{A}^2  \mathbf{B} & \mathbf{C} \mathbf{A}^3  \mathbf{B} & \dots & \mathbf{C} \mathbf{A}^{m_p + 1}  \mathbf{B} \\
               \vdots & \dots & \ddots & \vdots \\
                \mathbf{C} \mathbf{A} ^{m_o} \mathbf{B} & \mathbf{C} \mathbf{A}^{m_o + 1}  \mathbf{B} & \dots & \mathbf{C} \mathbf{A}^{m_p + m_o - 1}  \mathbf{B} \\
\end{bmatrix},
\end{aligned}
\ee
which is the element that rules out the need for adjoint system simulations in ERA. 
Note that the direct modes,
\be
\label{discdirmodes}
\mathbf{T}_r = \mathscr{P} \mathbf{V}_r \boldsymbol{\Sigma}_r^{-1/2},
\ee
and the adjoint modes,
\be
\label{discadjmodes}
\mathbf{T}_r^{-1} = \boldsymbol{\Sigma}_r^{-1/2} \mathbf{U}_r^{*} \mathscr{O}^{*},
\ee
can be computed by the reachability ($\mathscr{P}$) and observability ($\mathscr{O}$) matrices of the discrete-time system, where, $\mathbf{U}$ and $\mathbf{V}$ are obtained by SVD of the Hankel matrix and subscript $r$ denotes the retained singular values and singular vectors. Combining (\ref{BROMmatrices}), (\ref{discdirmodes}) and (\ref{discadjmodes}), balanced ROM matrices in ERA are obtained as \citep{Ma:2011, Brunton:19bk},
\be
\begin{aligned}
& \mathbf{A}_r = \boldsymbol{\Sigma}_r^{-1/2} \mathbf{U}_r^{*} \mathbf{H}^{\prime}\mathbf{V}_r \boldsymbol{\Sigma}_r^{-1/2},\ \
& \mathbf{B}_r = \boldsymbol{\Sigma}_r^{1/2} \mathbf{V}_r^{*} \begin{bmatrix}
\mathbf{I}_p & \mathbf{0} \\
\mathbf{0} & \mathbf{0} 
\end{bmatrix},\ \
& \mathbf{C}_r = \begin{bmatrix}
\mathbf{I}_q & \mathbf{0} \\
\mathbf{0} & \mathbf{0} 
\end{bmatrix} \mathbf{U}_r  \boldsymbol{\Sigma}_r^{1/2},
\end{aligned}
\ee
without the need for adjoint simulations to compute the inverse transformation matrix (\ref{discadjmodes}). Here, $\mathbf{I}_p \in \mathbb{R}^{p \times p}$ and $\mathbf{I}_q \in \mathbb{R}^{q \times q}$ are identity matrices, $p$ is the number of inputs and $q$ is the number of outputs.

\subsection{Tangential Interpolation}\label{tanint}
Systems with lightly-damped impulse response, multiple input channels, and full-state output require excessive storage and a high computational cost to take the SVD of a high-dimensional Hankel matrix. Tangential interpolation is employed as a way to reduce the offline cost of ERA in multi-input multi-output (MIMO) systems by projecting the impulse response onto the leading left and right tangential directions \citep{Kramer:2016}. 

The impulse response snapshots are first assembled in matrix $\mathbf{Q}_L$ as,
\be
\mathbf{Q}_L = \begin{bmatrix}
\mathbf{y}_1 & \mathbf{y}_2 & \dots & \mathbf{y}_{m_o + m_p -1}
\end{bmatrix}.
\ee
To identify the left tangential directions, we seek the solution to the optimization problem,
\be
\mathbf{P}_1 = argmin_{rank(\tilde{\mathbf{P}}_1) = l_1} \norm{\tilde{\mathbf{P}}_1 \mathbf{Q}_L - \mathbf{Q}_L}^2_F, 
\ee
where, $\mathbf{P}_1 = \mathbf{W}_1 \mathbf{W}_1^*$, and $\mathbf{W}_1$ contains the first $l_1$ columns of $\mathbf{U}_L$ given by SVD of the impulse response snapshots $\mathbf{Q}_L = \mathbf{U}_L \boldsymbol{\Sigma}_L \mathbf{V}_L^*$. Similarly, to obtain the right tangential directions we form,
$
\mathbf{Q}_R = [
\mathbf{y}_1^T \ \
\mathbf{y}_2^T \ \
\vdots \ \
\mathbf{y}_{m_o + m_p -1}^T
]^T,
$
and solve the optimization problem,
\be
\mathbf{P}_2 = argmin_{rank(\tilde{\mathbf{P}}_2) = l_2} \norm{\mathbf{Q}_R \tilde{\mathbf{P}}_2 - \mathbf{Q}_R}^2_F,
\ee
that is equivalent to SVD of the sequence of Markov parameters $\mathbf{Q}_R = \mathbf{U}_R \boldsymbol{\Sigma}_R \mathbf{V}_R^*$, where, $\mathbf{P}_2 = \mathbf{W}_2 \mathbf{W}_2^*$ and $\mathbf{W}_2 = \mathbf{V}_R (:, 1:l_2)$ \citep{Kramer:2016}. The new sequence of Markov parameters for ERA is obtained by projection of the original impulse response sequence $\hat{\mathbf{y}}_i = \mathbf{W}_1^* \mathbf{y}_i \mathbf{W}_2$. However, the balanced ROM trained with this sequence has $l_1$ outputs and $l_2$ inputs. The original input-output dimensions are recovered by lifting the system back to the high-dimensional input-output space as $\mathbf{A}_r = \hat{\mathbf{A}}$, $\mathbf{B}_r = \hat{\mathbf{B}} \mathbf{W}_2^*$, and $\mathbf{C}_r = \mathbf{W}_1 \hat{\mathbf{C}}$, where, $\hat{\mathbf{A}}$, $\hat{\mathbf{B}}$, and $\hat{\mathbf{C}}$ are the balanced ROM matrices obtained by applying ERA to the projected impulse response sequence $\hat{\mathbf{y}}_i$ \citep{Kramer:2016}.

\subsection{Markov Sequence Reconstruction using Low-fidelity Gappy POD}
\label{gappy}
ERA ROMs are constructed from a Hankel matrix that contains the time series generated from the response of the system when each of the input channels are separately activated by an impulsive input or a Gaussian-shaped input. This requires simulating the FOM $p$ times, where $p$ is the total number of input channels. Therefore, generating the training snapshots becomes infeasible in systems with many input channels (actuators). Acoustic response prediction under freestream perturbation is a perfect example of this class of systems, where hundreds of points are perturbed along the far-field boundary. 
To address this issue we propose using the time series of the impulse responses corresponding to a few input channels to reconstruct the entire sequence of Markov parameters. The idea is inspired by flow field reconstruction from sparse sensor measurements using gappy POD \citep{Willcox:06, Manohar:2018}, except for that here we are dealing with sparse actuator responses instead of sensor measurements. Consider the SVD of the training data matrix $\mathbf{Q} = \mathbf{U} \boldsymbol{\Sigma} \mathbf{V}^{*}$, where $\mathbf{Q} \in \mathbb{R}^{p \times n_t}$, and $n_t = m_p + m_o - 2$ is the total number of samples required to build the Hankel matrix. Note that $\mathbf{Q}$ contains the Markov sequence collected at one probe location in the flow field. This can be a desired sensor location or a probe located in a dynamically rich area of the computational domain. For example, in the case of the flow over an airfoil we choose a probe close to the surface of the airfoil that is affected by the acoustic response of the airfoil.

Let us define an observation matrix $\mathbf{S} = \mathbf{P} \mathbf{Q}$, where $\mathbf{S} \in \mathbb{R}^{p_s \times n_t}$, $p_s$ is the number of input channels for which we afford to compute the full sequence of Markov parameters, and $\mathbf{P} \in \mathbb{R}^{p_s \times p}$ is a sampling matrix that identifies the location of these channels. Therefore, the observation matrix can be approximated using a linear combination of the dominant POD modes of the training snapshots matrix as $\mathbf{S}  \approx \mathbf{P} \boldsymbol{\Phi} \mathbf{a}$. Here, $\boldsymbol{\Phi}$ contains the first $r$ column vectors of $\mathbf{U}$, and $r$ is identified using a threshold based on the cumulative energy of the POD modes,
\be
E_k = \frac{\sum_{i=1}^k \sigma_i}{\sum_{j=1}^K \sigma_j},
\ee
where, $\sigma_i$ is the i\textsuperscript{th} singular value of $\mathbf{Q}$, and $K$ is the total number of singular values.
The generalized coordinates are obtained by solving the least-squares optimization problem,
\be
\mathbf{a} = min_{\mathbf{a}^{*}} \norm{\mathbf{S} - \mathbf{P} \boldsymbol{\Phi} \mathbf{a}^{*}}_2^2.
\ee
The solution to this optimization problem is $\mathbf{a} = (\mathbf{P} \boldsymbol{\Phi})^{+} \mathbf{S}$, where the superscript $+$ denotes the Moore-Penrose psuedo-inverse. The full sequence of Markov parameters for the entire set of input channels is then reconstructed as $\hat{\mathbf{Q}} = \boldsymbol{\Phi} \mathbf{a}$. For sensor placement applications, Erichson et al. \cite{Erichson2019} proposed an alternative approach, where a decoder neural network is used to reconstruct the flow field from sparse sensor measurements. This approach can be easily adapted here for reconstruction of the sequence of Markov parameters from a few impulse response time series. 

The goal here is to identify the input channels that result in the most accurate POD coefficients for reconstruction of the Markov sequence. 
In the context of sensor placement, Willcox \cite{Willcox:06} proposed placing sensors at locations that minimize the condition number of matrix $\boldsymbol{\Psi}^T \boldsymbol{\Psi}$ using a greedy approach, where $\boldsymbol{\Psi} = \mathbf{P} \boldsymbol{\Phi}$.
Here, we take the approach employed by Manohar et al. \cite{Manohar:2018} using QR factorization pivots of the basis modes for sensor/actuator placement \citep{Drmac:2016}. To maximize the accuracy of Markov sequence reconstruction, considering the over-determined case when $p_s > r$, we compute the QR factorization of $\boldsymbol{\Phi} \boldsymbol{\Phi}^T$ with column pivoting as $(\boldsymbol{\Phi} \boldsymbol{\Phi}^T) \tilde{\mathbf{P}} = \tilde{\mathbf{Q}} \tilde{\mathbf{R}}$, where $\tilde{\mathbf{Q}} \in \mathbb{R}^{p \times p}$ is a unitary matrix, $\tilde{\mathbf{P}} \in \mathbb{R}^{p \times p}$ is a column permutation matrix, and $\tilde{\mathbf{R}} \in \mathbb{R}^{p \times p}$ is an upper triangular matrix.
We take the pivot locations given by the first $p_s$ columns of $\tilde{\mathbf{P}}$ (i.e., $\mathbf{P}^T = \tilde{\mathbf{P}}(:,1:p_s)$) as the input channels for which we query the FOM to compute the sequence of Markov parameters. The gappy POD method is then used to reconstruct the impulse response for the entire (p-dimensional) set of input channels. The choice of the over-determined problem instead of setting $p_s = r$ follows the study of Peherstorfer et al. \cite{Peherstorfer2020} that demonstrated the impact of over-sampling on the stability of the gappy POD approach.
We use the \verb#qr# function in MATLAB for computing the QR factorization with column pivoting. 

The training data in the proposed gappy POD approach for Markov sequence reconstruction includes the impulse response (alternatively, the Gaussian input response) corresponding to each of the input channels. To reduce the training cost, 
we propose a multi-fidelity approach for training the ERA ROMs. Since computing the full Markov sequence using the high-fidelity model is not computationally affordable, we use the Markov sequence obtained by solving the Euler equations on a coarser grid as the training data for the gappy POD method.
The data from this lower-fidelity model is used in the QR factorization to obtain the optimal actuator locations for maximum reconstruction accuracy. Next, we query the high-fidelity model to compute the impulse response only for the channels identified in the first step with the QR factorization. The resulting Markov sequence generated with $p_s \ll p$ channels is then used to build the ERA ROMs. Since the high-fidelity system is perturbed through the most effective channels, the multi-input ERA ROM is expected to approximately model the behavior of the original system when all $p$ input channels are perturbed.

\section{Numerical Experiments}
\label{sec:result}
\begin{figure}[h!]
  \centering
  \begin{minipage}[a]{0.6\textwidth}
    \includegraphics[trim=4 4 4 4, clip, width=\textwidth]{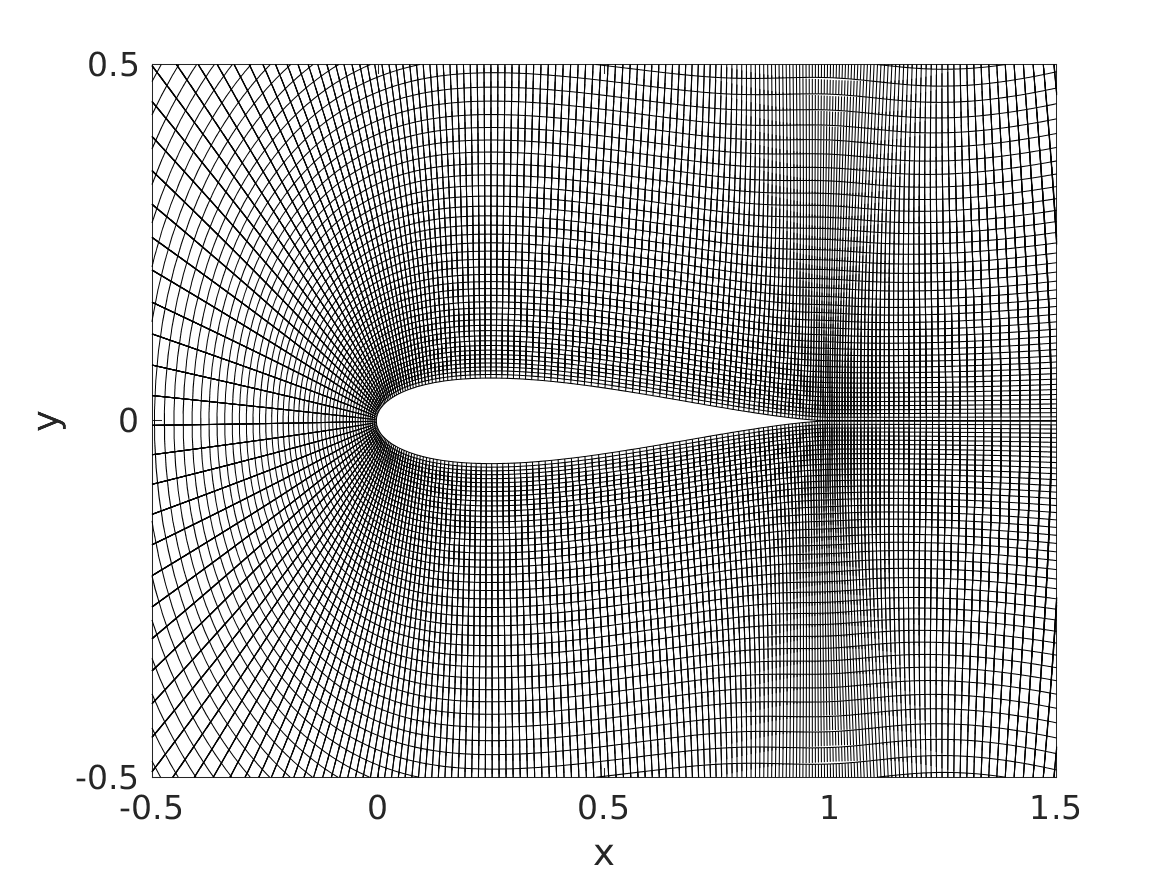}
  \end{minipage}
  \centering
  \caption{Zoomed-in view of the C-type grid around the airfoil.} 
   \label{grid}
\end{figure}

The two-dimensional Euler equations are solved to compute the flow field over a NACA0021 airfoil with the angle of attack $\alpha=0^{\circ}$ subject to a sinusoidal gust in a $401 \times 101$ computational domain with a C-type grid shown in figure~\ref{grid}. The grid extends more than 4 chord lengths in each direction. Far-field boundary values are set to the freestream conditions. Ghost cells are used along the far-field boundary and to apply the slip boundary condition at the surface of the airfoil. Pressure and density at the surface of the airfoil are computed by extrapolation of the interior boundary cell values.

Subsonic flow is considered with the freestream Mach number $M_{\infty}=0.5$, where the steady-state solution is obtained with the freestream quantities $u_{\infty} = M_{\infty}$, $v_{\infty}=0$, $\rho_{\infty}=1$ and $p_{\infty}=\frac{1}{\gamma}$. This steady-state solution is used in two high-fidelity solvers: in the linearized solver, the Jacobian and boundary flux terms are computed by the steady-state solution, and in the nonlinear solver the simulation is initialized with the steady-state solution. Here, the freestream is perturbed by,
\be
\label{ug}
u_g= - \frac{\epsilon w}{\sqrt{2}} sin \left(k \mathbf{x} - \omega t \right);
\ \ v_g= \frac{\epsilon w}{\sqrt{2}} sin(k \mathbf{x} - \omega t),
\ee
where, $k=1$ is the reduced frequency, $\epsilon=0.02$, $w=\sqrt{u_{\infty}^2+v_{\infty}^2}$, $\omega=\frac{2kw}{c}$ is the angular frequency, and $c$ is the chord length. The far-field boundary conditions are defined as $u_{f} = u_{\infty} + u_g$, $v_{f} = v_{\infty} + v_g$, $p_{f} = p_{\infty}$, and $\rho_{f} = \rho_{\infty}$ in the nonlinear solver. For the linearized solver, the far-field conditions are centered about the steady-state solution, therefore, we have $u'_{f} = u_g$, $v'_{f} = v_g$, $p'_{f} = 0$, and $\rho'_{f} = 0$.

\begin{figure}[h!]
  \centering
  \begin{minipage}[a]{0.5\textwidth}
    \includegraphics[trim=4 -0.1 4 4, clip, width=\textwidth]{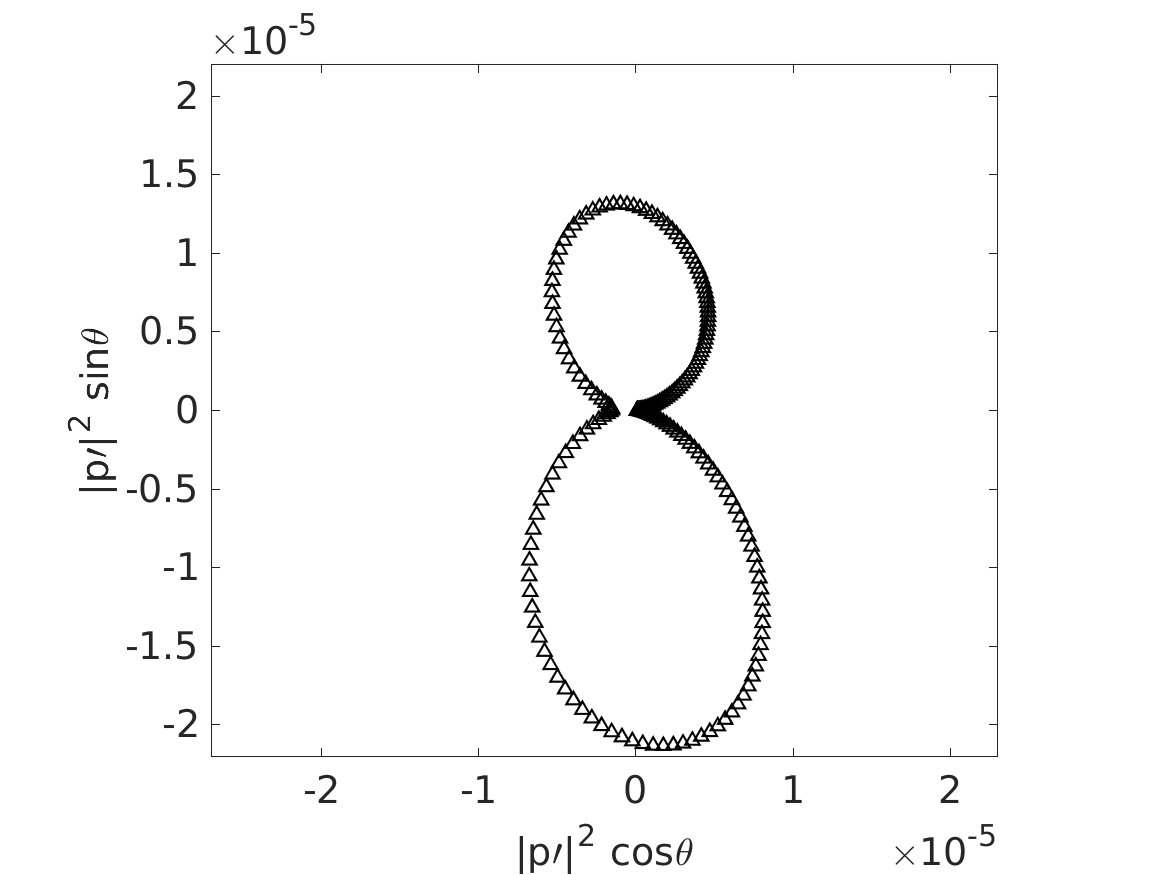}
  \end{minipage}
  \centering
  \caption{Acoustic intensity computed by the nonlinear Euler solver with a reduced frequency of $k=1$ along a circle of radius $r=c$.}
   \label{intensity}
\end{figure}

Figure~\ref{intensity} shows the acoustic intensity $\overline{(p')^2}$ computed by the nonlinear solver along a circle of radius $r=c$ centered at $x=\frac{c}{2}$ for a reduced frequency of $k=1$, and can be compared against the results obtained by Wang et al. \cite{Wang:2010}, although minor differences can be attributable to the 2nd order accurate numerical method used here in contrast to higher order discontinuous Galerkin approaches. 

ERA ROMs are constructed separately for each variable, and trained with high-fidelity snapshots of the system excited with a Gaussian input with zero mean and a standard deviation of 325. The amplitude of the Gaussian-shaped input is chosen to be about $1\%$ of the freestream velocity, and the standard deviation is adjusted according to the reconstruction accuracy of ROMs.
The Gaussian input is applied to both velocity components u and v. We should note that training ROMs with the snapshots of the response of the system to impulsive input resulted in ROMs that demonstrated order of magnitude deviation from the FOM in the prediction regime. This can be attributed to inadequate excitation of the system states in response to the instantaneous impulsive input. Therefore, we train ROMs with the snapshots collected from the response of the system to a Gaussian-shaped input that appears to better capture the dynamics of the system in this application.

To construct ERA ROMs based on the snapshots generated with the $401 \times 101$ grid, 604 points along the far-field boundary need to be separately excited and the corresponding Gaussian input responses should be collected and stacked as described in section~\ref{era} to build the Hankel matrix and compute the ROM matrices as a MIMO system with 604 input channels and full-state output (i.e., $40,000$ output channels). The memory and computation time required to complete this process, defeats the purpose of model reduction. Therefore, in sections~\ref{result_nonlinear} and \ref{result_linear} we demonstrate the ROM results based on a coarser $101 \times 51$ grid with 204 input channels in this section for two scenarios: when the training snapshots are generated by the nonlinear solver, and when the training snapshots are computed by the linearized solver. 
Later in section~\ref{gappy_results} we use the actuator selection by the gappy POD approach described in section~\ref{gappy} to build ERA ROMs based on the finer $401 \times 101$ grid with a reduced number of input channels and a fraction of the cost of the original setup.

We emphasize on the importance of the accuracy of ROMs in the prediction regime in this work. We train ERA ROMs using the Gaussian-shaped input and test them with unseen input signals. Therefore, the results demonstrated in this section evaluate the performance of ROMs in a purely predictive sense. We choose three different input signals to evaluate ROM predictions. The first input is a smooth sinusoidal signal,
\be
\label{ugtest}
u_g=- \frac{\epsilon w}{\sqrt{2}} \sin(\omega t);
\ \ v_g=\frac{\epsilon w}{\sqrt{2}} \sin(\omega t),
\ee
that is simultaneously applied to both velocity components at 204 far-field boundary cells following the same procedure as described for equation~\ref{ug}.

\begin{figure}[h!]
  \centering
  \begin{minipage}[a]{0.50\textwidth}
    \includegraphics[trim=4 4 4 4, clip, width=\textwidth]{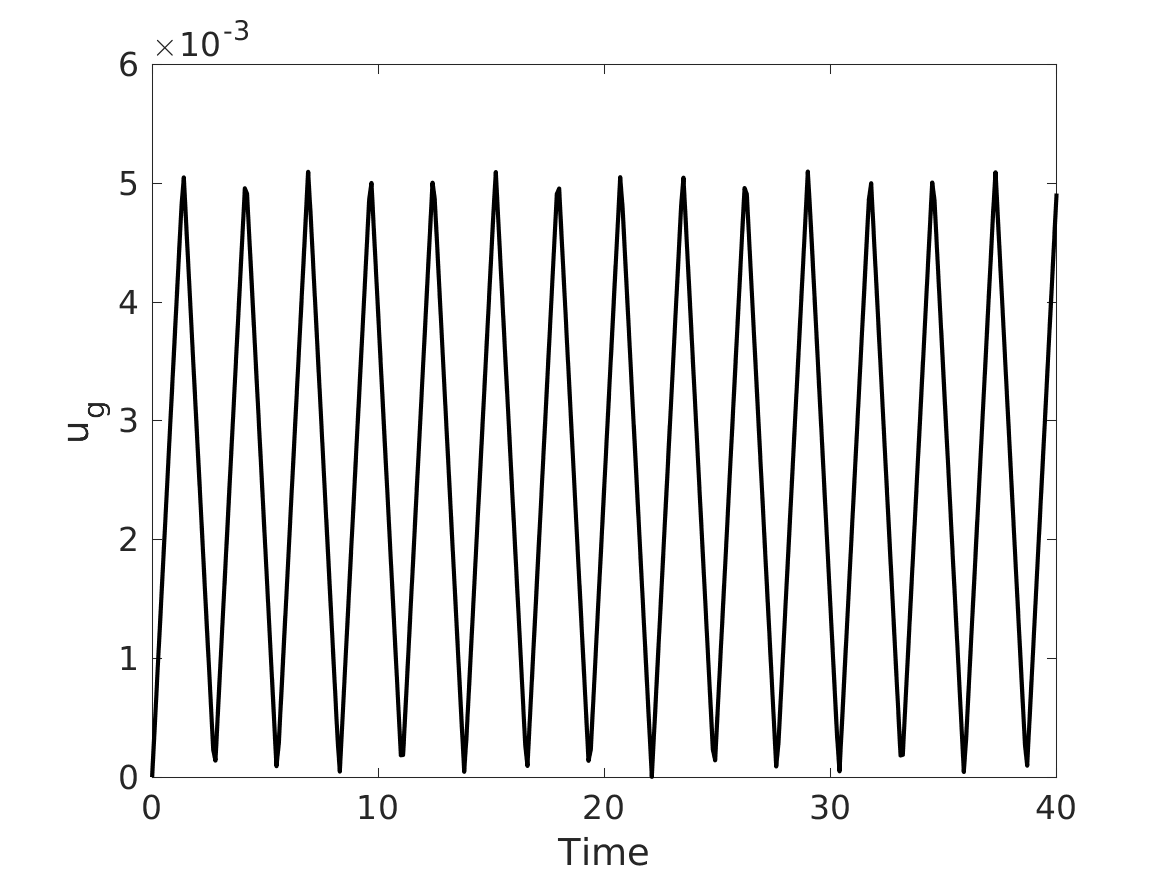}
  \end{minipage}
  \centering
  \caption{The triangular wave signal applied to the entire far-field boundary.}
   \label{tri_input}
\end{figure}

The second signal is a periodic triangular wave signal with sharp gradients,
\be
\label{ug_tri}
u_g = -\frac{2\epsilon w}{\sqrt{2}} \left|\omega t - \left\lfloor \omega t + \frac{1}{2} \right\rfloor \right|;
\ \ v_g = \frac{2\epsilon w}{\sqrt{2}} \left|\omega t - \left\lfloor \omega t + \frac{1}{2} \right\rfloor \right|,
\ee
where, parameters $\epsilon$, $w$ and $\omega$ take the same values as in equation~\ref{ug}. The far-field boundary values are similarly evaluated as $u_{f} = u_{\infty} + u_g$, $v_{f} = v_{\infty} + v_g$, $p_{f} = p_{\infty}$, and $\rho_{f} = \rho_{\infty}$. The input signal is shown in Figure~\ref{tri_input}.

The third input is a non-periodic square wave signal,
\be
\label{ug_square}
u_g = \begin{cases}
-0.005 \ \ 0.5 \leq t<1 \\
0 \ \ \qquad otherwise
\end{cases}; \ \ 
v_g = \begin{cases}
0.005 \ \ 0.5 \leq t<1 \\
0 \ \ \qquad otherwise
\end{cases}, 
\ee
applied to the far-field velocity components $u_f$ and $v_f$, respectively. 

We define the following relative error to evaluate the accuracy of ROM predictions,
\be
\label{rel_e}
e^k = \frac{\norm{\mathbf{q}^k - \tilde{\mathbf{q}}^k}_2}{\norm{\mathbf{q}^k}_2}, \qquad k = 1, \dots, n_t,
\ee
where, $\mathbf{q}^k$ and $\tilde{\mathbf{q}}^k$ are the FOM and ROM solutions at the $k^{th}$ time step, respectively, and $n_t$ is the total number of time steps.

\subsection{ROMs based on the Nonlinear Euler Solver}\label{result_nonlinear}
\subsubsection{ROM Training}
The nonlinear high-fidelity simulations based on the coarse grid are performed with a time step of $\Delta t = 1 \times 10^{-3}$, that results in a CFL number of 0.156. The training snapshots are collected every 100 time steps and a total of $5 \times 10^{4}$ time steps are computed to generate the training data, which results in 500 snapshots (Markov parameters) per input channel for each variable.
The tangential interpolation method is used to project the training snapshots onto the leading left and right tangential modes to reduce the number of input and output channels before computing SVD of the Hankel matrix. Table~\ref{t:tanint} shows the number of the retained left and right tangential modes for each variable. The retained tangential modes capture $80\%$ of the energy in the training snapshots, while, the retained Hankel singular vectors (i.e., the last column in Table~\ref{t:tanint}) capture $75\%$ of the input-output energy. The remaining modes are hardly controllable and observable, and therefore, they are discarded without loss of accuracy. 
\begin{table}[h!]
 \begin{center}
  \caption{The number of the retained tangential modes for the ROM of each variable out of a total of 5000 left and 204 right tangential modes and the dimension of the balanced ROMs. }
  \label{t:tanint}
  \begin{tabular}{lllll}\hline
        & Variable & Left modes & Right modes & ROM dimension \\\hline
        & Pressure & 183 & 45 & 109 \\\hline
        & Velocity (u) & 157 & 33 & 89  \\\hline
        & Velocity (v) & 190 & 45 & 118  \\\hline
        & Density & 187 & 45 & 109  \\\hline
  \end{tabular}
 \end{center}
\end{table}

\begin{figure}[h!]
  \centering
  \begin{minipage}[a]{0.43\textwidth}
    \includegraphics[trim=4 4 4 4, clip, width=\textwidth]{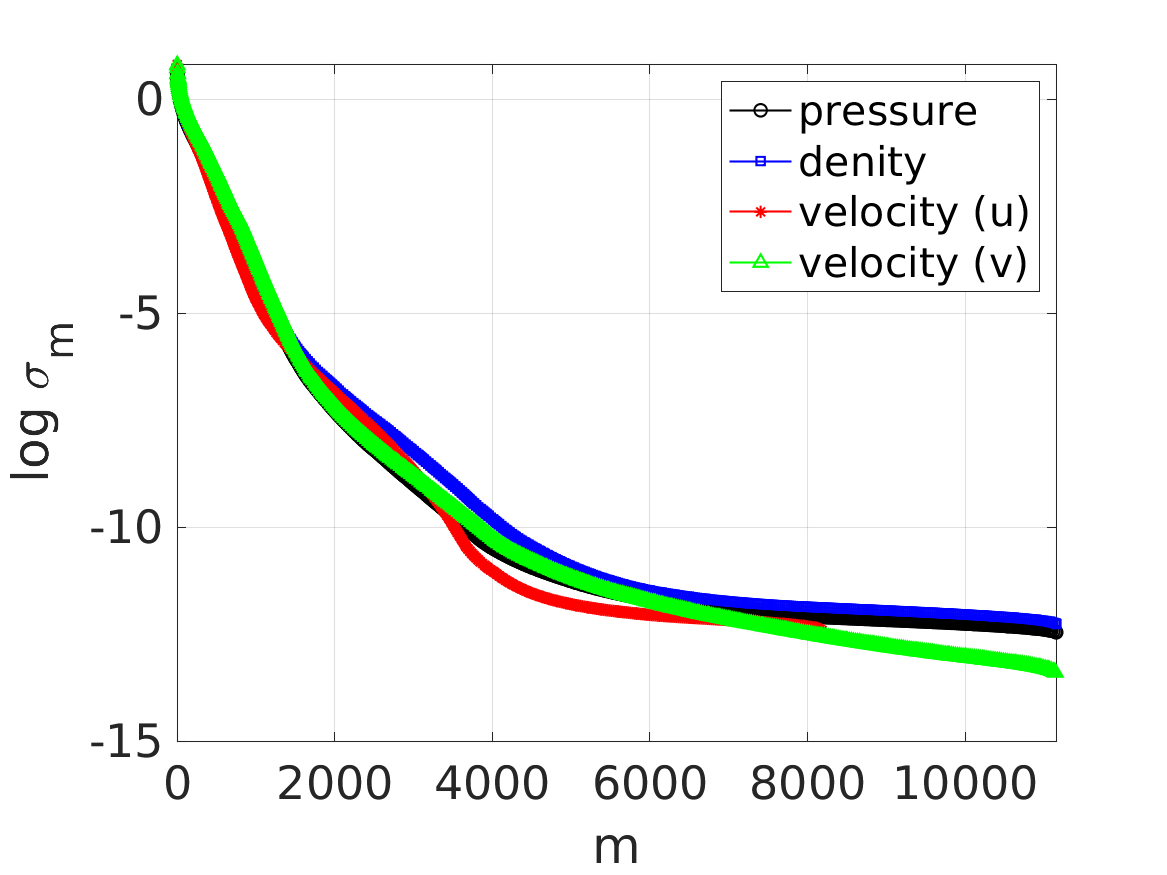}
    \rput(0.5,0.1){\psscalebox{0.5}{\color{black} \textbf{a)}}}
  \end{minipage}
  \centering
  \begin{minipage}[a]{0.43\textwidth}
    \includegraphics[trim=4 4 4 4, clip, width=\textwidth]{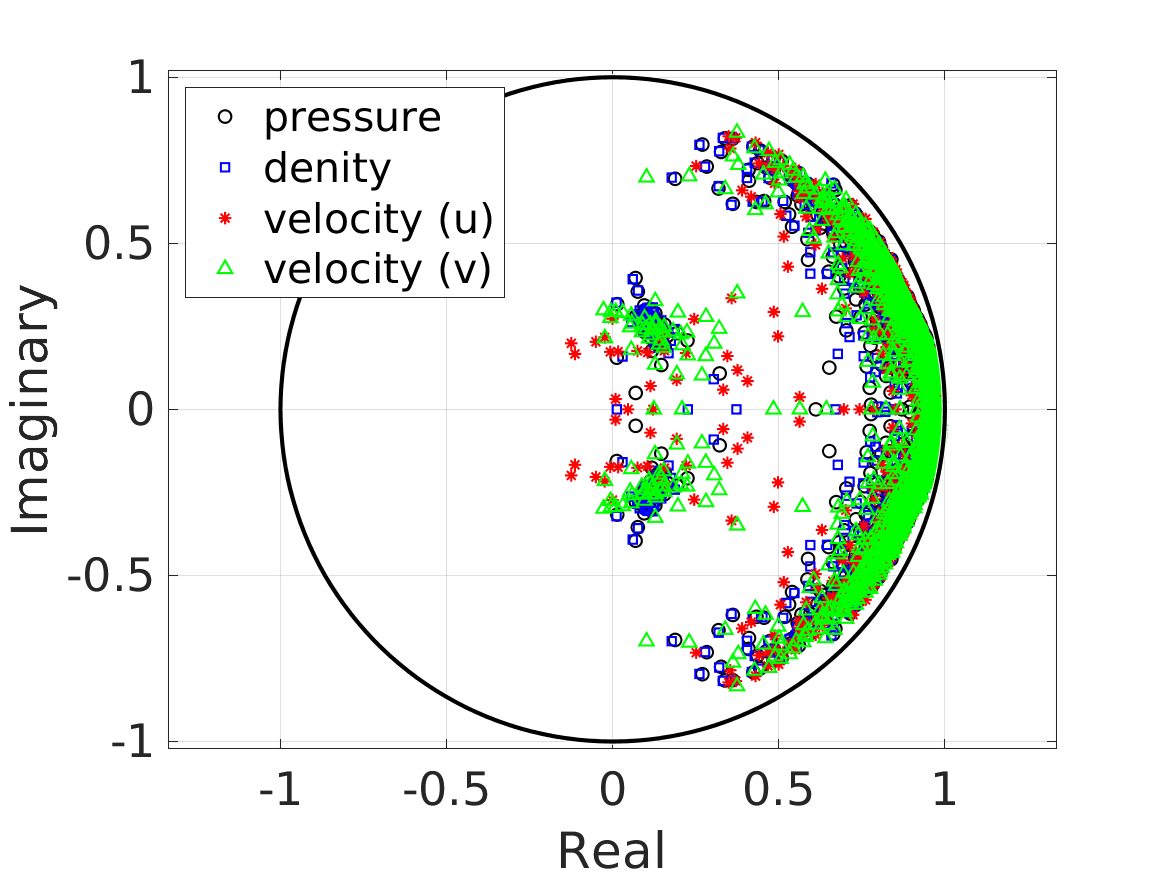}
    \rput(0.5,0.1){\psscalebox{0.5}{\color{black} \textbf{b)}}}
  \end{minipage}
  \centering
  \begin{minipage}[a]{0.125\textwidth}
    \includegraphics[trim=7cm 4 7cm 4, clip, width=\textwidth]{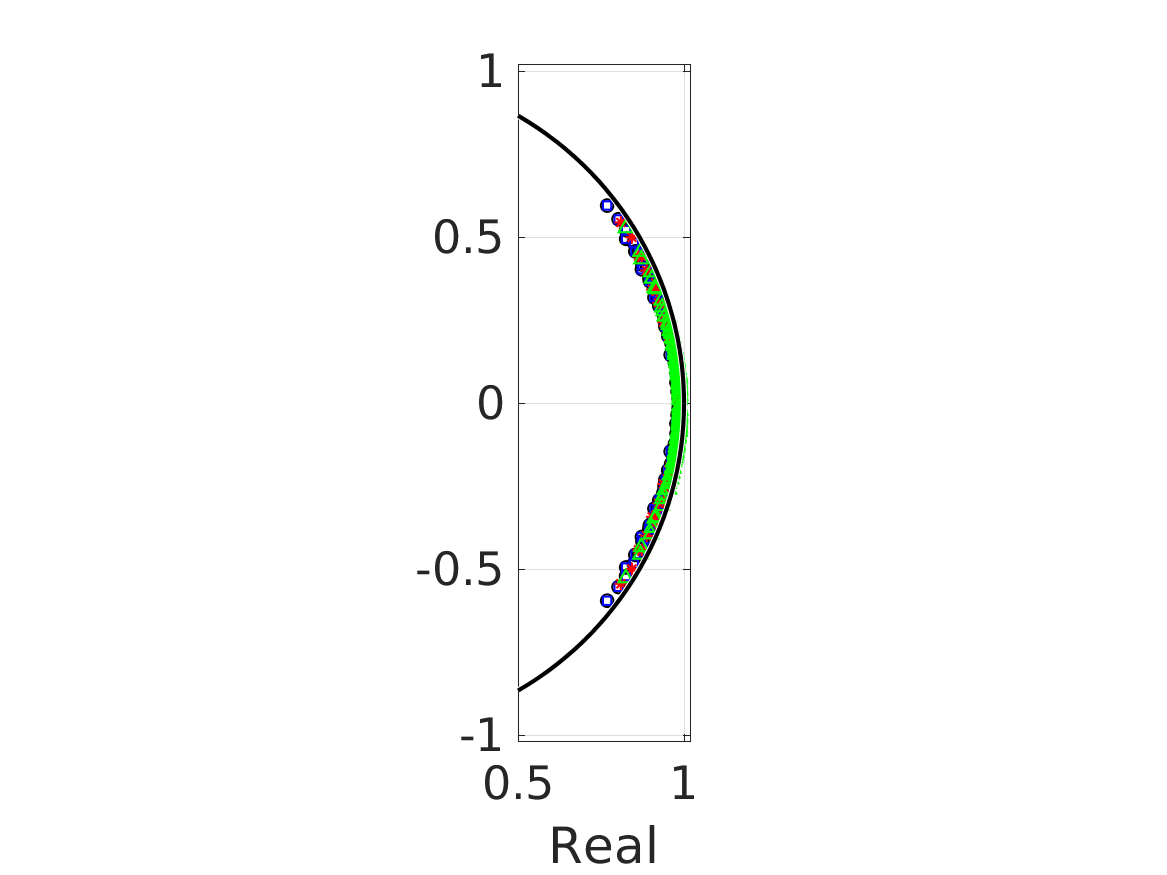}
    \rput(0.5,0.1){\psscalebox{0.5}{\color{black} \textbf{c)}}}
  \end{minipage}
  \centering
  \caption{a) Singular values of the Hankel matrix. b) and c) Eigenvalues of the ERA ROMs. The balancing modes used to construct ROMs capture $99.99\%$ (b) and $75\%$ (c) of the input-output energy.}
   \label{sv_rom_eigs}
\end{figure}

Figure~\ref{sv_rom_eigs} (a) shows the decay of Hankel singular values for different variables.
The constructed ROMs for all four variables are stable with eigenvalues located inside the unit circle (i.e., the stability margin of discrete-time systems). Figure~\ref{sv_rom_eigs} (b) and (c) show the eigenvalues for ROMs of different variables when the retained balancing modes capture $99.99\%$ and $75\%$ of the input-output energy, respectively. The black circle shows the discrete-time system stability margin.

\begin{figure}[h!]
  \centering
  \begin{minipage}[a]{0.49\textwidth}
    \includegraphics[trim=4 4 4 4, clip, width=\textwidth]{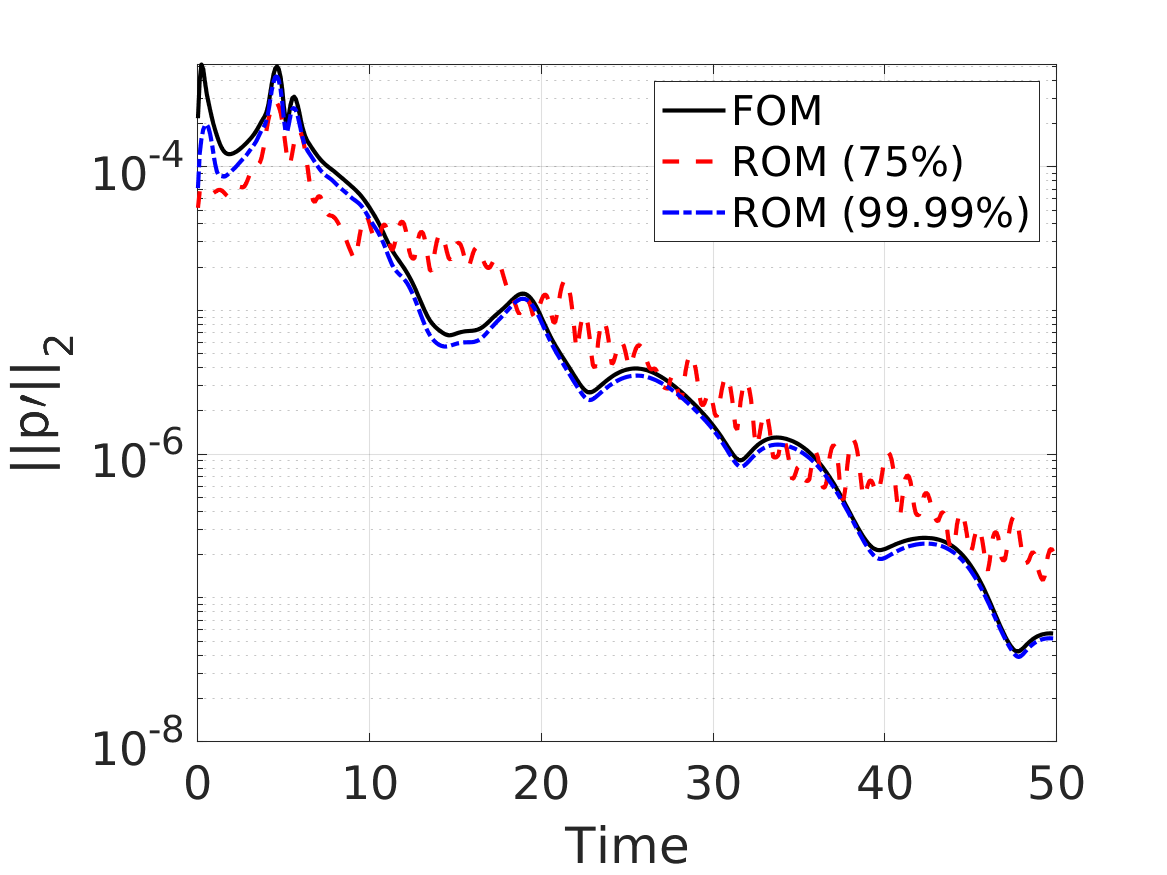}
    \rput(0.5,0.1){\psscalebox{0.5}{\color{black} \textbf{a)}}}
    \vspace{0.1cm}
  \end{minipage}
  \centering
  \begin{minipage}[a]{0.49\textwidth}
    \includegraphics[trim=4 4 4 4, clip, width=\textwidth]{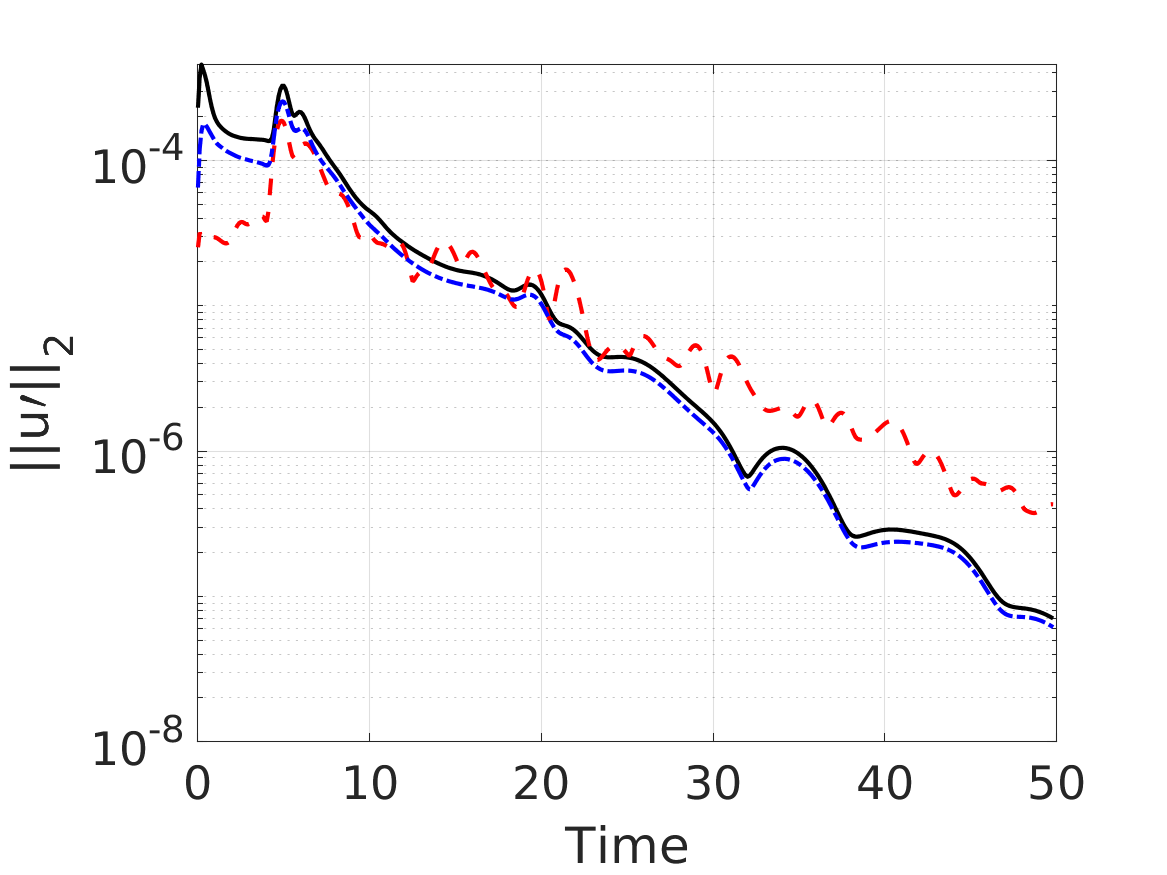}
    \rput(0.5,0.1){\psscalebox{0.5}{\color{black} \textbf{b)}}}
    \vspace{0.1cm}
  \end{minipage}
  \centering
  \begin{minipage}[a]{0.49\textwidth}
    \includegraphics[trim=4 4 4 4, clip, width=\textwidth]{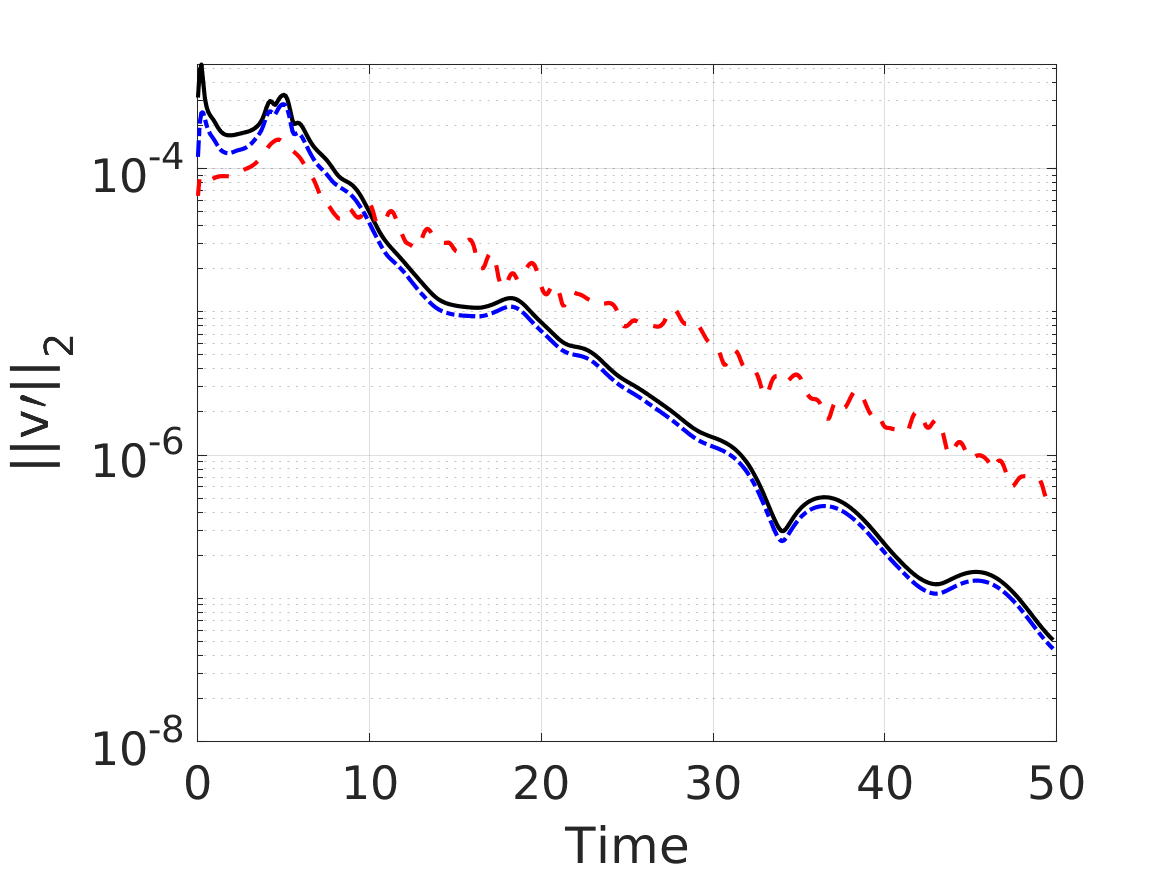}
    \rput(0.5,0.1){\psscalebox{0.5}{\color{black} \textbf{c)}}}
  \end{minipage}
  \centering
  \begin{minipage}[a]{0.49\textwidth}
    \includegraphics[trim=4 4 4 4, clip, width=\textwidth]{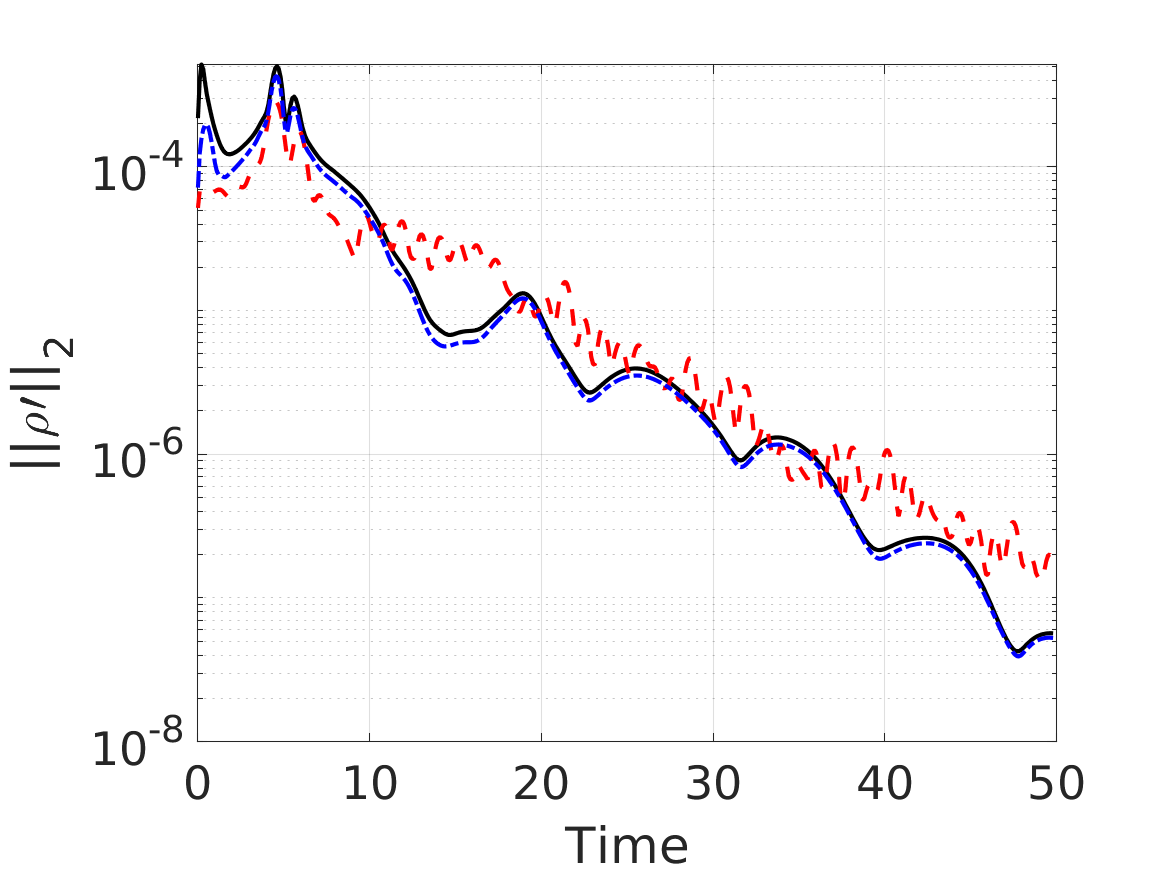}
    \rput(0.5,0.1){\psscalebox{0.5}{\color{black} \textbf{d)}}}
  \end{minipage}
  \centering
  \caption{Norm of pressure (a), u-velocity (b), v-velocity (c), and density (d) computed by the FOM and ERA ROMs in response to the Gaussian-shaped perturbation applied to a channel at $(x, y) = (-1.4660, -4.5892)$.}
   \label{rom_reconst}
\end{figure}

In Figure~\ref{rom_reconst} we demonstrate the accuracy of the ROM reconstruction in response to the Gaussian input for each variable. Here, the norm of the state is computed for each ROM and compared against the FOM. Two ROMs are evaluated: a larger ROM in which the Hankel singular values capture $99.99\%$ of the input-output energy. This ROM is clearly very accurate. All four ROMs closely follow the solution of the FOM in the training interval. But the larger ROMs are also more expensive, requiring more than 800 balancing modes. The smaller ROM on the other hand, is constructed with enough modes to capture $75\%$ of the input-output energy. This ROM follows a similar trend as the FOM in reconstruction of the norm of the state, but it demonstrates some numerical artifacts. 

\subsubsection{ROM Prediction}
\begin{figure}[h!]
  \centering
  \begin{minipage}[a]{0.49\textwidth}
    \includegraphics[trim=4 4 4 4, clip, width=\textwidth]{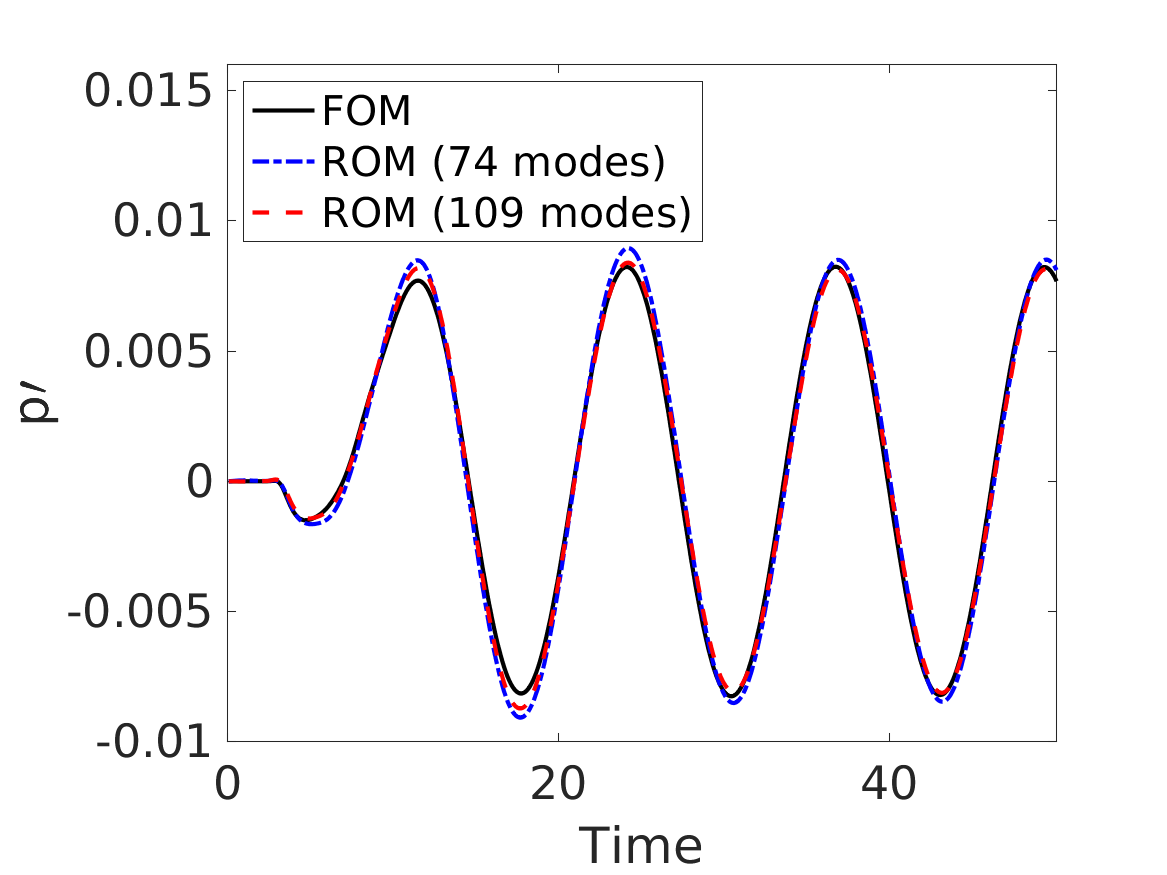}
    \rput(0.5,0.1){\psscalebox{0.5}{\color{black} \textbf{a)}}}
  \end{minipage}
  \vspace{0.1cm}
  \centering
  \begin{minipage}[a]{0.49\textwidth}
    \includegraphics[trim=4 4 4 4, clip, width=\textwidth]{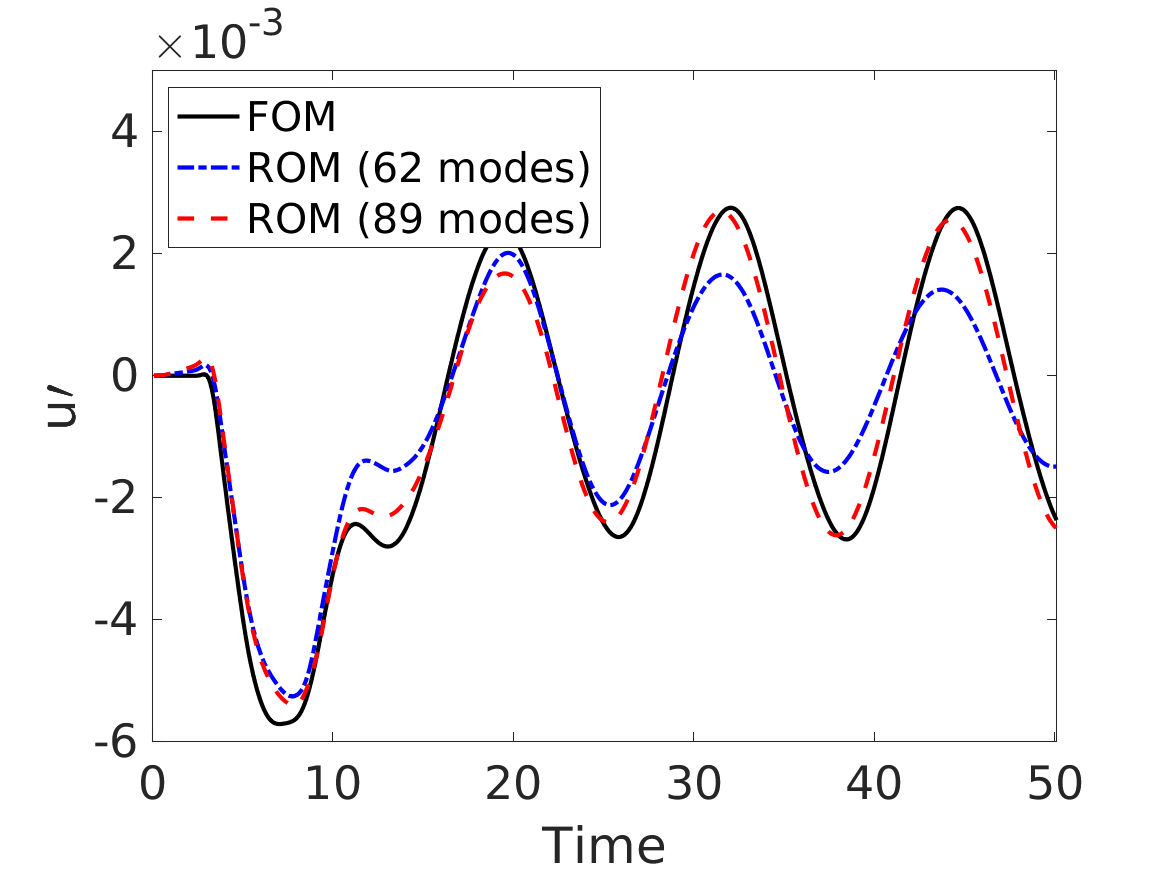}
    \rput(0.5,0.1){\psscalebox{0.5}{\color{black} \textbf{b)}}}
  \end{minipage}
  \vspace{0.1cm}
  \centering
  \begin{minipage}[a]{0.49\textwidth}
    \includegraphics[trim=4 4 4 4, clip, width=\textwidth]{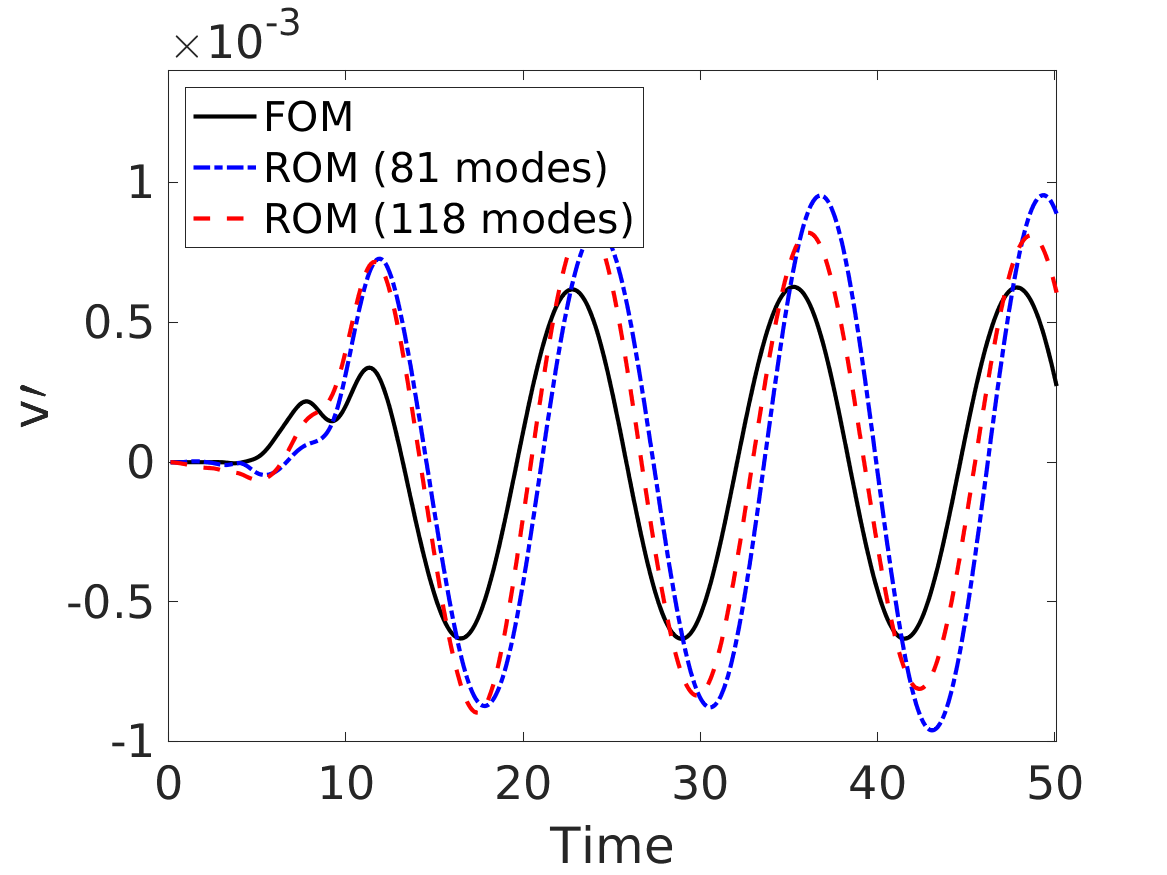}
    \rput(0.5,0.1){\psscalebox{0.5}{\color{black} \textbf{c)}}}
  \end{minipage}
  \centering
  \begin{minipage}[a]{0.49\textwidth}
    \includegraphics[trim=4 4 4 4, clip, width=\textwidth]{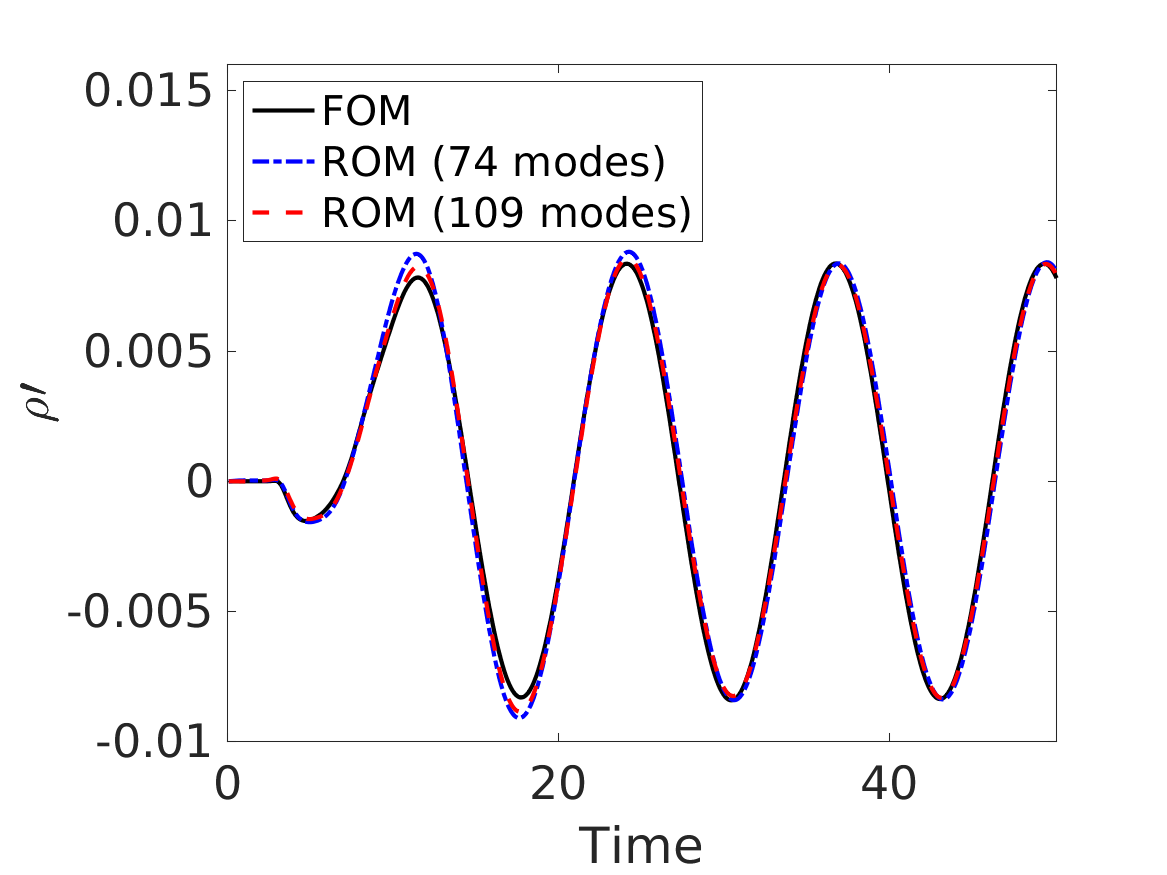}
    \rput(0.5,0.1){\psscalebox{0.5}{\color{black} \textbf{d)}}}
  \end{minipage}
  \centering
  \caption{Comparison of pressure (a), u-velocity (b), v-velocity (c), and density (d) computed by the HFM and ERA ROMs at $(x,y) = (0.3510,-0.1096)$. The freestream is perturbed with the sinusoidal input.}
   \label{sine_probe}
\end{figure}

%\subsubsection{ROM Predictions for the Sinusoidal Input}
The predictive performance of the ERA ROM is first tested with the sinusoidal input in equation~\ref{ugtest}. A total of $5 \times 10^{4}$ FOM time steps are computed and high-fidelity snapshots are collected every 100 time steps to validate the ERA ROM predictions. Figure~\ref{sine_probe} compares the pressure, velocity, and density probes computed by two different ROMs against the FOM. The smaller ROMs are constructed with $65\%$ input-output energy content and the larger ROMs are generated with the Hankel singular vectors that capture $75\%$ of the input-output energy, where the subspace dimensions for this ROM are reported in Table~\ref{t:tanint}.
Both of the ROM solutions replicate the FOM response despite the input signal is different from the input used to generate the training snapshots for the ROMs. 

\begin{figure}[h!]
  \centering
  \begin{minipage}[a]{0.49\textwidth}
    \includegraphics[trim=4 4 -0.1 4, clip, width=\textwidth]{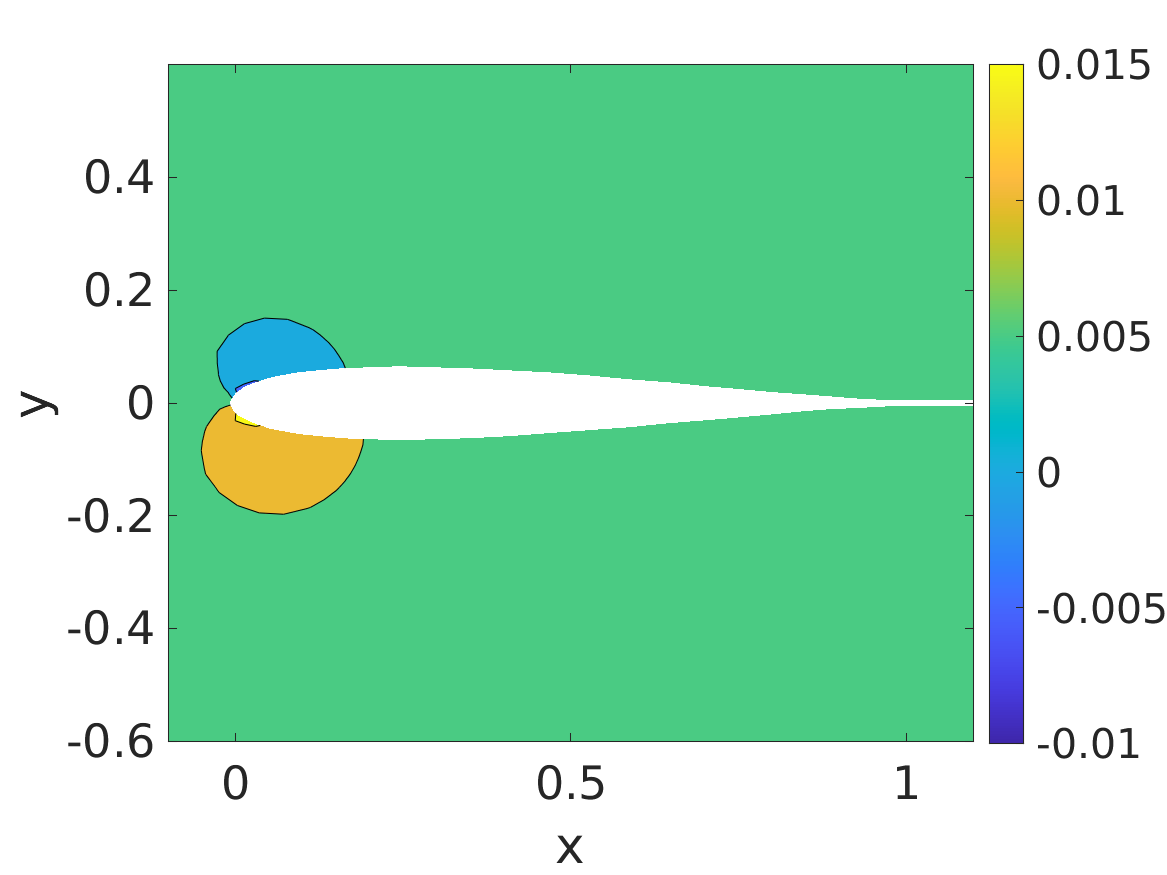}
    \rput(0.5,0.1){\psscalebox{0.5}{\color{black} \textbf{a)}}}
  \end{minipage}
  \centering
  \begin{minipage}[a]{0.49\textwidth}
    \includegraphics[trim=-0.1 4 -0.1 4, clip, width=\textwidth]{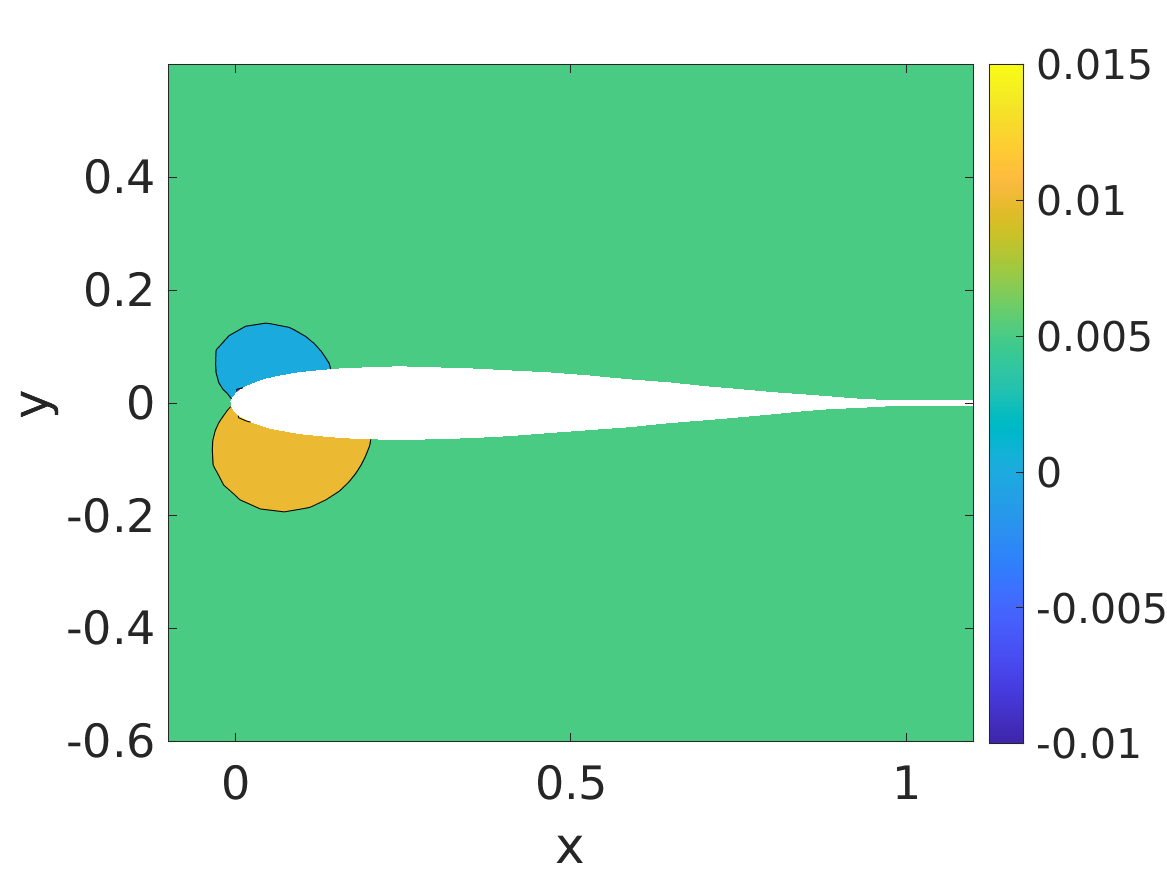}
    \rput(0.5,0.1){\psscalebox{0.5}{\color{black} \textbf{b)}}}
  \end{minipage}
  \centering
  \caption{Pressure contours computed by the a) FOM and b) ERA ROM at $t=49.2$. The freestream velocity is perturbed by the sinusoidal input.}
   \label{pcontour}
\end{figure}

Figure~\ref{pcontour} shows the pressure perturbation contours computed by the FOM and predicted by the ROM in Table~\ref{t:tanint} at $t = 49.2$, when 204 far-field points are simultaneously perturbed by the sinusoidal input.

%\subsubsection{ROM Predictions for the Triangular Wave Input}
To further demonstrate that the generated ROMs are purely predictive outside of the training setup (i.e., in response to unseen input signals), the system response is predicted when the entire far-field boundary is perturbed by the periodic triangular wave in equation~\ref{ug_tri}.

\begin{figure}[h!]
  \centering
  \begin{minipage}[a]{0.49\textwidth}
    \includegraphics[trim=4 -0.1 4 4, clip, width=\textwidth]{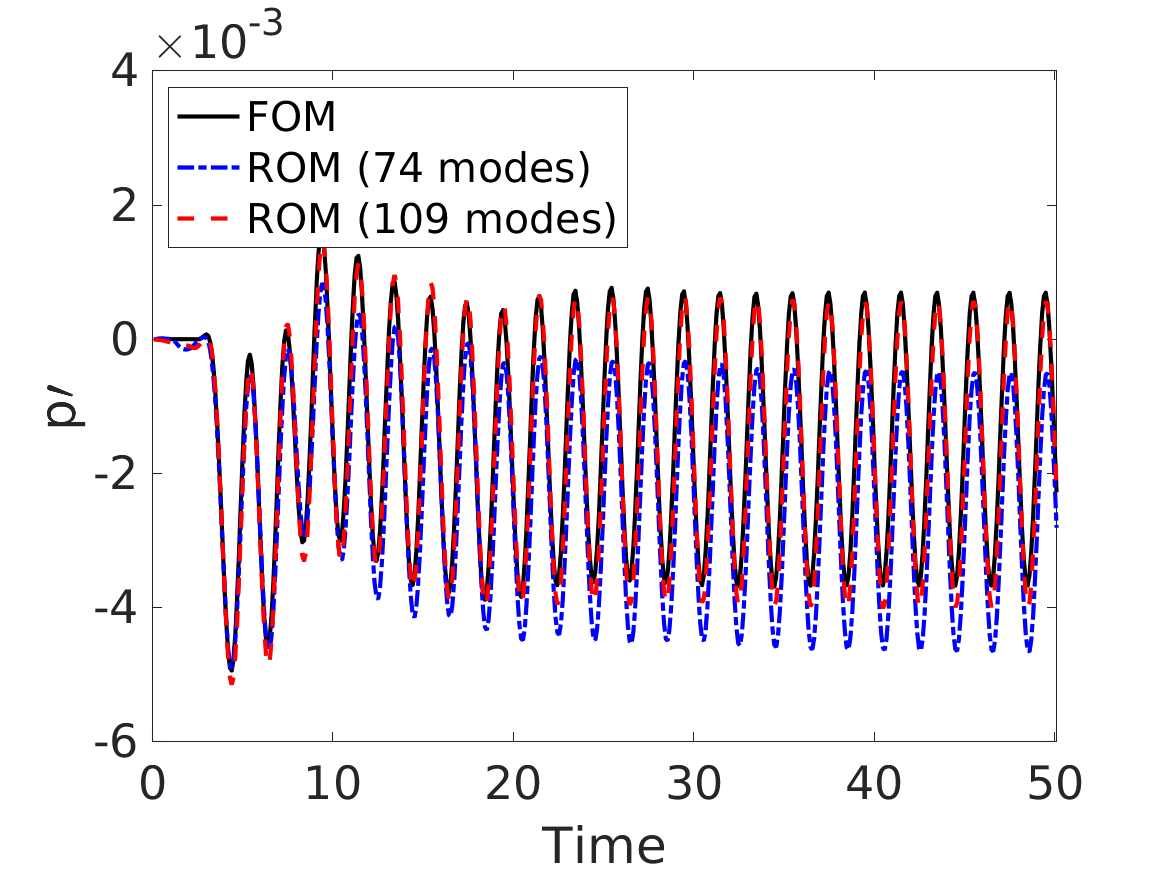}
    \rput(0.5,0.1){\psscalebox{0.5}{\color{black} \textbf{a)}}}
    \vspace{0.1cm}
  \end{minipage}
  \centering
  \begin{minipage}[a]{0.49\textwidth}
    \includegraphics[trim=4 -0.1 4 4, clip, width=\textwidth]{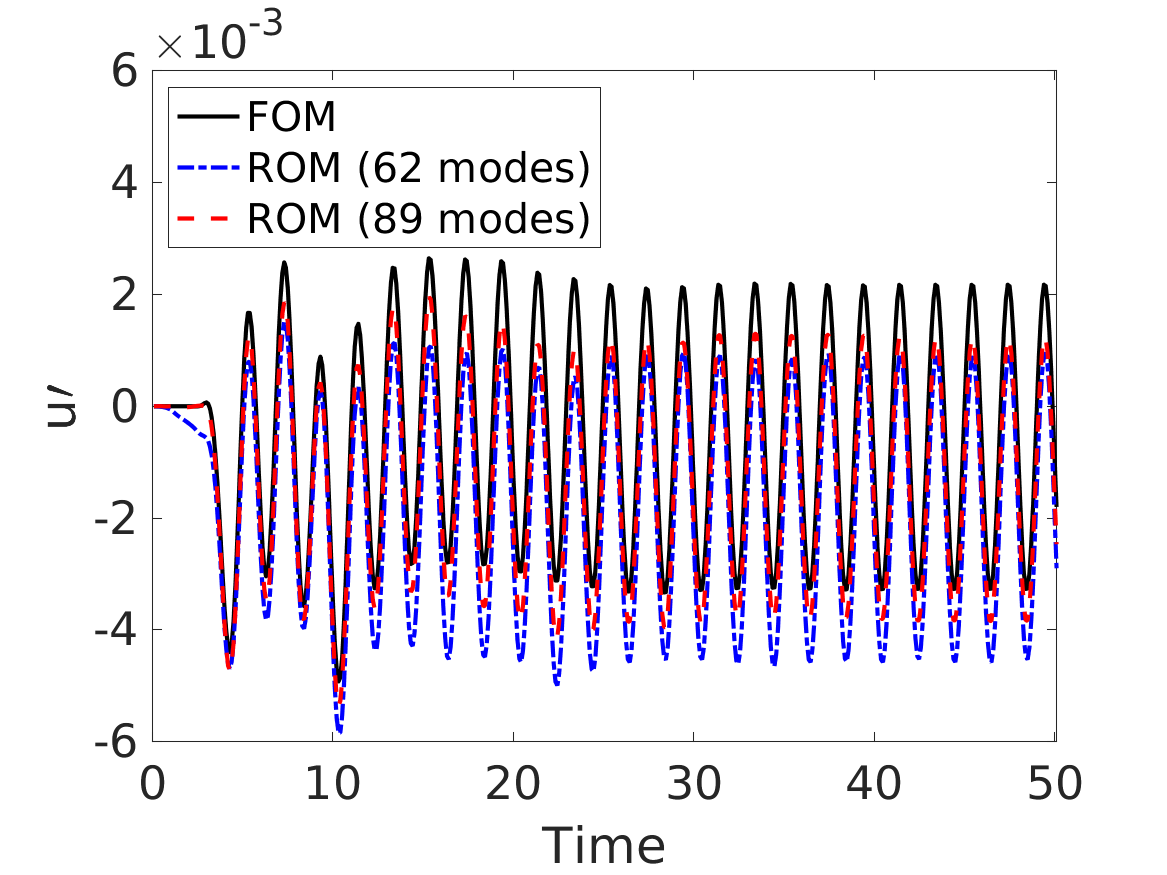}
    \rput(0.5,0.1){\psscalebox{0.5}{\color{black} \textbf{b)}}}
    \vspace{0.1cm}
  \end{minipage}
  \centering
  \begin{minipage}[a]{0.49\textwidth}
    \includegraphics[trim=4 -0.1 4 4, clip, width=\textwidth]{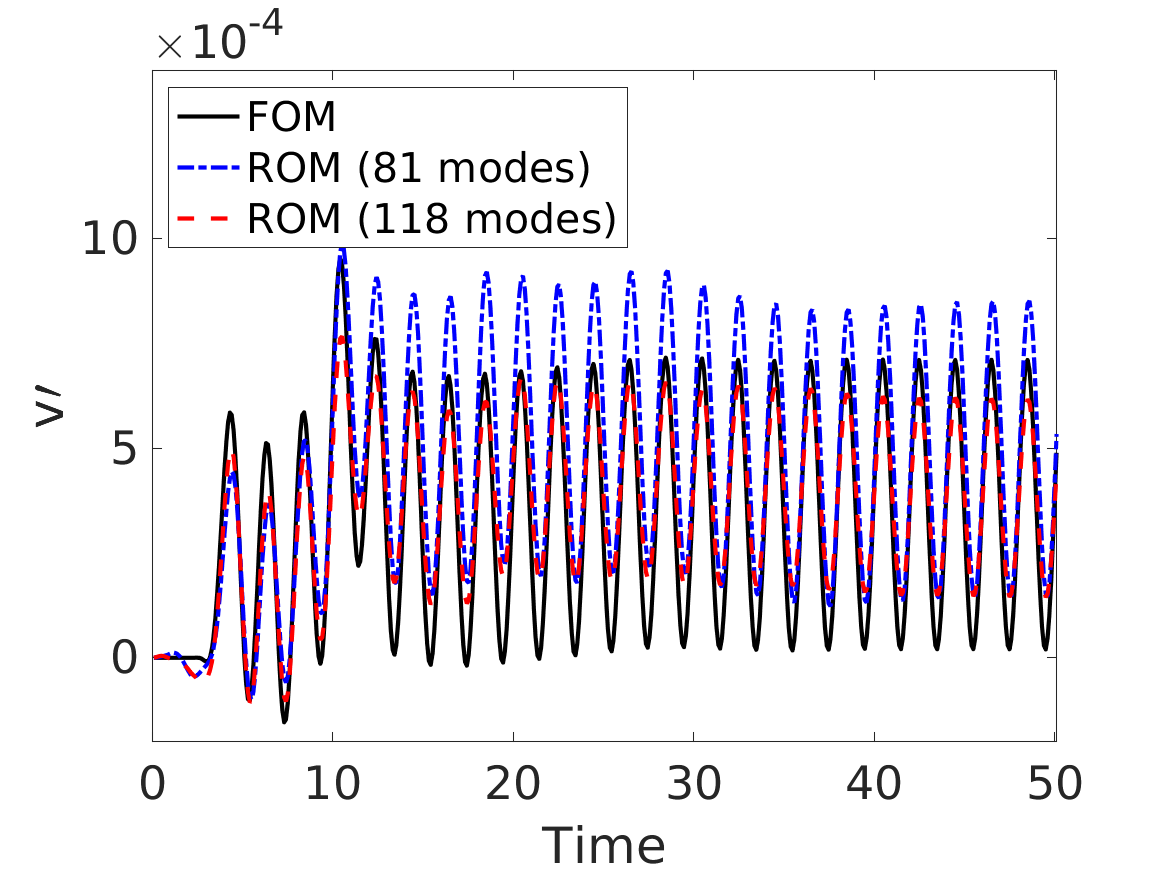}
    \rput(0.5,0.1){\psscalebox{0.5}{\color{black} \textbf{c)}}}
  \end{minipage}
  \centering
  \begin{minipage}[a]{0.49\textwidth}
    \includegraphics[trim=4 -0.1 4 4, clip, width=\textwidth]{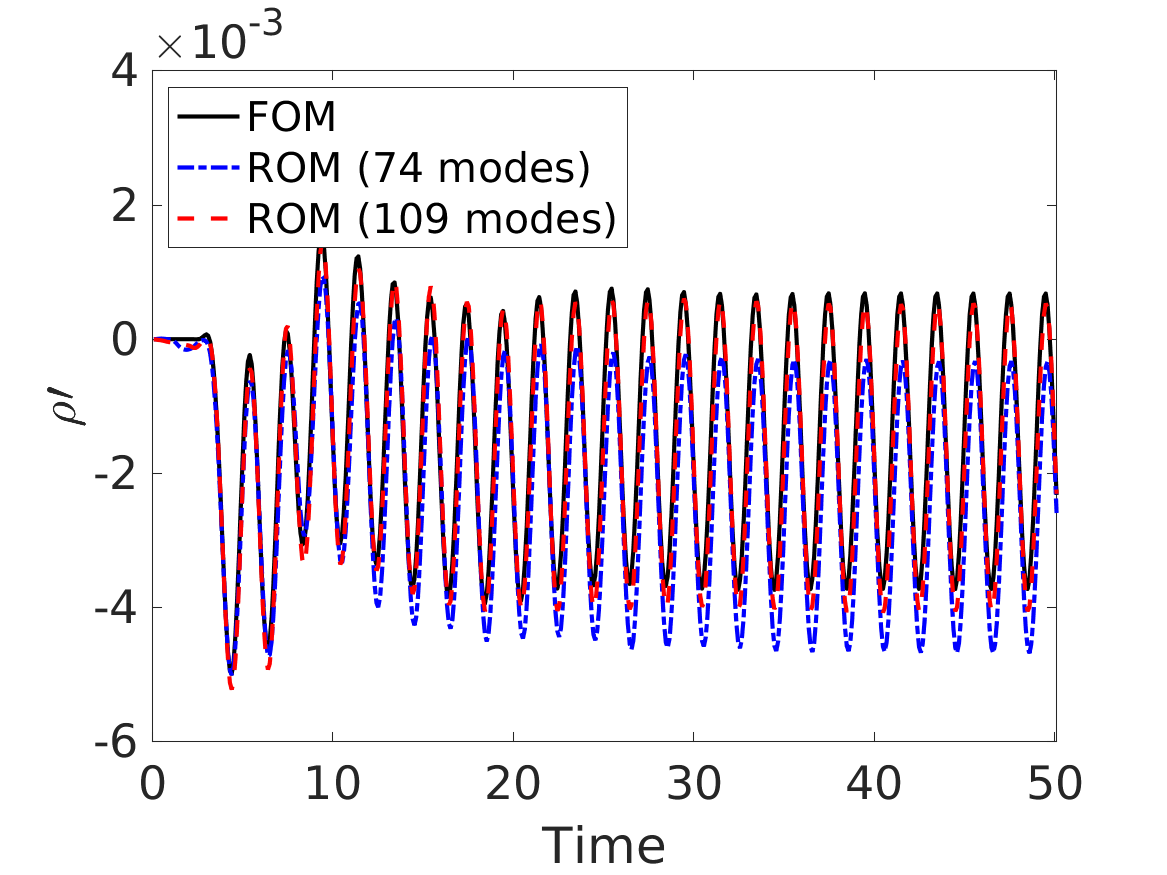}
    \rput(0.5,0.1){\psscalebox{0.5}{\color{black} \textbf{d)}}}
  \end{minipage}
  \centering
  \caption{Comparison of pressure (a), u-velocity (b), v-velocity (c), and density (d) computed by the HFM and ERA ROMs at $(x, y) = (0.5293, 0.096)$. The freestream is perturbed with the triangular wave.} 
   \label{tri_probe}
\end{figure}

Figure~\ref{tri_probe} compares the ERA ROM predictions against the FOM solution for a probe located at $(x, y) = (0.5293, 0.096)$. The ERA ROMs are clearly predictive for input signals different from the training data. Similar to the previous case, two ROMs are tested, where the larger ROM is slightly more accurate than the smaller ROM.

\begin{figure}[h!]
  \centering
  \begin{minipage}[a]{0.49\textwidth}
    \includegraphics[trim=4 -0.1 4 4, clip, width=\textwidth]{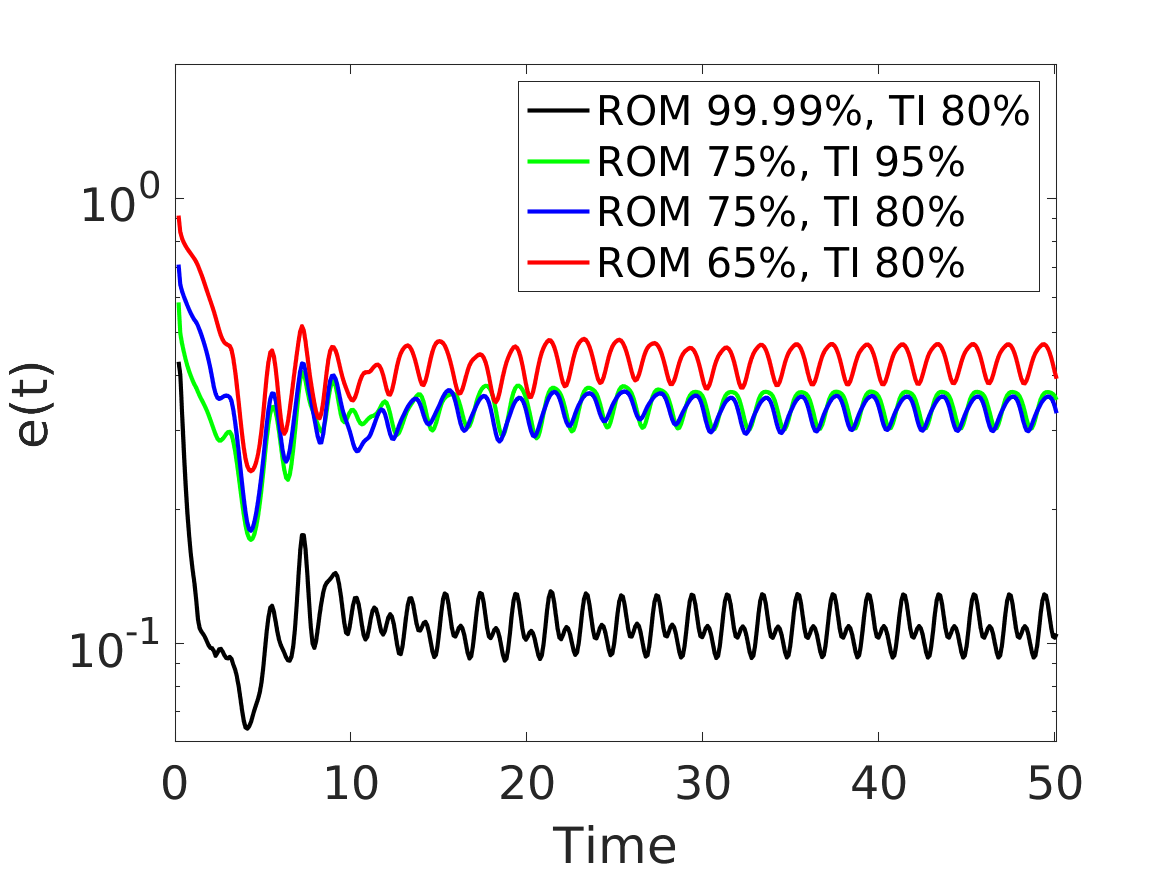}
    \rput(0.5,0.1){\psscalebox{0.5}{\color{black} \textbf{a)}}}
    \vspace{0.1cm}
  \end{minipage}
  \centering
  \begin{minipage}[a]{0.49\textwidth}
    \includegraphics[trim=4 -0.1 4 4, clip, width=\textwidth]{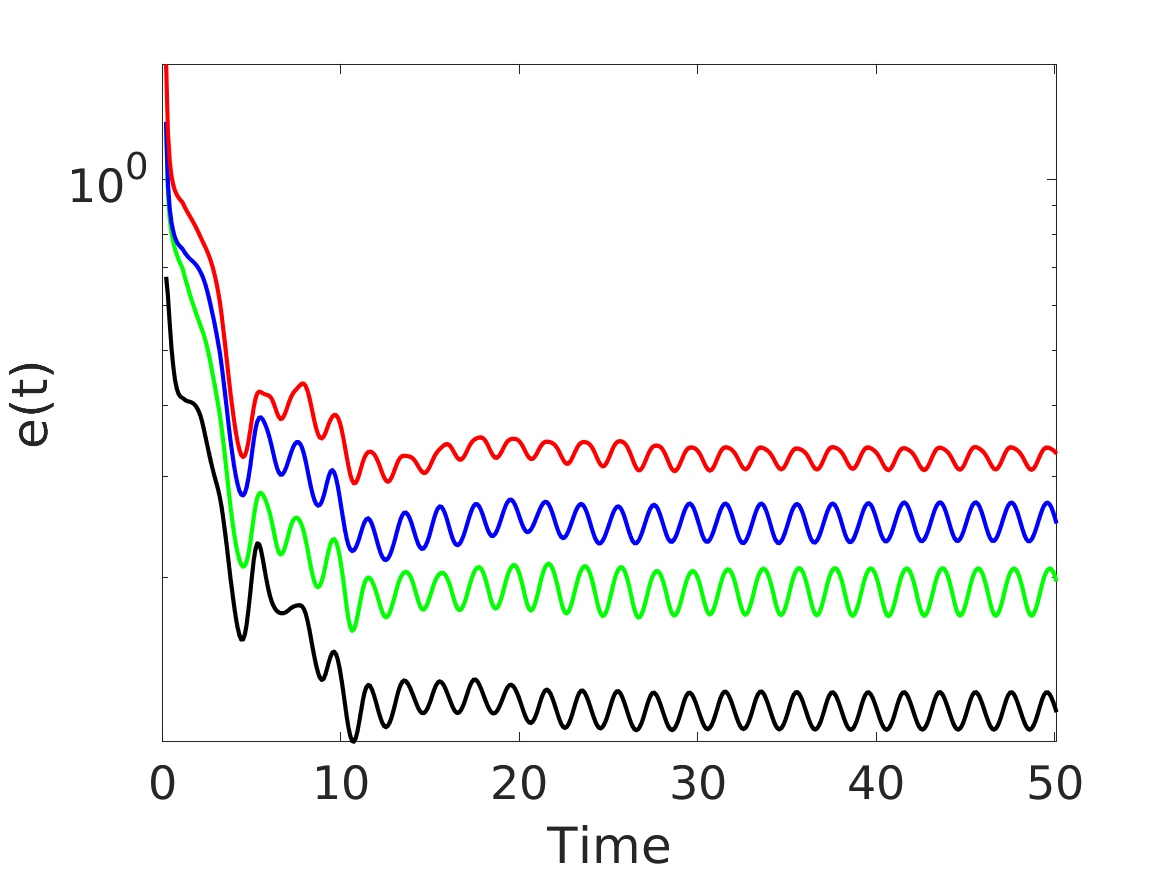}
    \rput(0.5,0.1){\psscalebox{0.5}{\color{black} \textbf{b)}}}
    \vspace{0.1cm}
  \end{minipage}
  \centering
  \begin{minipage}[a]{0.49\textwidth}
    \includegraphics[trim=4 -0.1 4 4, clip, width=\textwidth]{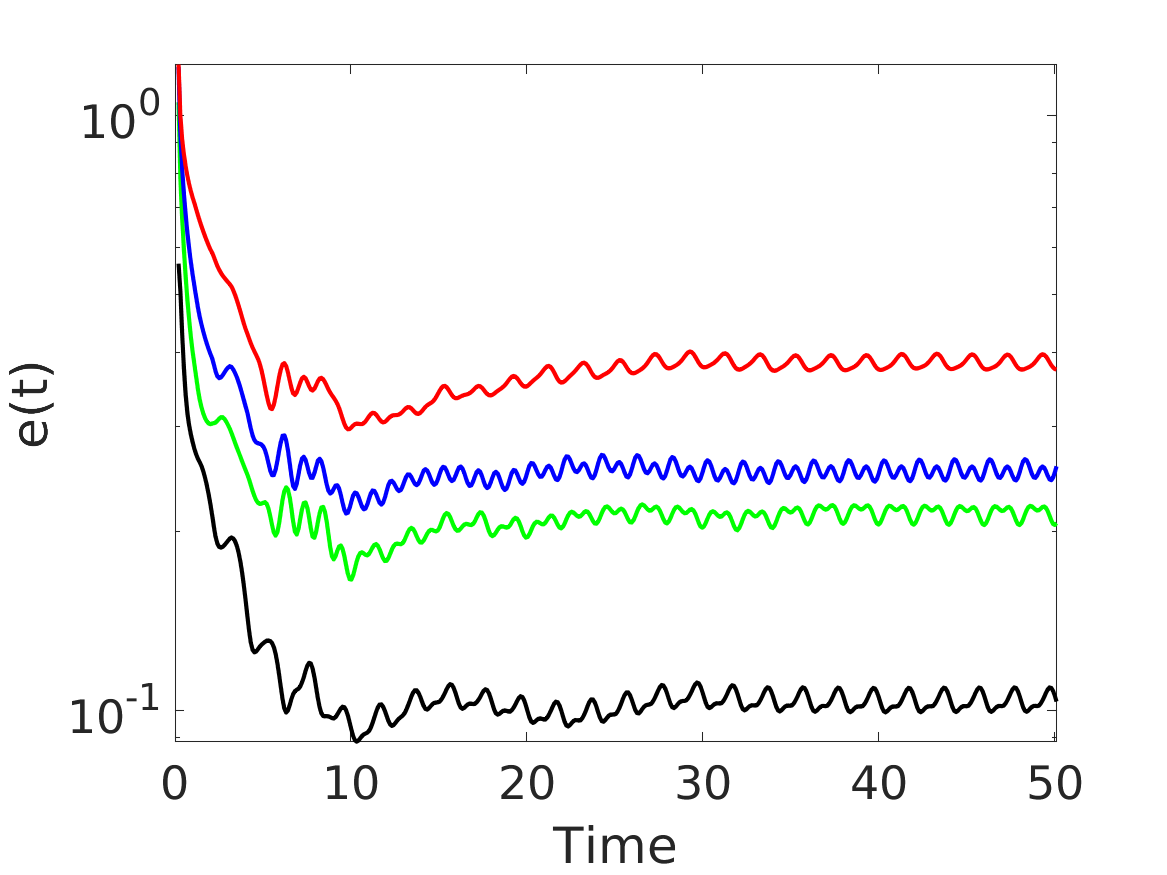}
    \rput(0.5,0.1){\psscalebox{0.5}{\color{black} \textbf{c)}}}
  \end{minipage}
  \centering
  \begin{minipage}[a]{0.49\textwidth}
    \includegraphics[trim=4 -0.1 4 4, clip, width=\textwidth]{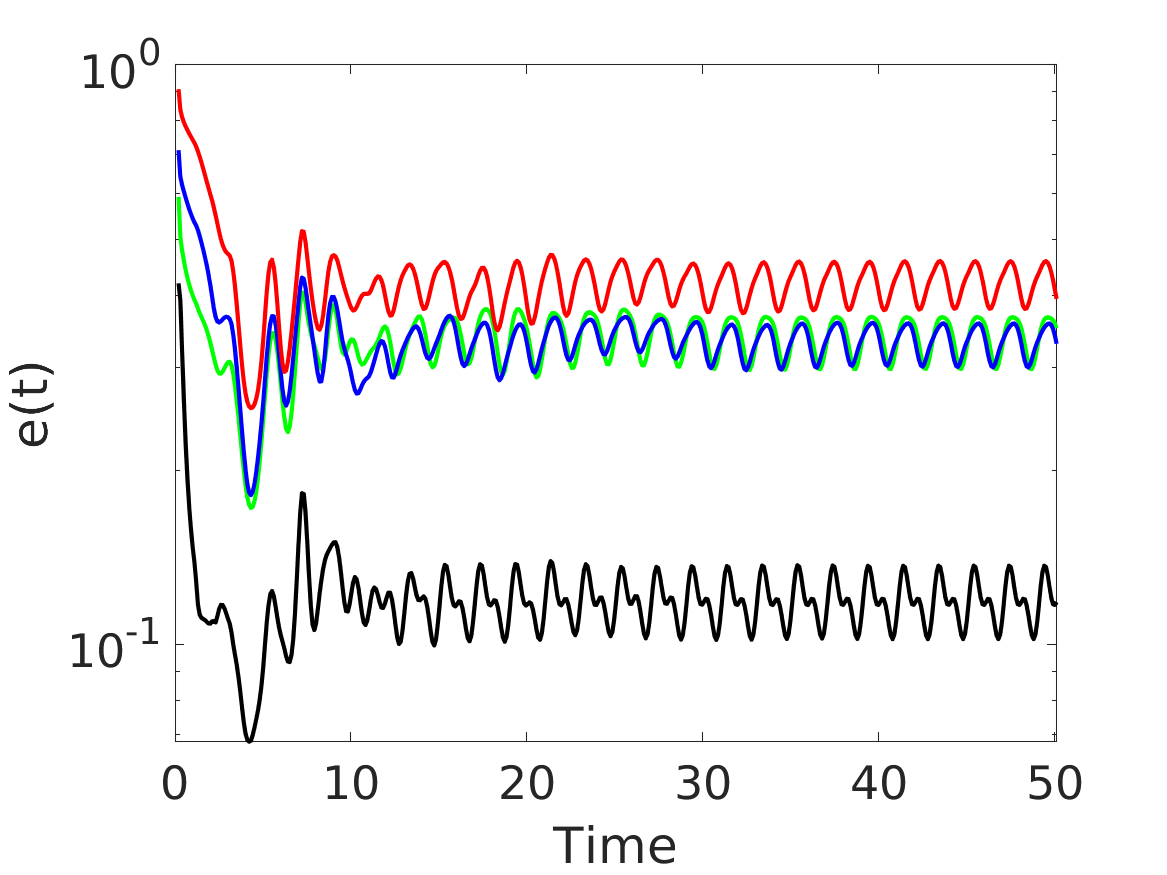}
    \rput(0.5,0.1){\psscalebox{0.5}{\color{black} \textbf{d)}}}
  \end{minipage}
  \centering
  \caption{Relative errors computed for pressure (a), u-velocity (b), v-velocity (c), and density (d). TI stands for tangential interpolation.
  The freestream is perturbed with the triangular wave.} 
   \label{tri_error}
\end{figure}

The impact of the size of the lower-dimensional space is next evaluated on the predictive performance of ROMs using the relative error in equation~\ref{rel_e}. Figure~\ref{tri_error} compares ERA ROM prediction errors in response to the triangular wave signal when the retained tangential modes capture $95\%$ and $80\%$ of the energy in the training snapshots matrix, and when the retained Hankel singular vectors capture $99.99\%$, $75\%$ and $65\%$ of the input-output energy. The predictive performance of ROMs is more sensitive to the number of balancing modes than the number of retained tangential modes during tangential interpolation. We chose to retain enough tangential modes to capture $80\%$ of the energy. In the meantime, the intermediate ROM with $75\%$ input-output energy content is sufficiently accurate and efficient for our application.

\begin{figure}[h!]
  \centering
  \begin{minipage}[a]{0.49\textwidth}
    \includegraphics[trim=4 -0.1 4 4, clip, width=\textwidth]{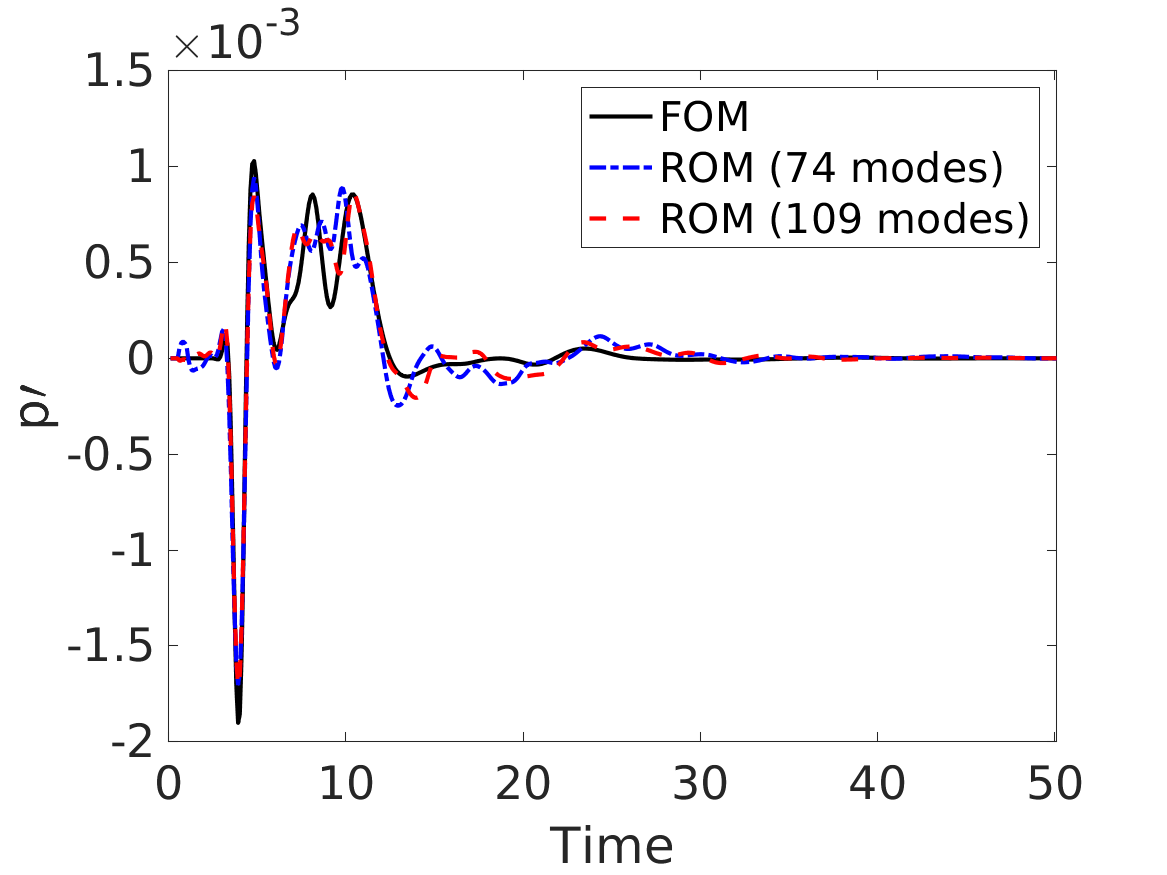}
    \rput(0.5,0.1){\psscalebox{0.5}{\color{black} \textbf{a)}}}
    \vspace{0.1cm}
  \end{minipage}
  \centering
  \begin{minipage}[a]{0.49\textwidth}
    \includegraphics[trim=4 -0.1 4 4, clip, width=\textwidth]{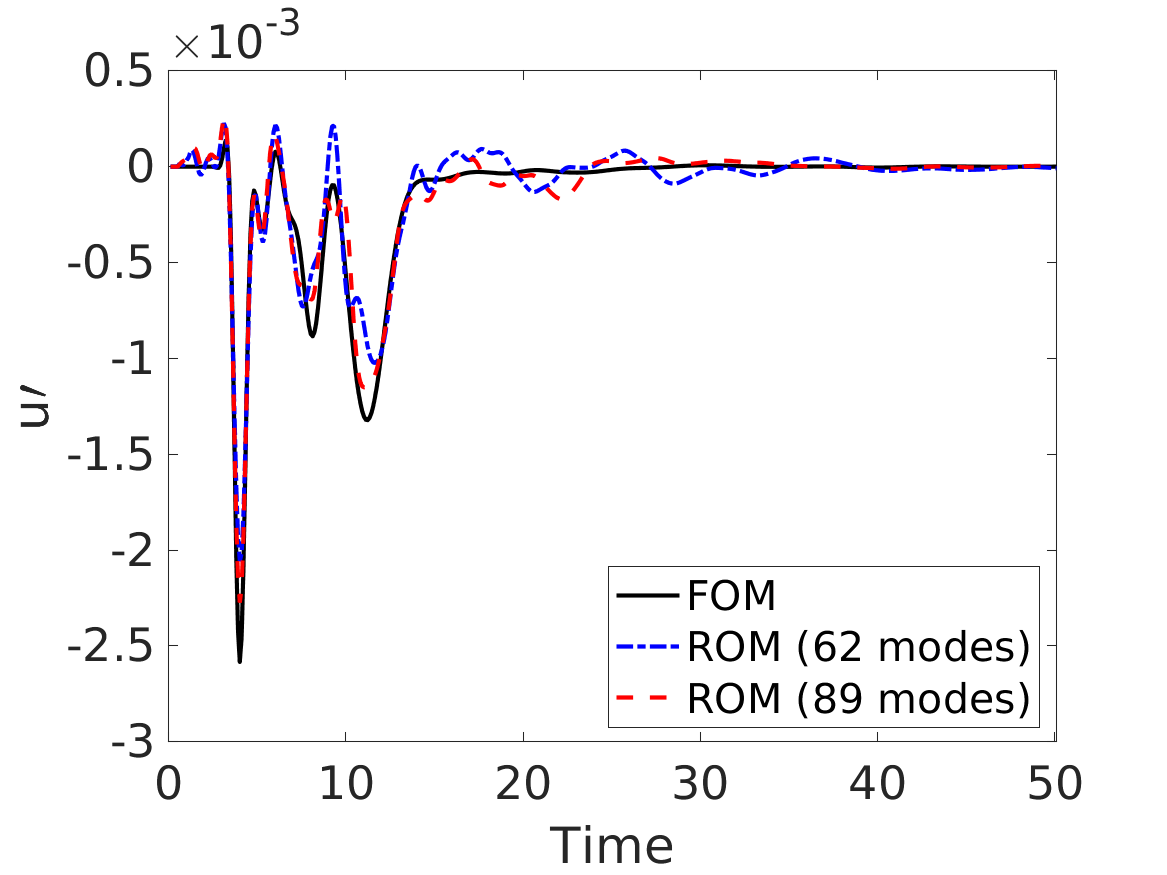}
    \rput(0.5,0.1){\psscalebox{0.5}{\color{black} \textbf{b)}}}
    \vspace{0.1cm}
  \end{minipage}
  \centering
  \begin{minipage}[a]{0.49\textwidth}
    \includegraphics[trim=4 -0.1 4 4, clip, width=\textwidth]{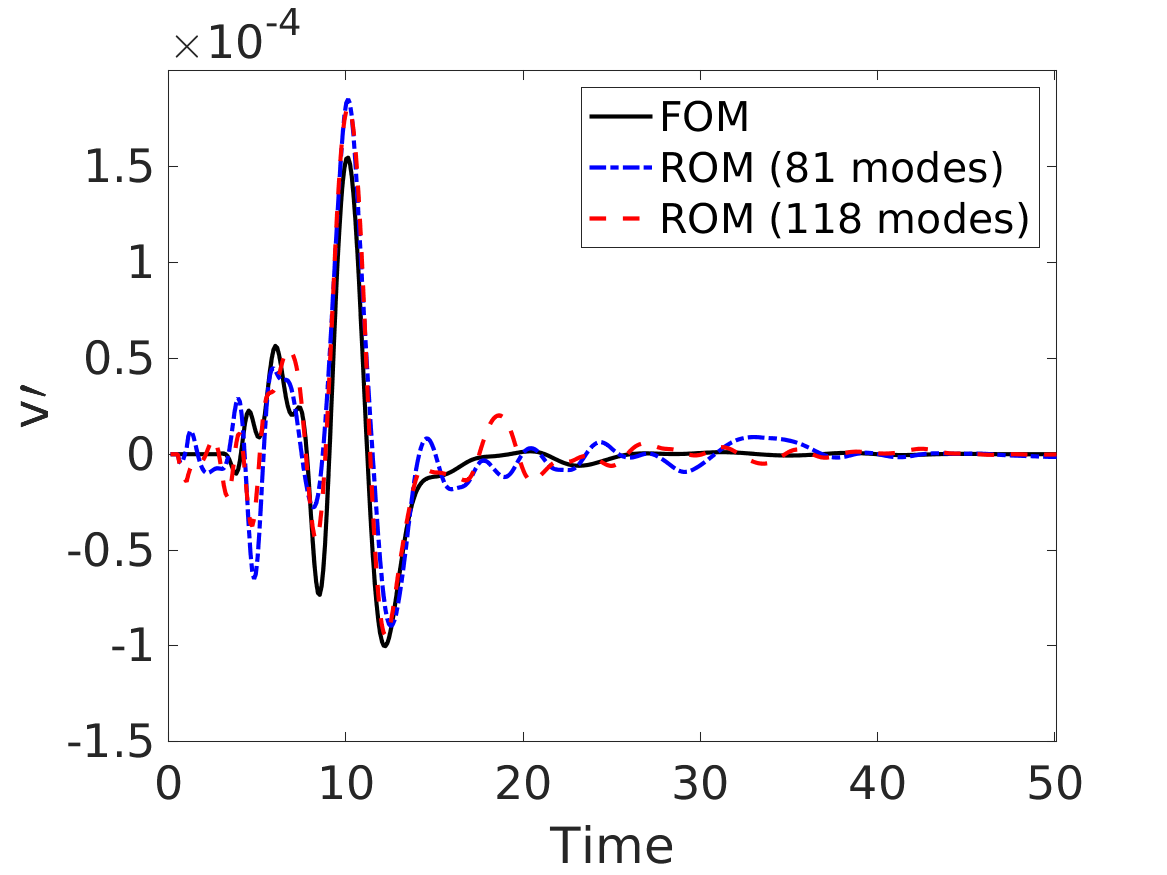}
    \rput(0.5,0.1){\psscalebox{0.5}{\color{black} \textbf{c)}}}
  \end{minipage}
  \centering
  \begin{minipage}[a]{0.49\textwidth}
    \includegraphics[trim=4 -0.1 4 4, clip, width=\textwidth]{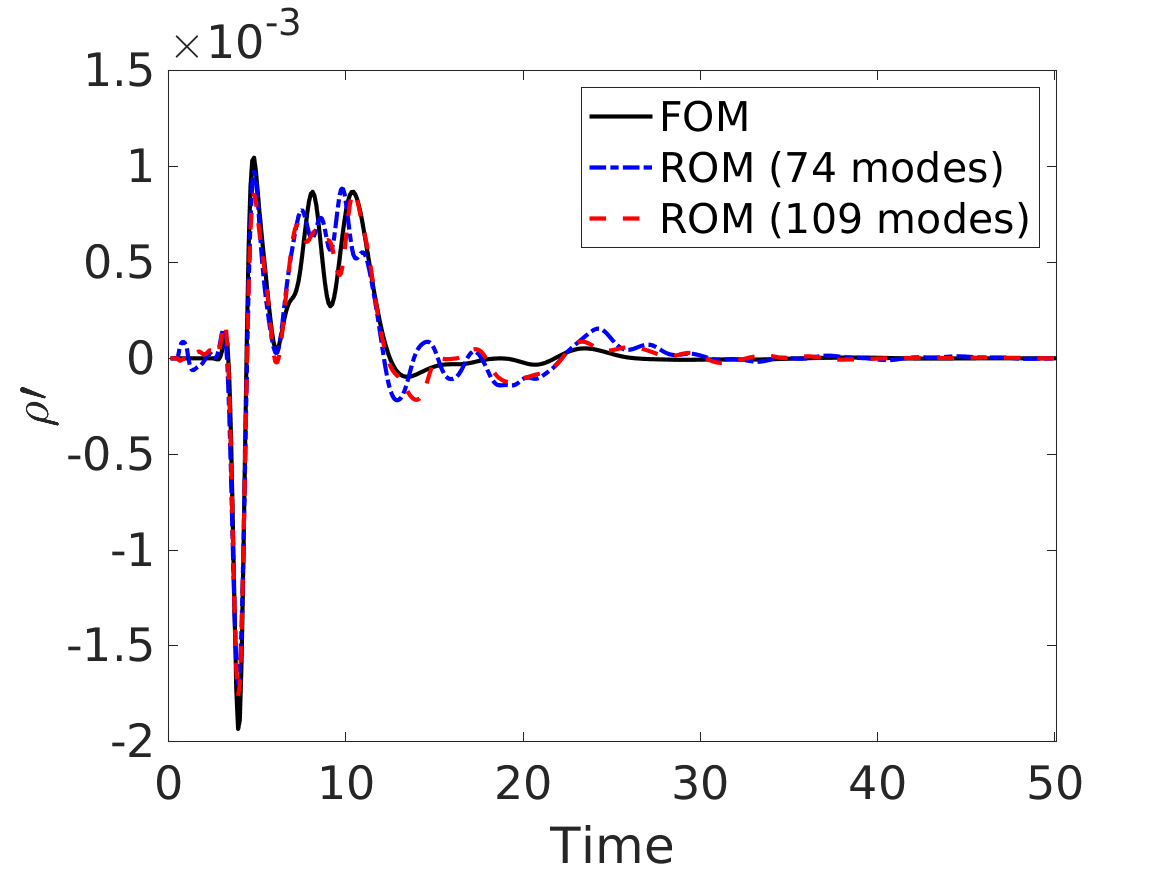}
    \rput(0.5,0.1){\psscalebox{0.5}{\color{black} \textbf{d)}}}
  \end{minipage}
  \centering
  \caption{Comparison of pressure (a), u-velocity (b), v-velocity (c), and density (d) computed by the HFM and ERA ROMs at $(x,y) = (0.3510,-0.1096)$. The freestream is perturbed with the non-periodic square wave.} 
   \label{step_probe}
\end{figure}

%\subsubsection{ROM Predictions for the Non-periodic Square Wave Input}
Next, we employ the ERA ROMs to predicted the acoustic response to the non-periodic square wave shown in equation~\ref{ug_square}. This type of input resembles transient dynamics that are more challenging for ROMs to capture. Figure~\ref{step_probe} compares the predictions by ERA ROMs with $65\%$ and $75\%$ input-output energy content against the FOM for a probe located at $(x,y) = (0.3510,-0.1096)$. Despite the non-periodic nature of the input signal, ROM predictions follow the FOM response closely. There are some numerical oscillations that are suppressed as we increase the number of modes. Other performance metrics are shown in \nameref{appB}.

In the context of ROMs, as enablers of rapid prediction of dynamics in many-query applications, computational efficiency is a key performance indicator. Although the offline cost of ROMs is typically dominant, but this portion of the computation cost is a one time price that is paid to achieve near real-time online computations. The online speed up factor determines whether a certain problem is amenable to model reduction. A comparison of the online computation time for the nonlinear FOM against the ROM specified in Table~\ref{t:tanint} indicates a speed up factor of 818.36. Note that specific applications may have lower thresholds on the accuracy of quantities of interest, which allows smaller ROM dimensions and higher online speed up factors.

\subsection{ROMs based on the Linearized Euler Solver} \label{result_linear}
\subsubsection{FOM}
\begin{figure}[h!]
  \centering
  \begin{minipage}[a]{0.49\textwidth}
    \includegraphics[trim=4 4 4 4, clip, width=\textwidth]{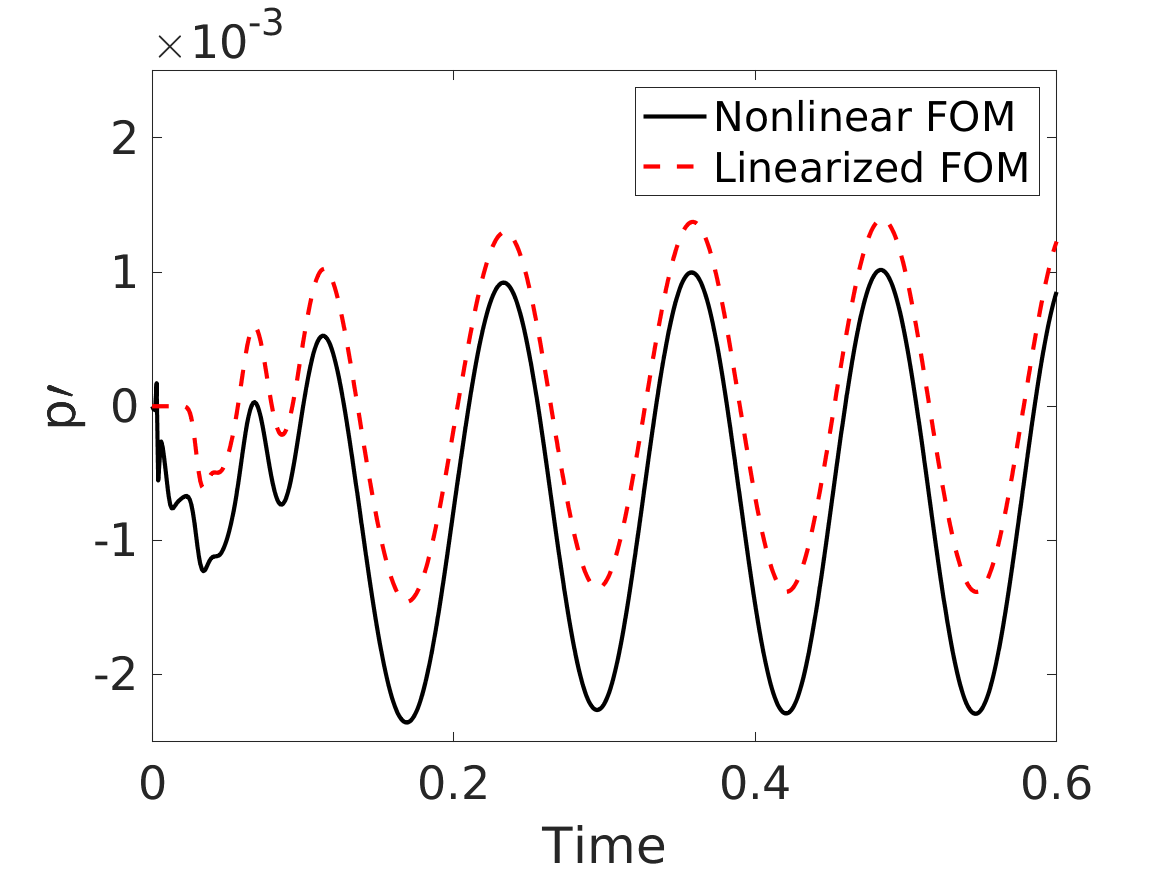}
    \rput(0.5,0.1){\psscalebox{0.01}{\color{black} \textbf{a)}}}
    \vspace{0.1cm}
  \end{minipage}
  \centering
  \begin{minipage}[a]{0.49\textwidth}
    \includegraphics[trim=4 4 4 4, clip, width=\textwidth]{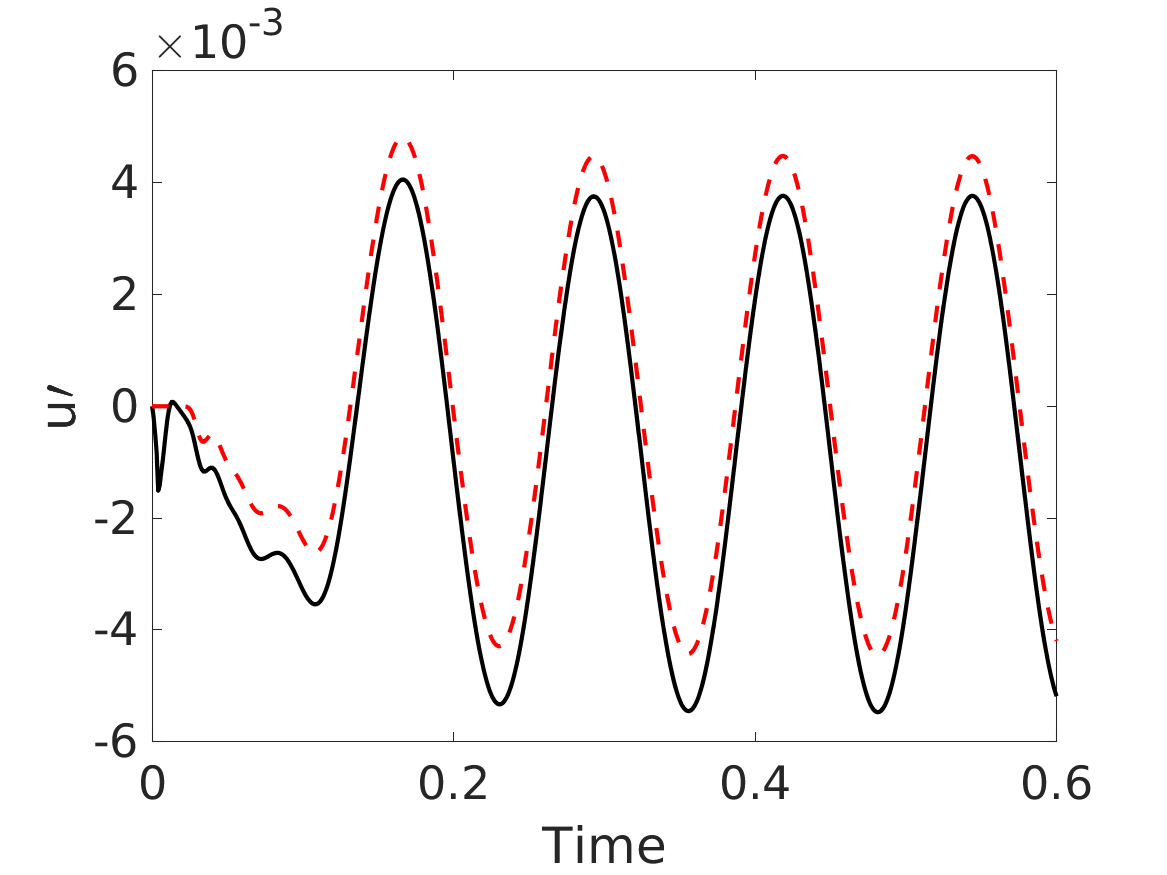}
    \rput(0.5,0.1){\psscalebox{0.01}{\color{black} \textbf{b)}}}
    \vspace{0.1cm}
  \end{minipage}
  \centering
  \begin{minipage}[a]{0.49\textwidth}
    \includegraphics[trim=4 4 4 0.1cm, clip, width=\textwidth]{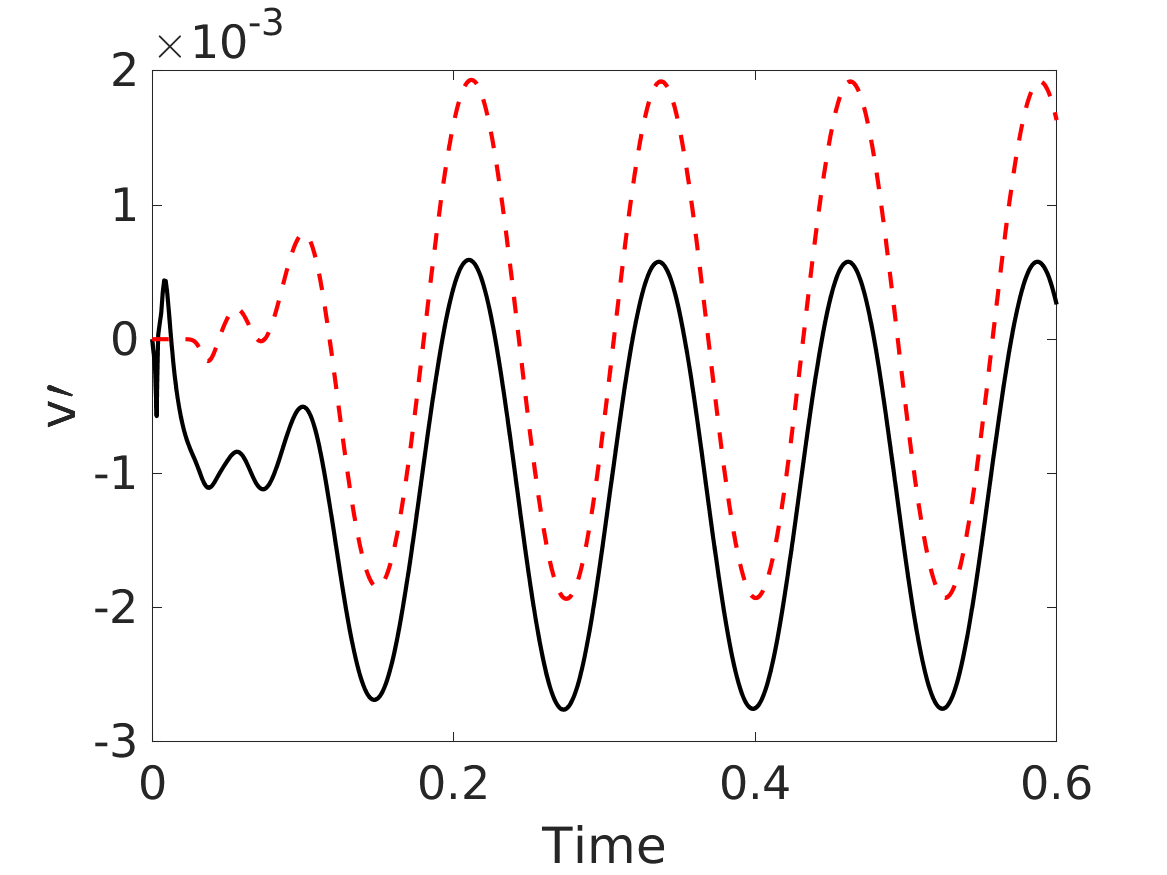}
    \rput(0.5,0.1){\psscalebox{0.5}{\color{black} \textbf{c)}}}
  \end{minipage}
  \centering
  \begin{minipage}[a]{0.49\textwidth}
    \includegraphics[trim=4 4 4 0.1cm, clip, width=\textwidth]{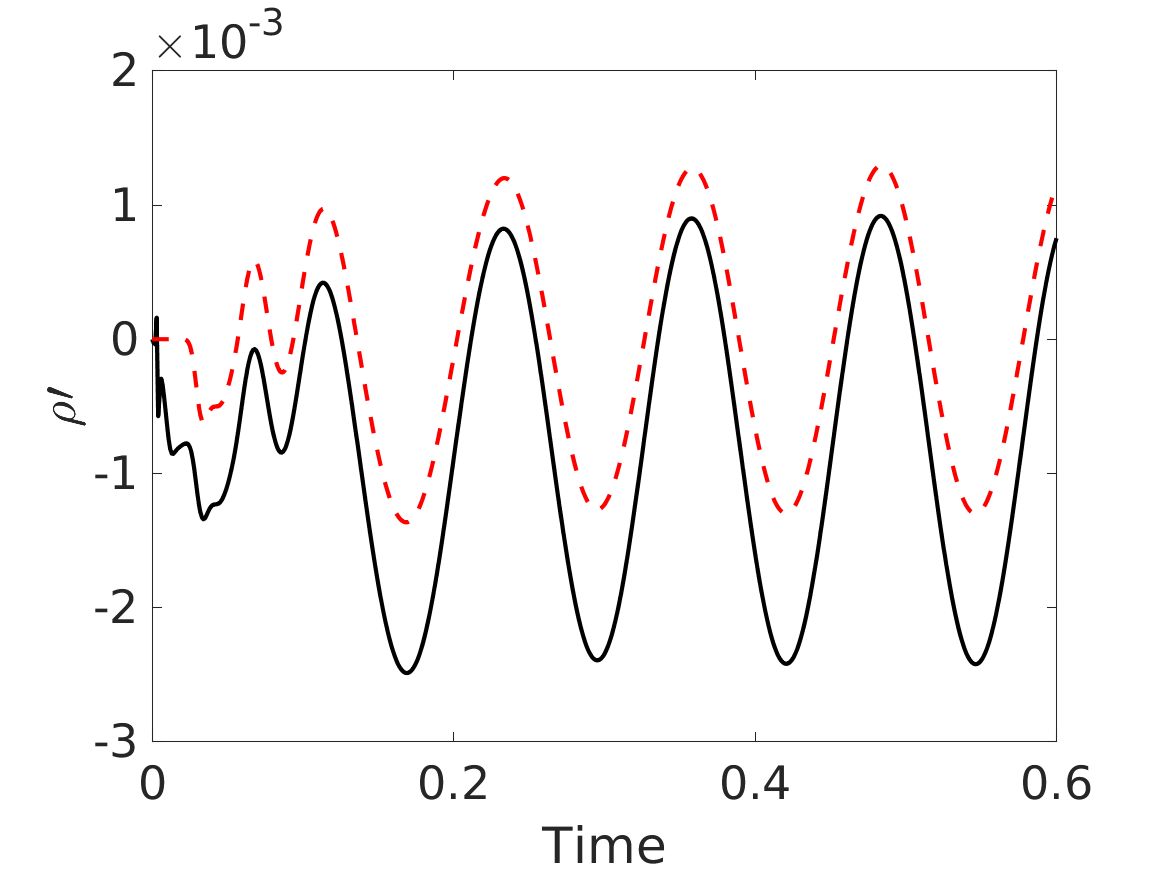}
    \rput(0.5,0.1){\psscalebox{0.5}{\color{black} \textbf{d)}}}
  \end{minipage}
  \centering
  \caption{Comparison of pressure (a), u-velocity (b), v-velocity (c), and density (d) computed by the linearized and nonlinear FOMs at $(x, y) = (0.4628, -0.3129)$. The freestream is perturbed with the sinusoidal input.}
   \label{sine_probe_lvsn_50_30}
\end{figure}

\begin{figure}[h!]
  \centering
  \begin{minipage}[a]{0.49\textwidth}
    \includegraphics[trim=4 4 4 4, clip, width=\textwidth]{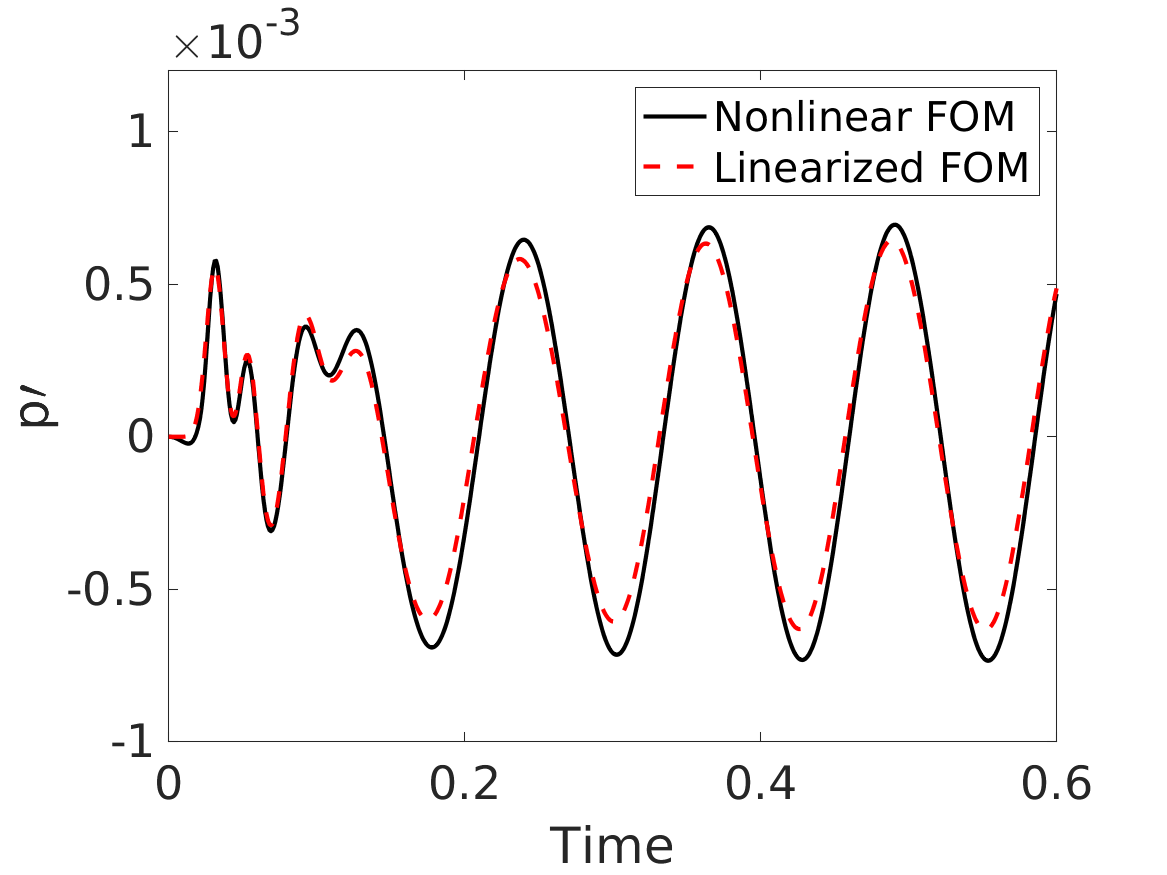}
    \rput(0.5,0.1){\psscalebox{0.01}{\color{black} \textbf{a)}}}
    \vspace{0.1cm}
  \end{minipage}
  \centering
  \begin{minipage}[a]{0.49\textwidth}
    \includegraphics[trim=4 4 4 4, clip, width=\textwidth]{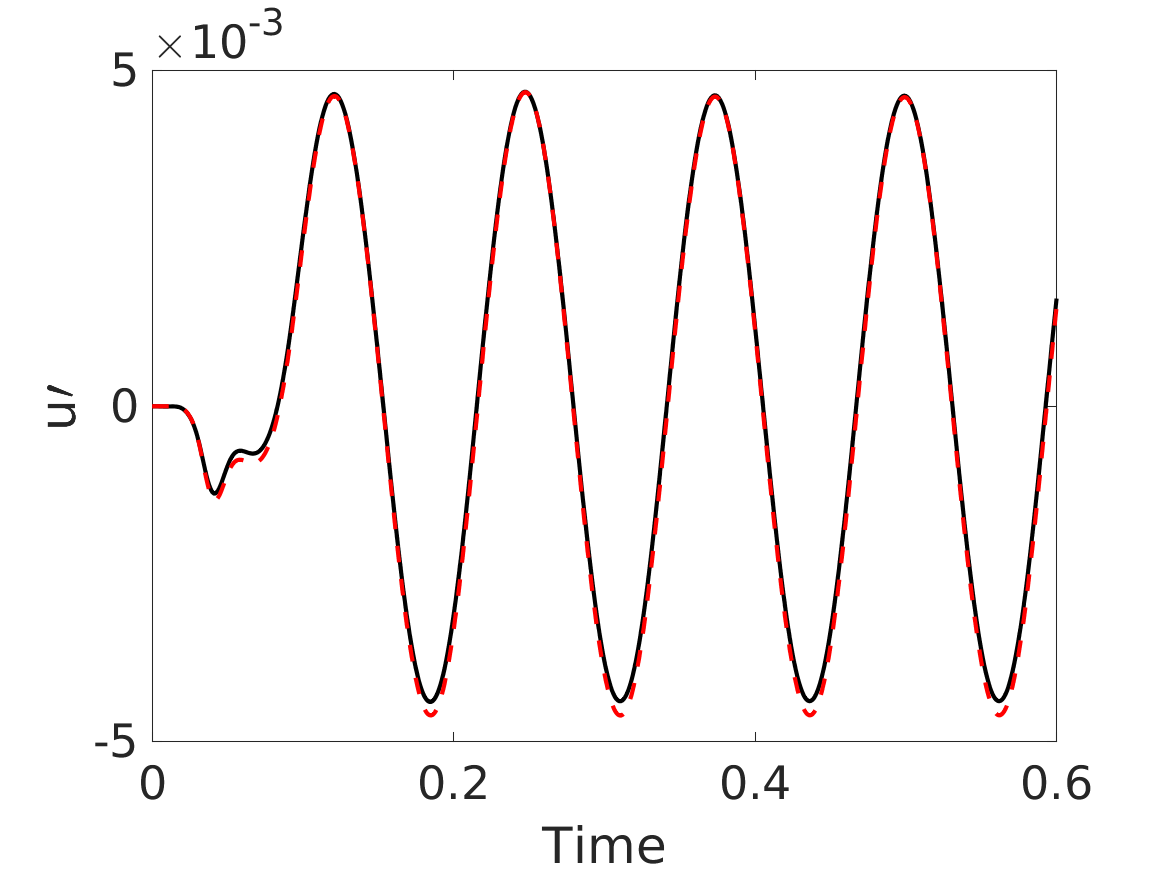}
    \rput(0.5,0.1){\psscalebox{0.01}{\color{black} \textbf{b)}}}
    \vspace{0.1cm}
  \end{minipage}
  \centering
  \begin{minipage}[a]{0.49\textwidth}
    \includegraphics[trim=4 4 4 0.1cm, clip, width=\textwidth]{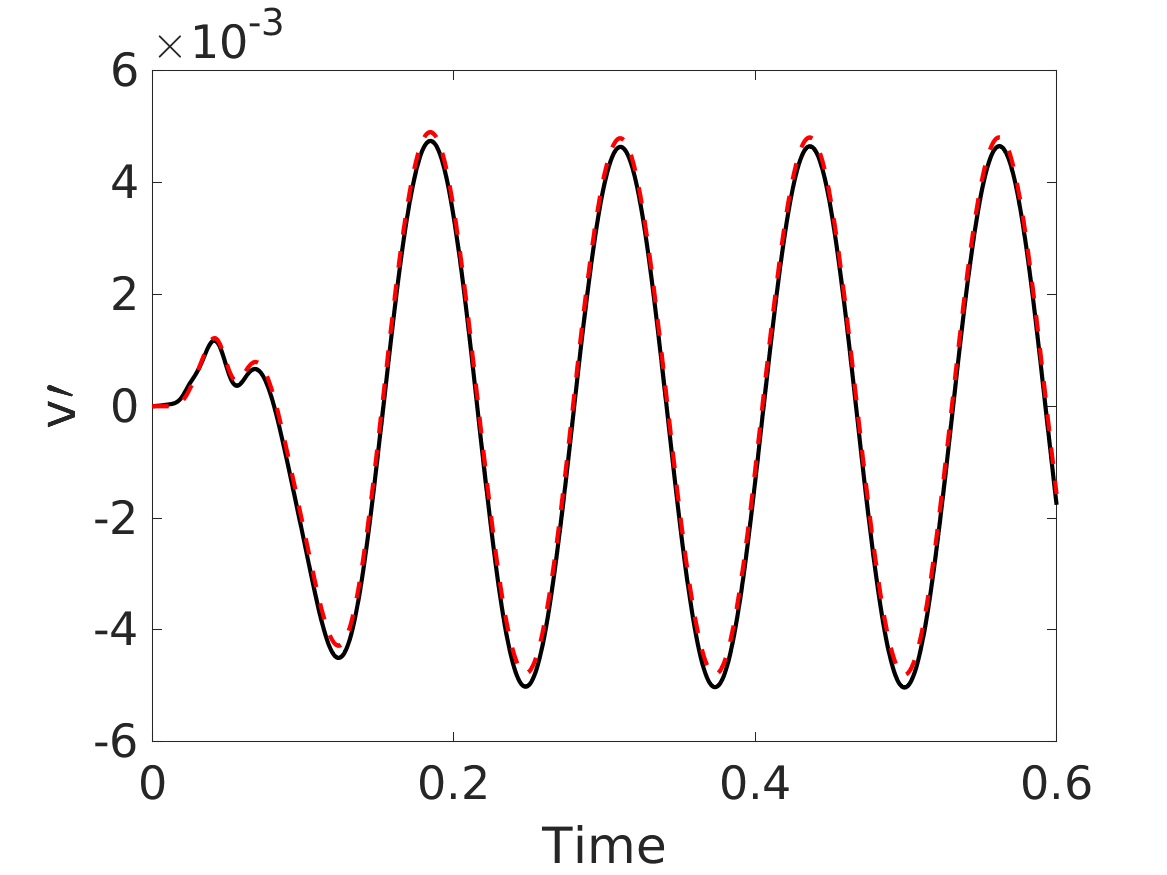}
    \rput(0.5,0.1){\psscalebox{0.5}{\color{black} \textbf{c)}}}
  \end{minipage}
  \centering
  \begin{minipage}[a]{0.49\textwidth}
    \includegraphics[trim=4 4 4 0.1cm, clip, width=\textwidth]{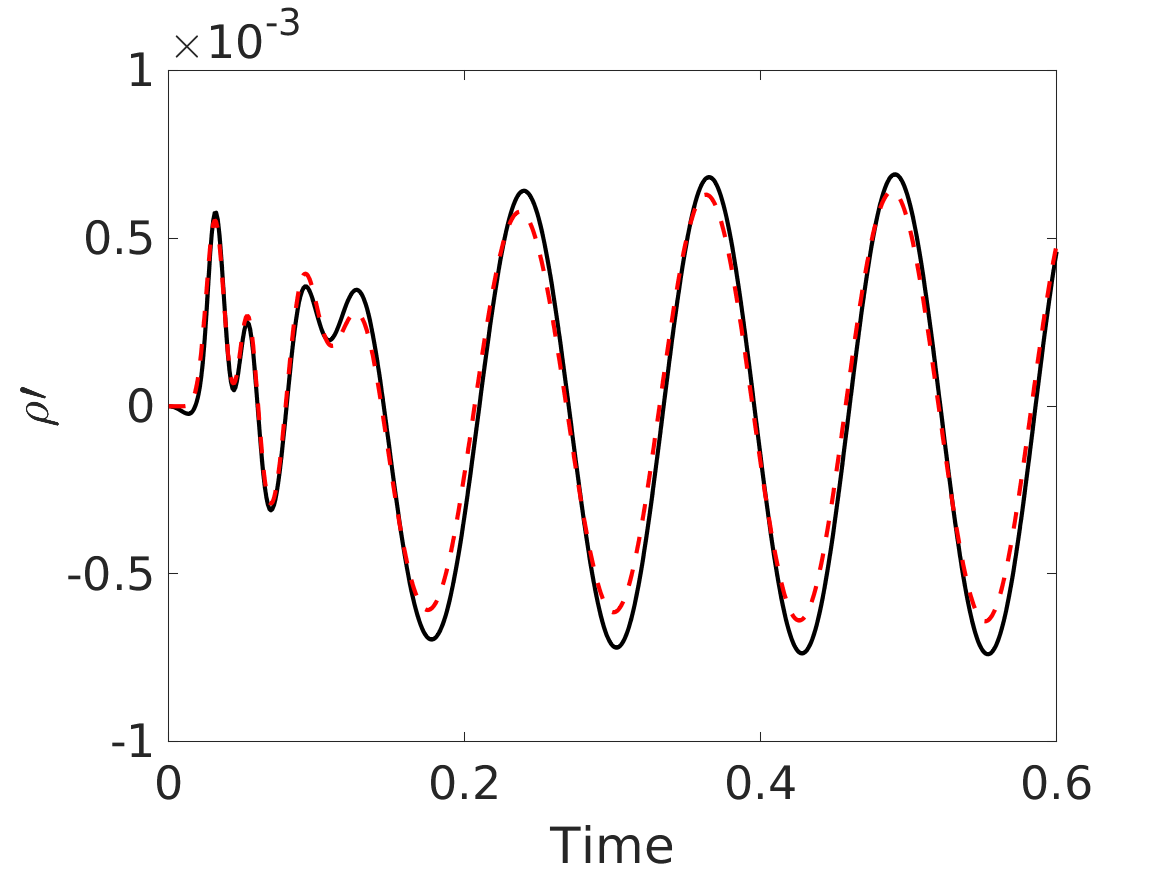}
    \rput(0.5,0.1){\psscalebox{0.5}{\color{black} \textbf{d)}}}
  \end{minipage}
  \centering
  \caption{Comparison of pressure (a), u-velocity (b), v-velocity (c), and density (d) computed by the linearized and nonlinear FOMs at $(x, y) = (0.7486, -2.6112)$. The freestream is perturbed with the sinusoidal input. }
   \label{sine_probe_lvsn_100_80}
\end{figure}

In this section, the two-dimensional linearized Euler equations are used to generate the training snapshots with the same training setup as in the nonlinear solver. Here, the linearized FOM Jacobians are computed using the steady-state solution obtained by the nonlinear solver. 
In figures~\ref{sine_probe_lvsn_50_30} and \ref{sine_probe_lvsn_100_80}, pressure, velocity, and density probes at (x, y)=(0.4628, -0.3129) and (x, y)=(0.7486, -2.6112) are shown, respectively, to compare the solutions obtained by solving the nonlinear and linearized Euler equations using a computational grid with $401 \times 101$ cells. Note that unlike the results shown in the previous sections, for which we used a second-order scheme, here, we use a first-order reconstruction for the two FOMs to compute Roe fluxes.
It can be seen that at location (x, y)=(0.4628, -0.3129), which is close to the bottom of the airfoil, the first-order nonlinear solver drifts away from the solution obtained by the linearized solver. We attribute this to the accumulation of numerical errors in the first-order nonlinear solver. 
Figure~\ref{sine_probe_lvsn_100_80} shows a better agreement between the first-order linearized and nonlinear solvers at (x, y)=(0.7486, -2.6112), which is close to the far-field boundary that is not affected by the the acoustic response of the airfoil.

\subsubsection{ROM Training}
For this case also the ERA ROMs are trained by the snapshots from the response of the system to the Gaussian-shaped input with zero mean and a standard deviation of 325. Following the strategy described at the beginning of the section, the coarse grid with $101 \times 51$ cells is used to evaluate the ROMs. The same sampling properties as the nonlinear solver are used here to collect the training snapshots. 
Table~\ref{t:tanintlin} shows the number of the retained modes used to build ERA ROMs for the four variables. The dimension of ROMs is chosen such that the remaining modes in the system capture $80\%$ of the input-output energy. Similarly, the retained tangential modes capture $80\%$ of the energy in the training snapshots.
\begin{table}[h!]
 \begin{center}
  \caption{The number of the retained tangential modes for the ROM of each variable out of a total of 5000 left and 204 right tangential modes and the dimension of the balanced ROMs.}
  \label{t:tanintlin}
  \begin{tabular}{lllll}\hline
        & Variable & Left modes & Right modes & ROM dimension \\\hline
        & Pressure & 118 & 46 & 67 \\\hline
        & Velocity (u) & 148 & 77 & 56  \\\hline
        & Velocity (v) & 129 & 52 & 73  \\\hline
        & Density & 121 & 47 & 68  \\\hline
  \end{tabular}
 \end{center}
\end{table}

\begin{figure}[h!]
  \centering
  \begin{minipage}[a]{0.43\textwidth}
    \includegraphics[trim=4 4 4 4, clip, width=\textwidth]{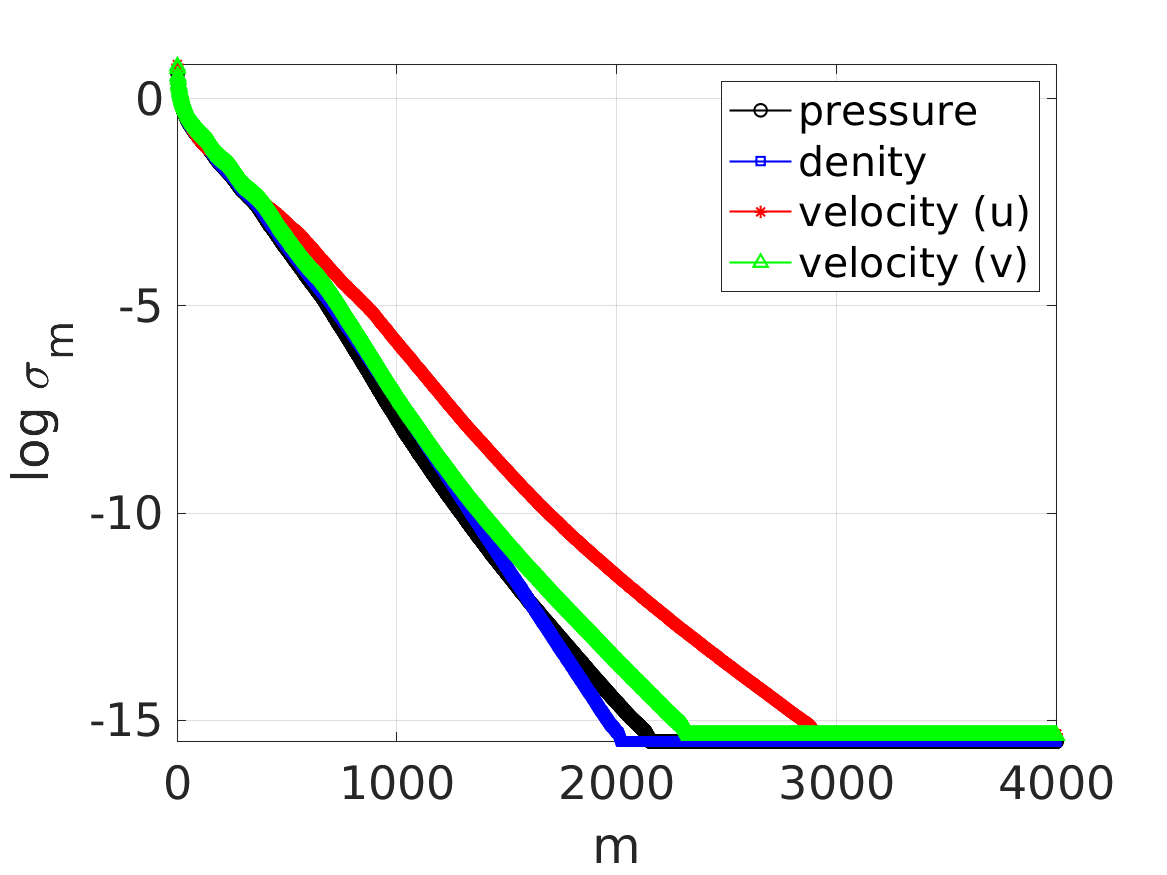}
    \rput(0.5,0.1){\psscalebox{0.5}{\color{black} \textbf{a)}}}
  \end{minipage}
  \centering
  \begin{minipage}[a]{0.43\textwidth}
    \includegraphics[trim=4 -0.1 4 4, clip, width=\textwidth]{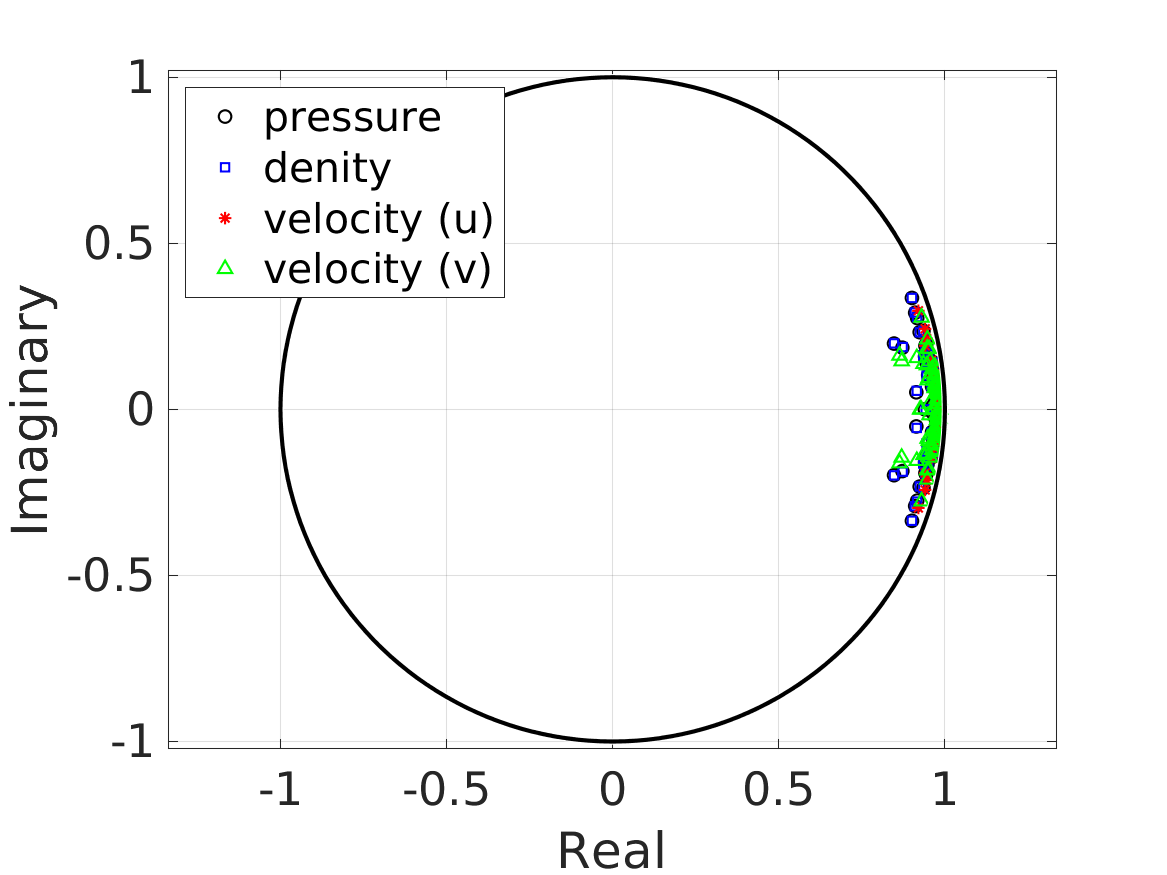}
    \rput(0.5,0.1){\psscalebox{0.5}{\color{black} \textbf{b)}}}
  \end{minipage}
  \centering
  \caption{a) Singular values of the Hankel matrix. b) Eigenvalues of the ERA ROMs. The ERA ROMs are constructed with enough balancing modes to capture $80\%$ of the input-output energy.}
   \label{sv_rom_eigs_linear}
\end{figure}

Figure~\ref{sv_rom_eigs_linear} shows the decay of the Hankel singular values for the different variables, along with the eigenvalues of the ERA ROMs. Clearly, when the Hankel matrix is constructed using the training snapshots from the linearized FOM, the singular values decay faster than in Figure~\ref{sv_rom_eigs} that was based on the training snapshots from the nonlinear FOM. This is also shown in Table~\ref{t:tanintlin}, where ROMs based on the linearized FOM need fewer modes to capture more energy compared to those based on the nonlinear FOM in section~\ref{result_nonlinear}.
For the ROMs in Table~\ref{t:tanintlin}, Figure~\ref{sv_rom_eigs_linear} (b) shows that the entire eigenvalue spectrum is inside the unit circle and the created ROMs are stable.

\subsubsection{ROM Prediction}
\begin{figure}[h!]
  \centering
  \begin{minipage}[a]{0.49\textwidth}
    \includegraphics[trim=4 0.1cm 4 4, clip, width=\textwidth]{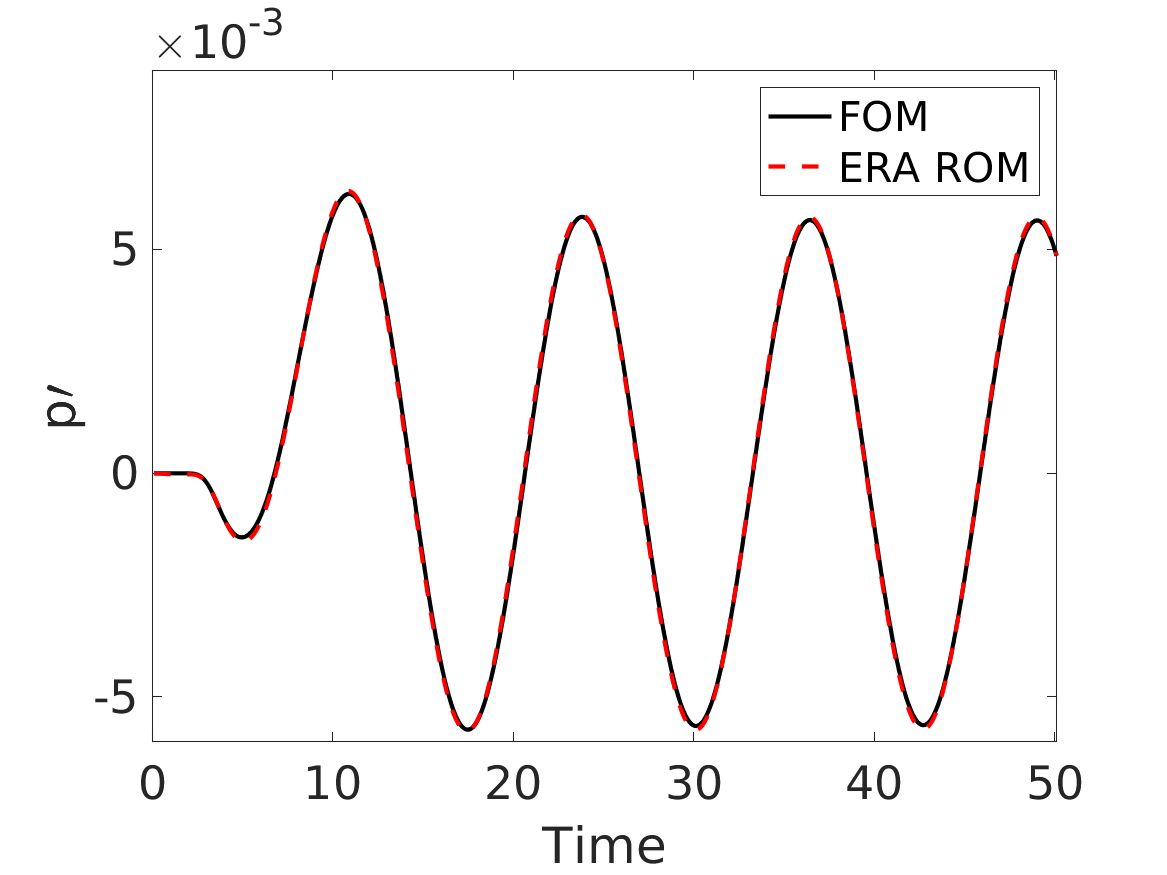}
    \rput(0.5,0.1){\psscalebox{0.5}{\color{black} \textbf{a)}}}
    \vspace{0.1cm}
  \end{minipage}
  \centering
  \begin{minipage}[a]{0.49\textwidth}
    \includegraphics[trim=4 0.1cm 4 4, clip, width=\textwidth]{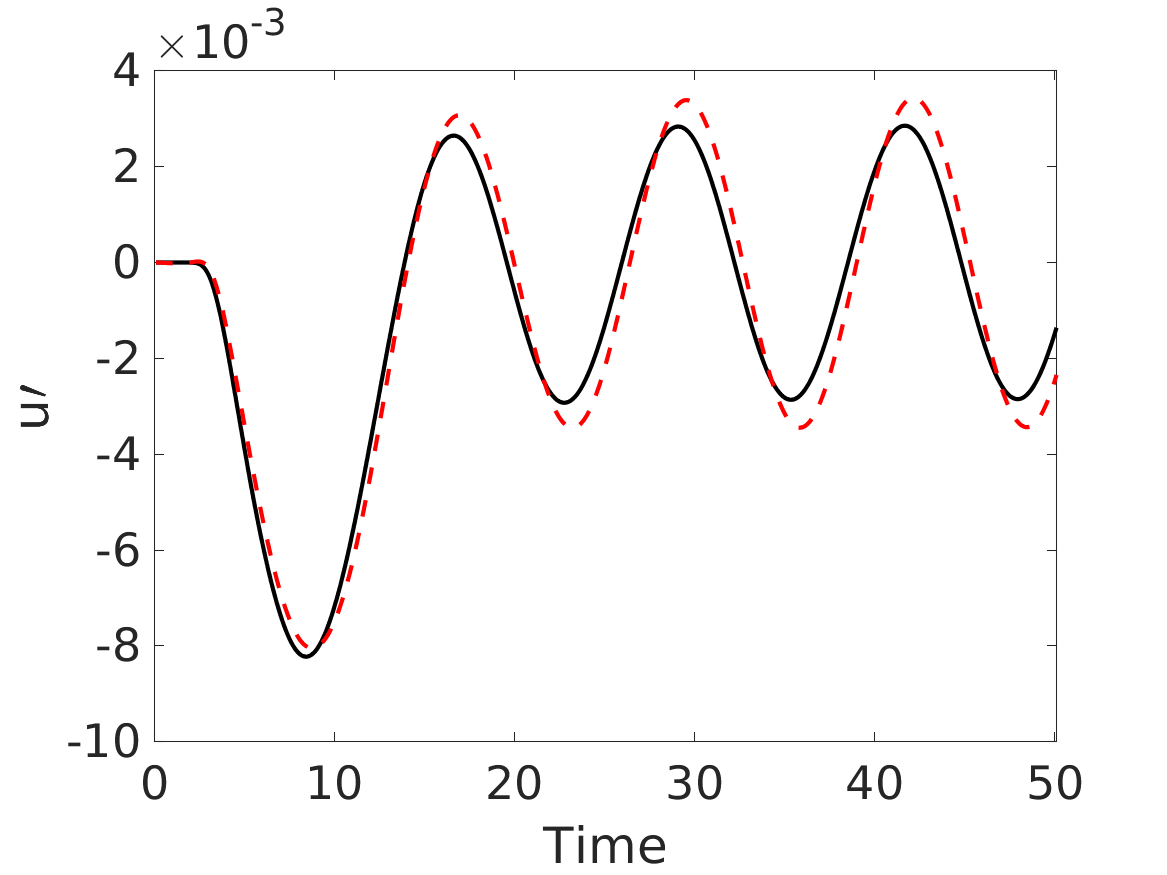}
    \rput(0.5,0.1){\psscalebox{0.5}{\color{black} \textbf{b)}}}
    \vspace{0.1cm}
  \end{minipage}
  \centering
  \begin{minipage}[a]{0.49\textwidth}
    \includegraphics[trim=4 0.1cm 4 4, clip, width=\textwidth]{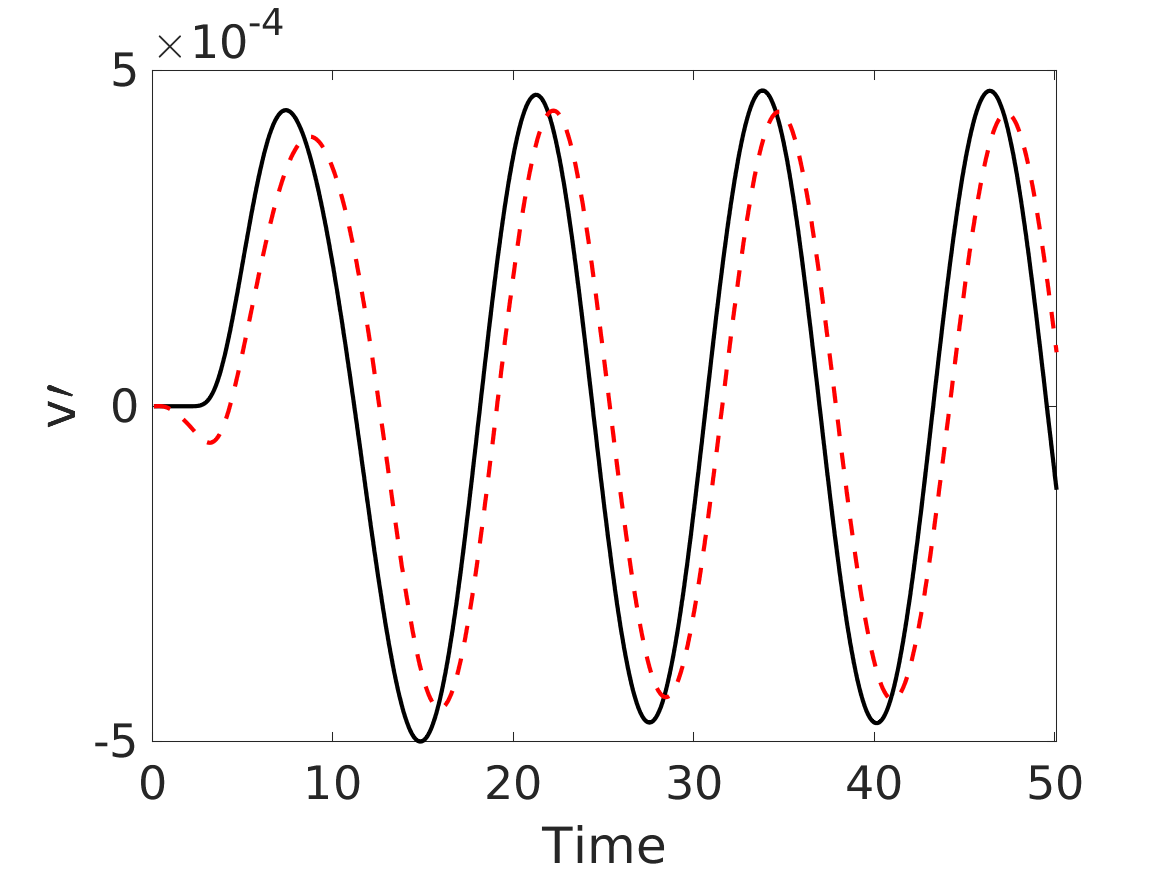}
    \rput(0.5,0.1){\psscalebox{0.5}{\color{black} \textbf{c)}}}
  \end{minipage}
  \centering
  \begin{minipage}[a]{0.49\textwidth}
    \includegraphics[trim=4 0.1cm 4 4, clip, width=\textwidth]{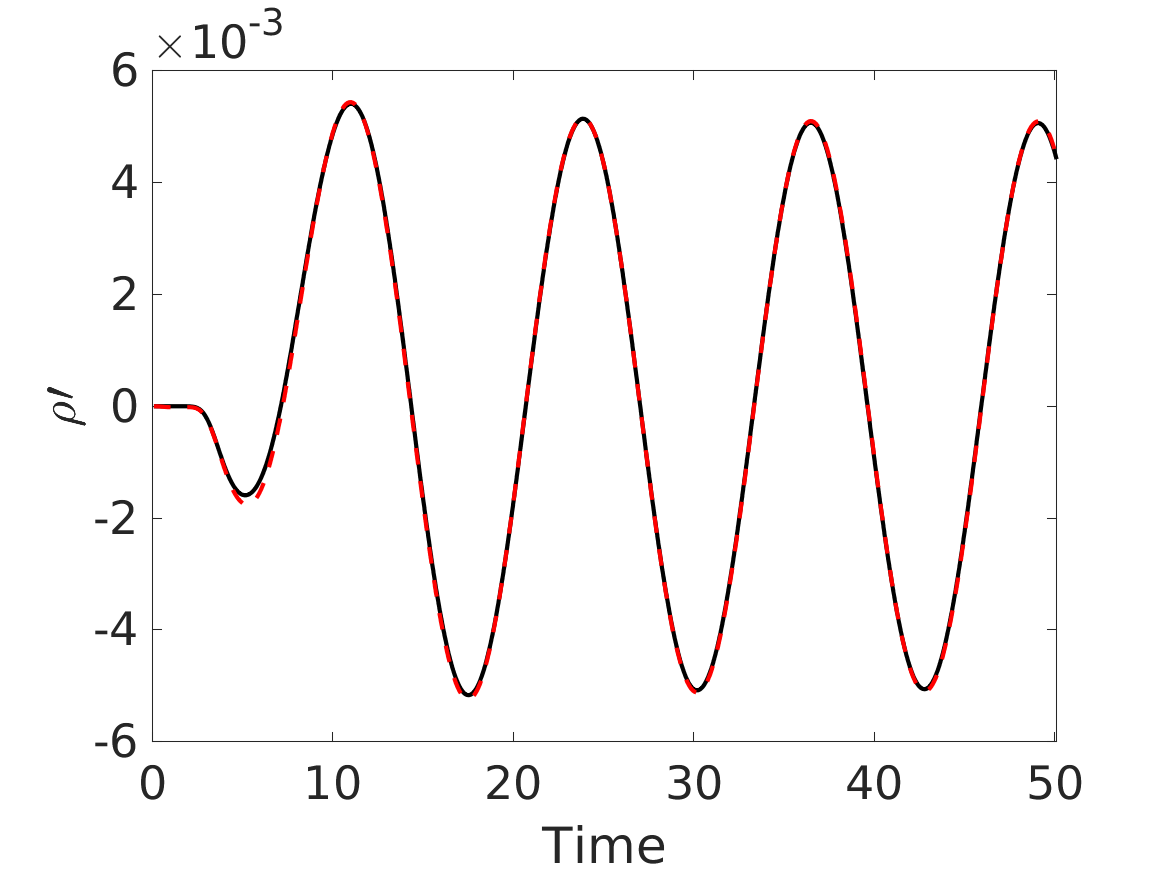}
    \rput(0.5,0.1){\psscalebox{0.5}{\color{black} \textbf{d)}}}
  \end{minipage}
  \centering
  \caption{Comparison of pressure (a), u-velocity (b), v-velocity (c), and density (d) computed by the linearized FOM and the ERA ROM at $(x,y) = (0.3510,-0.1096)$. The freestream is perturbed with the sinusoidal input.}
   \label{sine_probe_linear}
\end{figure}

%\subsubsection{ROM Predictions for the Sinusoidal Input}
The sinusoidal gust in equation~\ref{ugtest} is used to perturb the flow at the far-field boundary to evaluate the predictive performance of ROMs based on the linearized FOM. Figure~\ref{sine_probe_linear} compares the pressure, velocity, and density probes evaluated by the ERA ROMs against those computed by the linearized FOM. The probes are located at $(x,y) = (0.3510,-0.1096)$. The solution predicted by the ROM is in good agreement with the linearized FOM.

\begin{figure}[h!]
  \centering
  \begin{minipage}[a]{0.49\textwidth}
    \includegraphics[trim=4 4 4 -0.05, clip, width=\textwidth]{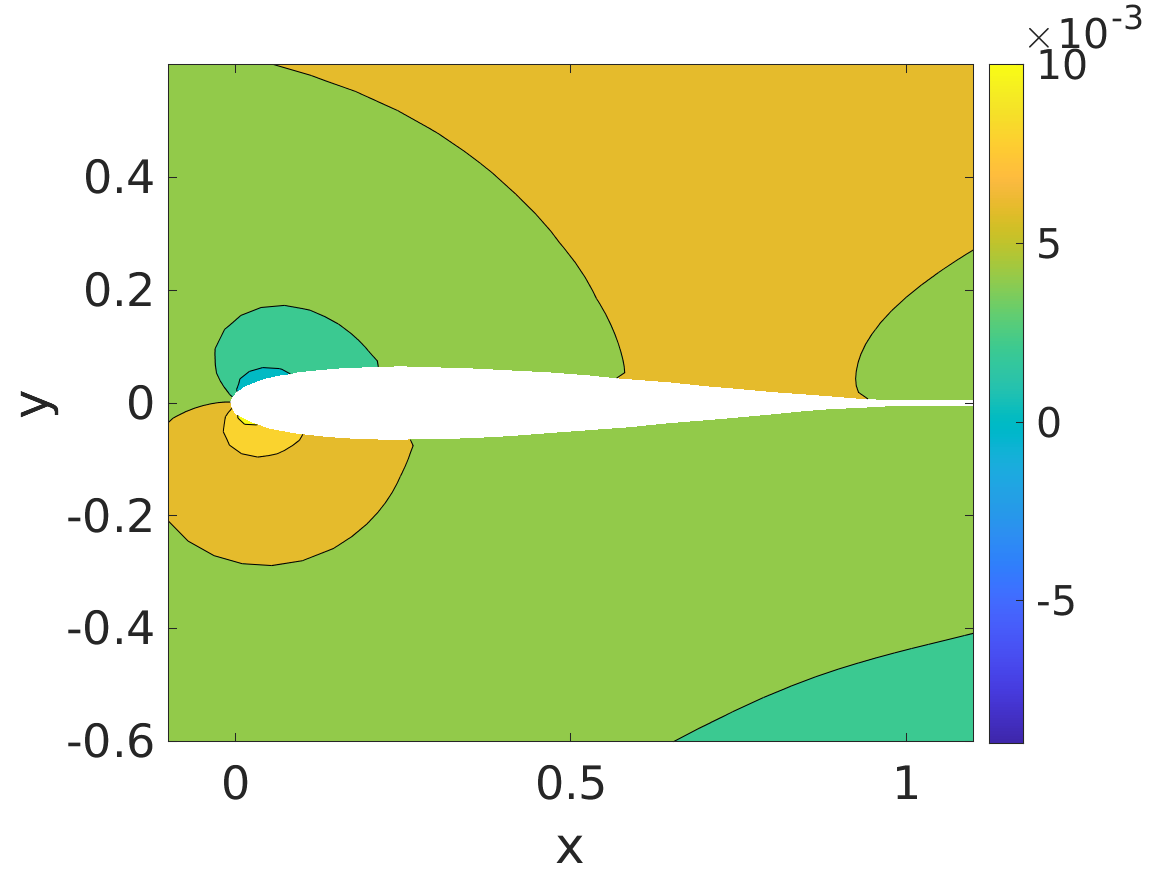}
    \rput(0.5,0.1){\psscalebox{0.5}{\color{black} \textbf{a)}}}
  \end{minipage}
  \centering
  \begin{minipage}[a]{0.49\textwidth}
    \includegraphics[trim=4 4 4 -0.05, clip, width=\textwidth]{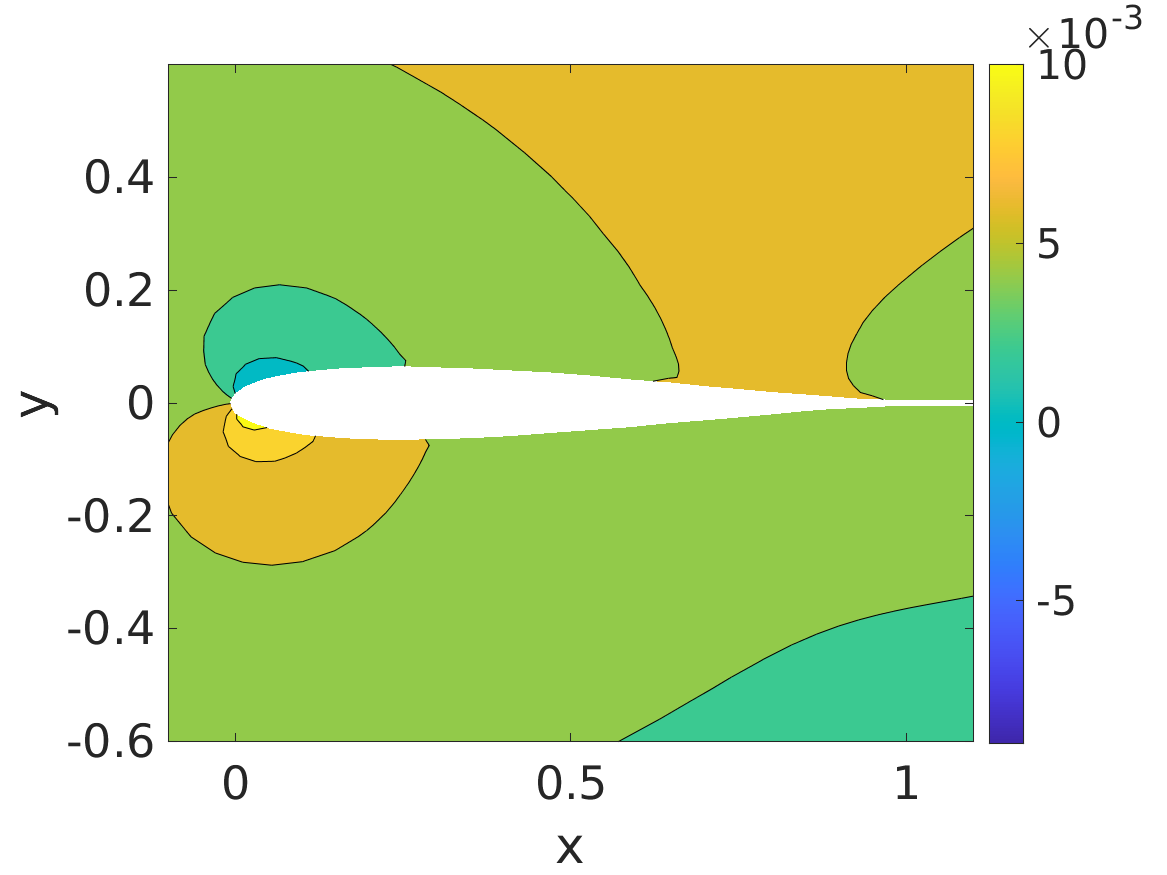}
    \rput(0.5,0.1){\psscalebox{0.5}{\color{black} \textbf{b)}}}
  \end{minipage}
  \centering
  \caption{Pressure contours computed by the linearized FOM (a) and ERA ROM (b) at $t=48.5$. The freestream velocity is perturbed by the sinusoidal input.}
   \label{pcontour_linear}
\end{figure}

A broader look at the pressure perturbation contours in Figure~\ref{pcontour_linear} at $t=48.5$ also demonstrates the accuracy of ROMs trained with the snapshots generated based on the linearized FOM.

\begin{figure}[h!]
  \centering
  \begin{minipage}[a]{0.49\textwidth}
    \includegraphics[trim=4 -0.1 4 4, clip, width=\textwidth]{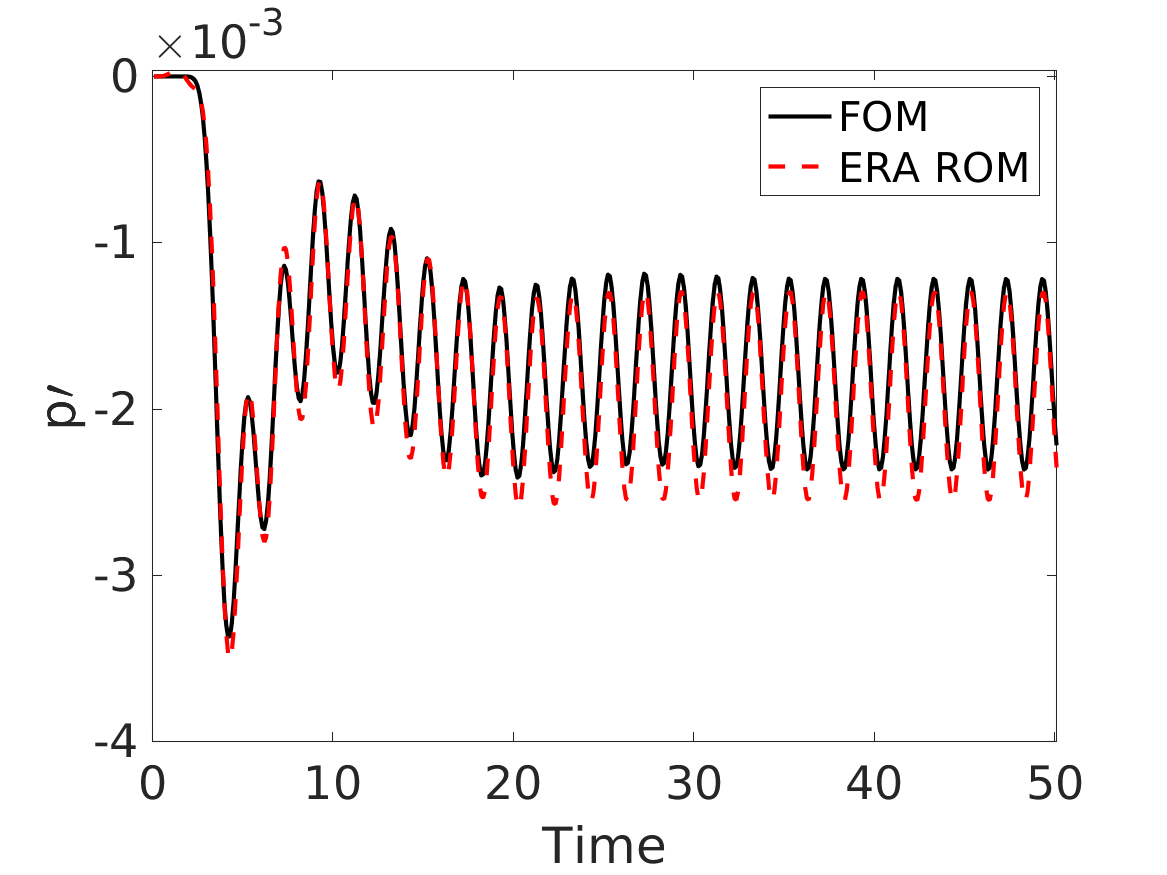}
    \rput(0.5,0.1){\psscalebox{0.5}{\color{black} \textbf{a)}}}
    \vspace{0.1cm}
  \end{minipage}
  \centering
  \begin{minipage}[a]{0.49\textwidth}
    \includegraphics[trim=4 -0.1 4 4, clip, width=\textwidth]{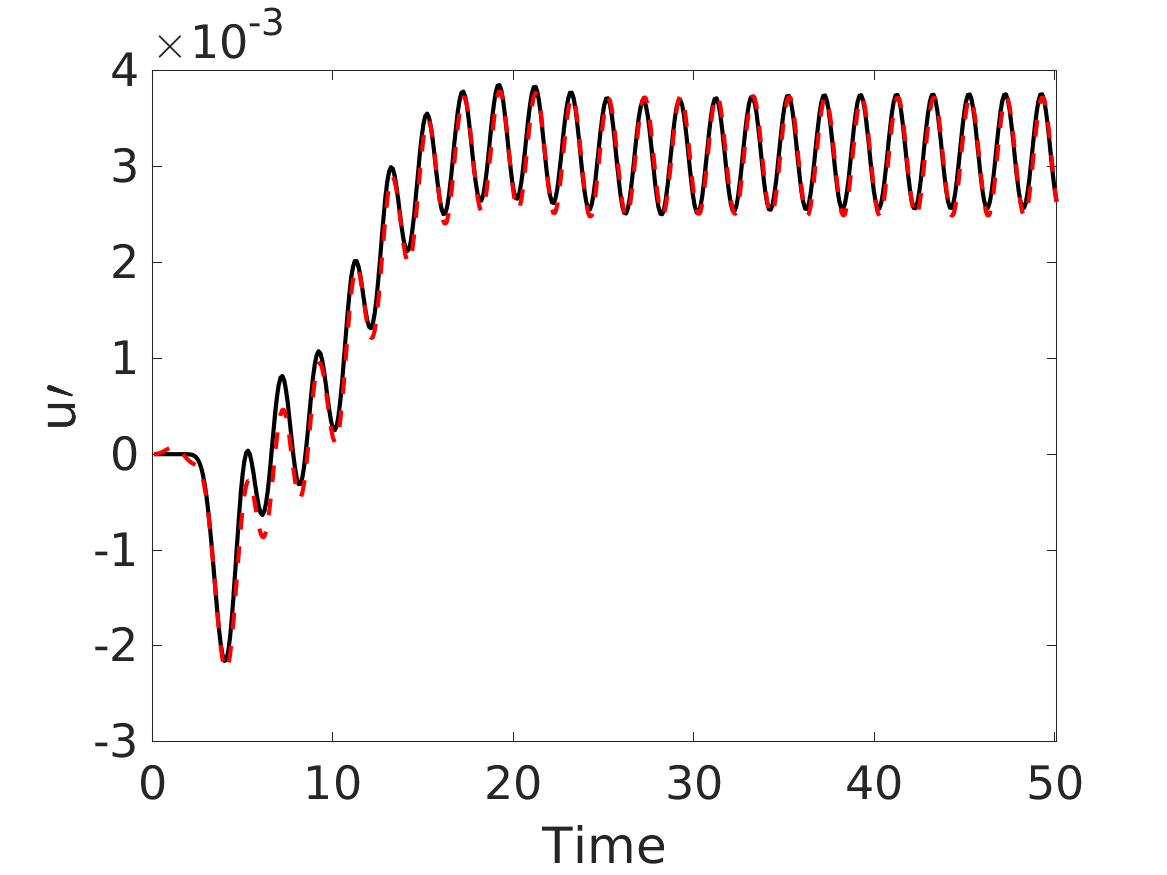}
    \rput(0.5,0.1){\psscalebox{0.5}{\color{black} \textbf{b)}}}
    \vspace{0.1cm}
  \end{minipage}
  \centering
  \begin{minipage}[a]{0.49\textwidth}
    \includegraphics[trim=4 -0.1 4 4, clip, width=\textwidth]{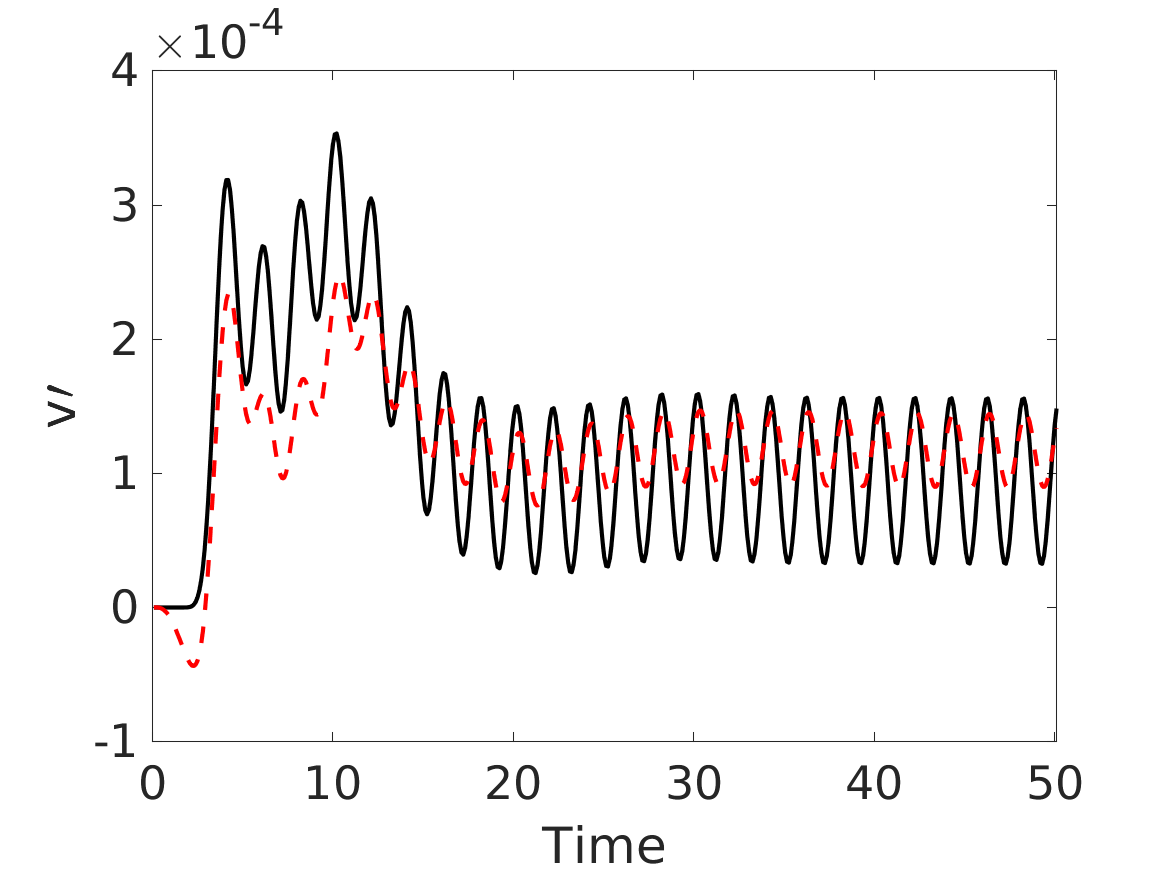}
    \rput(0.5,0.1){\psscalebox{0.5}{\color{black} \textbf{c)}}}
  \end{minipage}
  \centering
  \begin{minipage}[a]{0.49\textwidth}
    \includegraphics[trim=4 -0.1 4 4, clip, width=\textwidth]{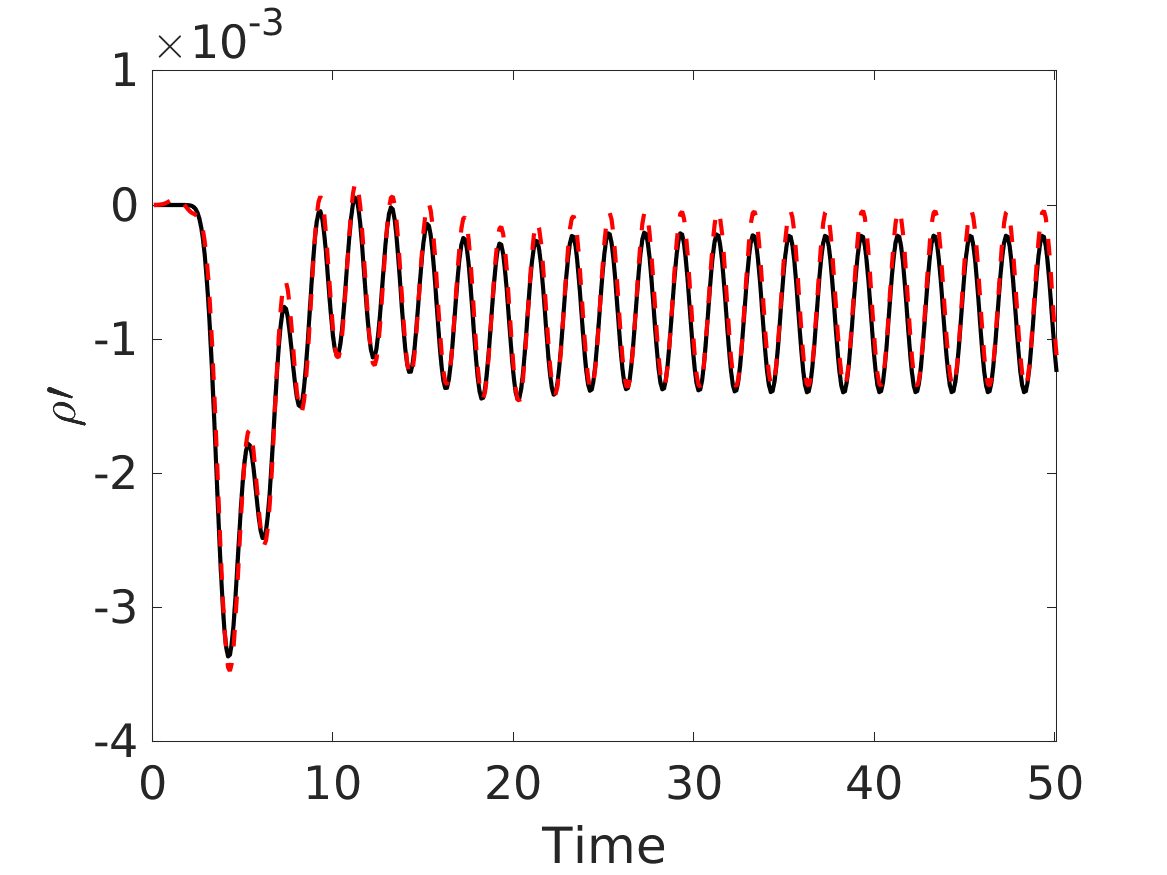}
    \rput(0.5,0.1){\psscalebox{0.5}{\color{black} \textbf{d)}}}
  \end{minipage}
  \centering
  \caption{Comparison of pressure (a), u-velocity (b), v-velocity (c), and density (d) computed by the linearized FOM and the ERA ROM at $(x, y) = (0.5293, 0.096)$. The freestream is perturbed with the triangular wave.} 
   \label{tri_probe_linear}
\end{figure}

%\subsubsection{ROM Predictions for the Triangular Wave Input}
The predictive performance of ROMs is now tested with the triangular wave input in equation~\ref{ug_tri} that is more difficult to capture because of the sharp gradients. Figure~\ref{tri_probe_linear} shows ROM predictions measured by a probe located at $(x, y) = (0.5293, 0.096)$. Except for the amplitude deviations in the ROM for the v component of velocity, ROMs based on the linearized solver predict the dynamics in response to the unseen input signal accurately.

\begin{figure}[h!]
  \centering
  \begin{minipage}[a]{0.49\textwidth}
    \includegraphics[trim=4 -0.1 4 4, clip, width=\textwidth]{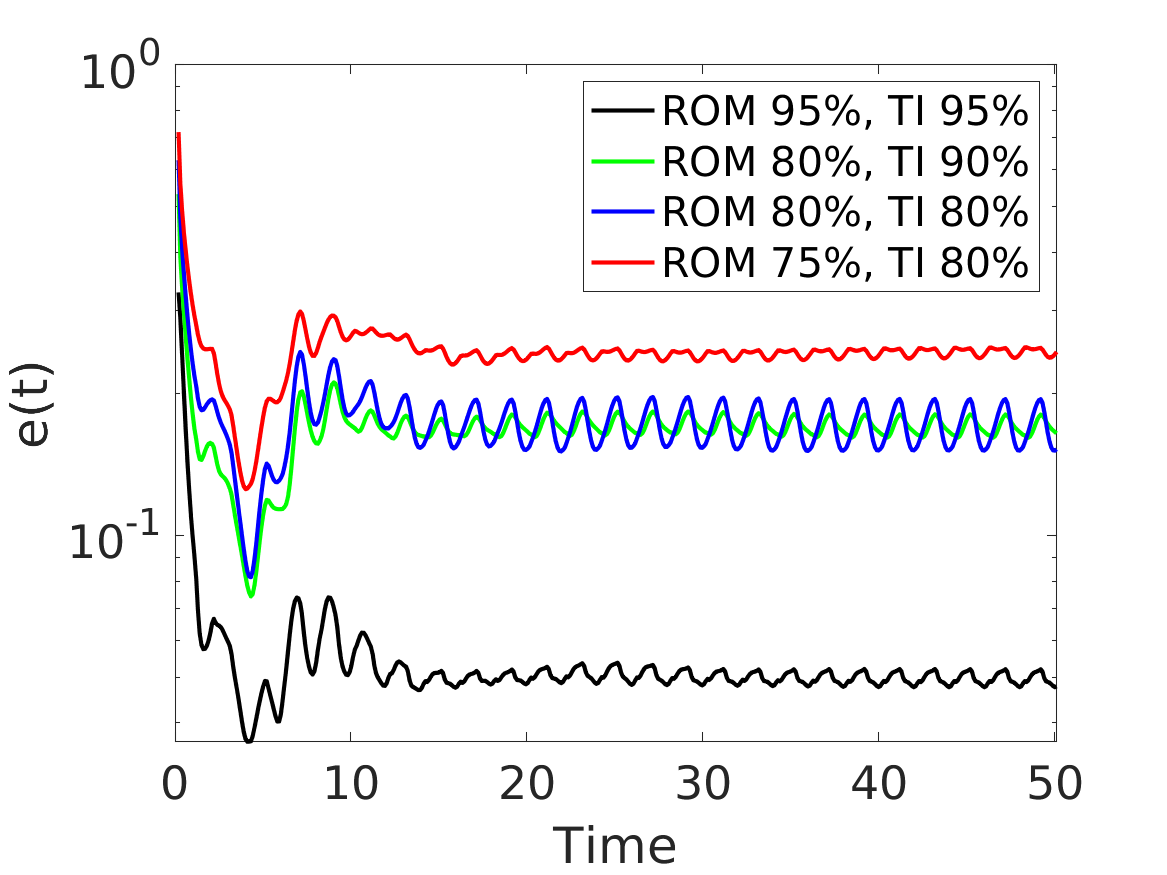}
    \rput(0.5,0.1){\psscalebox{0.5}{\color{black} \textbf{a)}}}
    \vspace{0.1cm}
  \end{minipage}
  \centering
  \begin{minipage}[a]{0.49\textwidth}
    \includegraphics[trim=4 -0.1 4 4, clip, width=\textwidth]{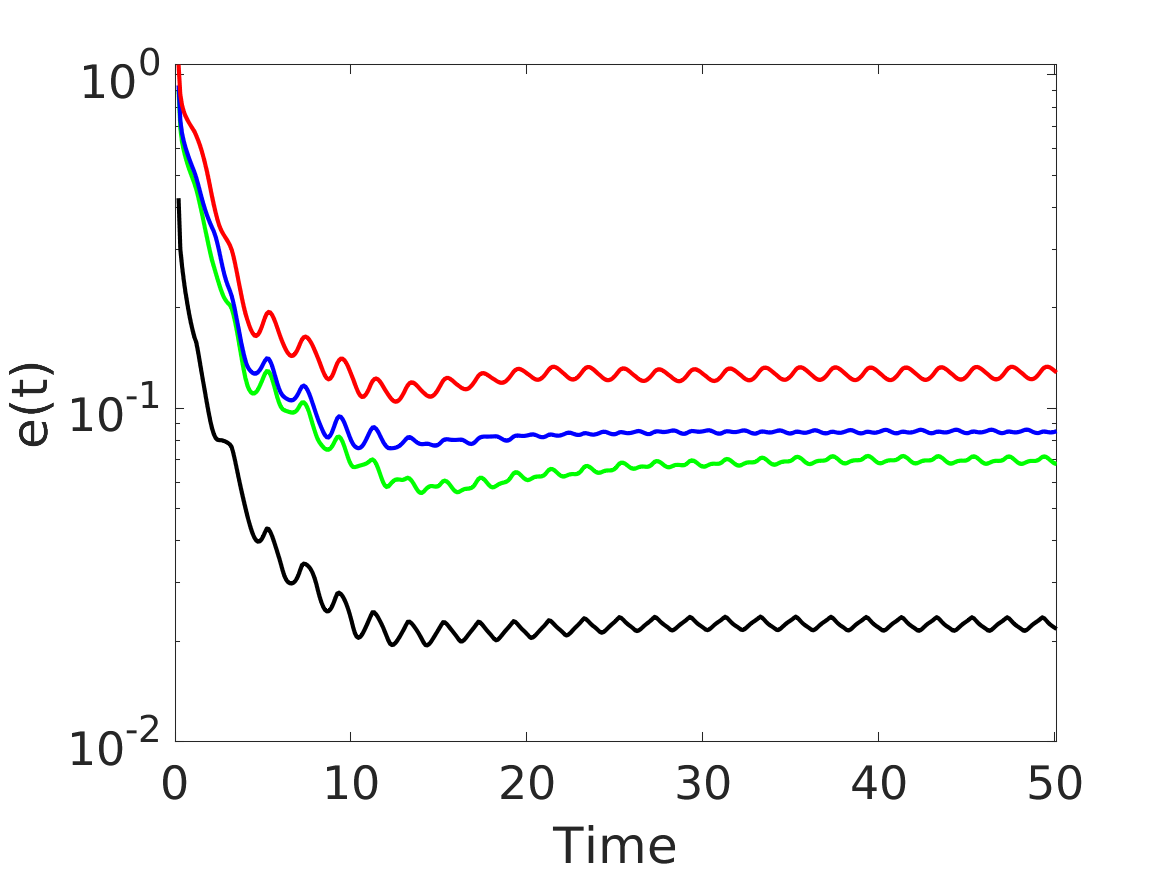}
    \rput(0.5,0.1){\psscalebox{0.5}{\color{black} \textbf{b)}}}
    \vspace{0.1cm}
  \end{minipage}
  \centering
  \begin{minipage}[a]{0.49\textwidth}
    \includegraphics[trim=4 -0.1 4 4, clip, width=\textwidth]{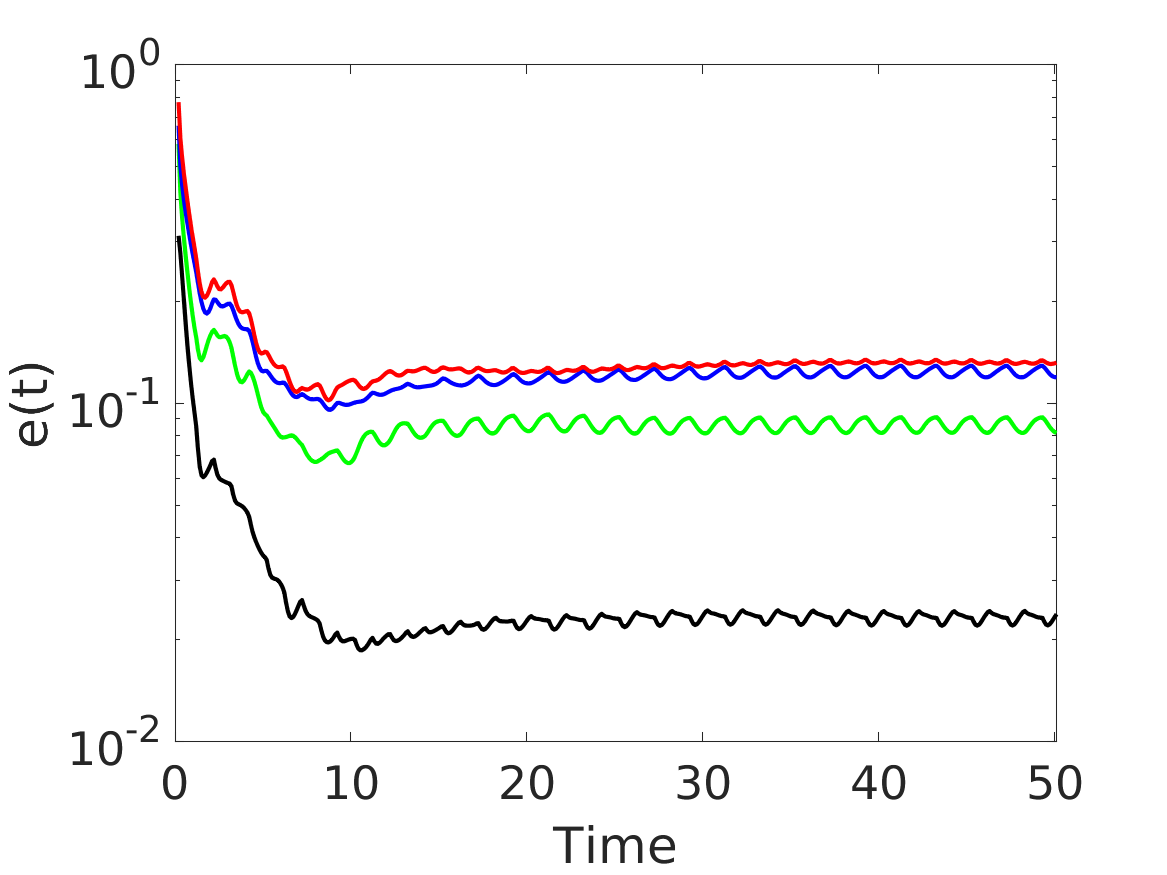}
    \rput(0.5,0.1){\psscalebox{0.5}{\color{black} \textbf{c)}}}
  \end{minipage}
  \centering
  \begin{minipage}[a]{0.49\textwidth}
    \includegraphics[trim=4 -0.1 4 4, clip, width=\textwidth]{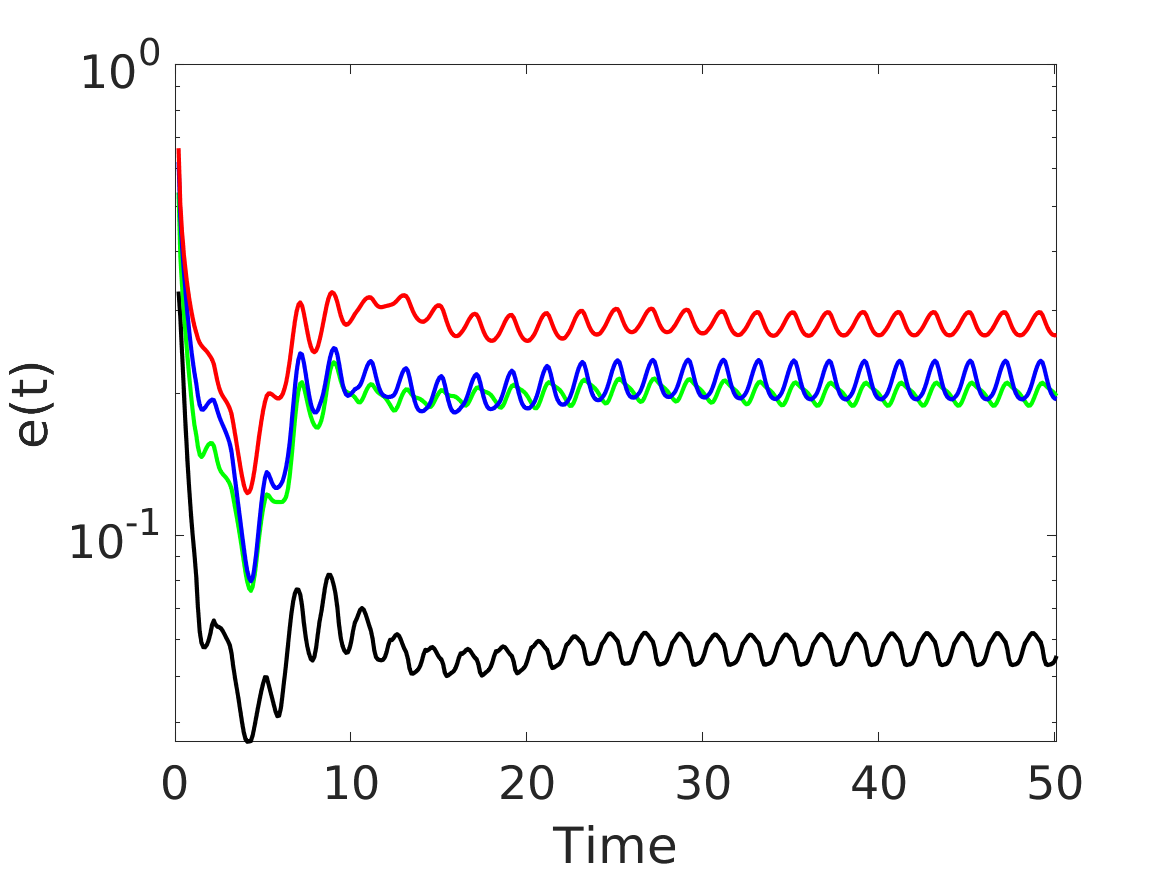}
    \rput(0.5,0.1){\psscalebox{0.5}{\color{black} \textbf{d)}}}
  \end{minipage}
  \centering
  \caption{Relative errors computed for pressure (a), u-velocity (b), v-velocity (c), and density (d). 
  The freestream is perturbed with the triangular wave, while the ROM is trained with the snapshots of the Gaussian input response generated by the linearized FOM.} 
   \label{tri_error_linear}
\end{figure}

The relative error in equation~\ref{rel_e} is plotted in Figure~\ref{tri_error_linear} with respect to time for the pressure, velocity, and density ROMs. ROMs of different dimensions are compared while retaining different numbers of modes in the tangential interpolation step. Considering that ROMs with $80\%$ and $90\%$ energy capture in the tangential interpolation step are almost equally accurate for most of the variables, we choose the lower energy level to reduce the size of ROMs. On the other hand, the ROM generated by retaining balancing modes that capture $95\%$ of the input-output energy requires more than 200 modes for each variable, which makes predictions more expensive. Therefore, we used the ROM with $80\%$ energy content in both Hankel singular vectors and tangential modes to obtain the results demonstrated in this section. This ROM is sufficiently accurate, and the dimension of subspaces are sufficiently low as shown in Table~\ref{t:tanintlin} to ensure computational efficiency. Hence, the ROM trained by the Gaussian input response of the linearized FOM achieves an online speed up factor of 158.33 during prediction compared to the linearized FOM. Note that in addition, the linearized FOM delivers a speedup factor of 5.39 compared to the nonlinear FOM, which contributes to computational savings in the offline phase in which ROM matrices are constructed using the training data from the FOM.

\begin{figure}[h!]
  \centering
  \begin{minipage}[a]{0.49\textwidth}
    \includegraphics[trim=4 -0.1 4 4, clip, width=\textwidth]{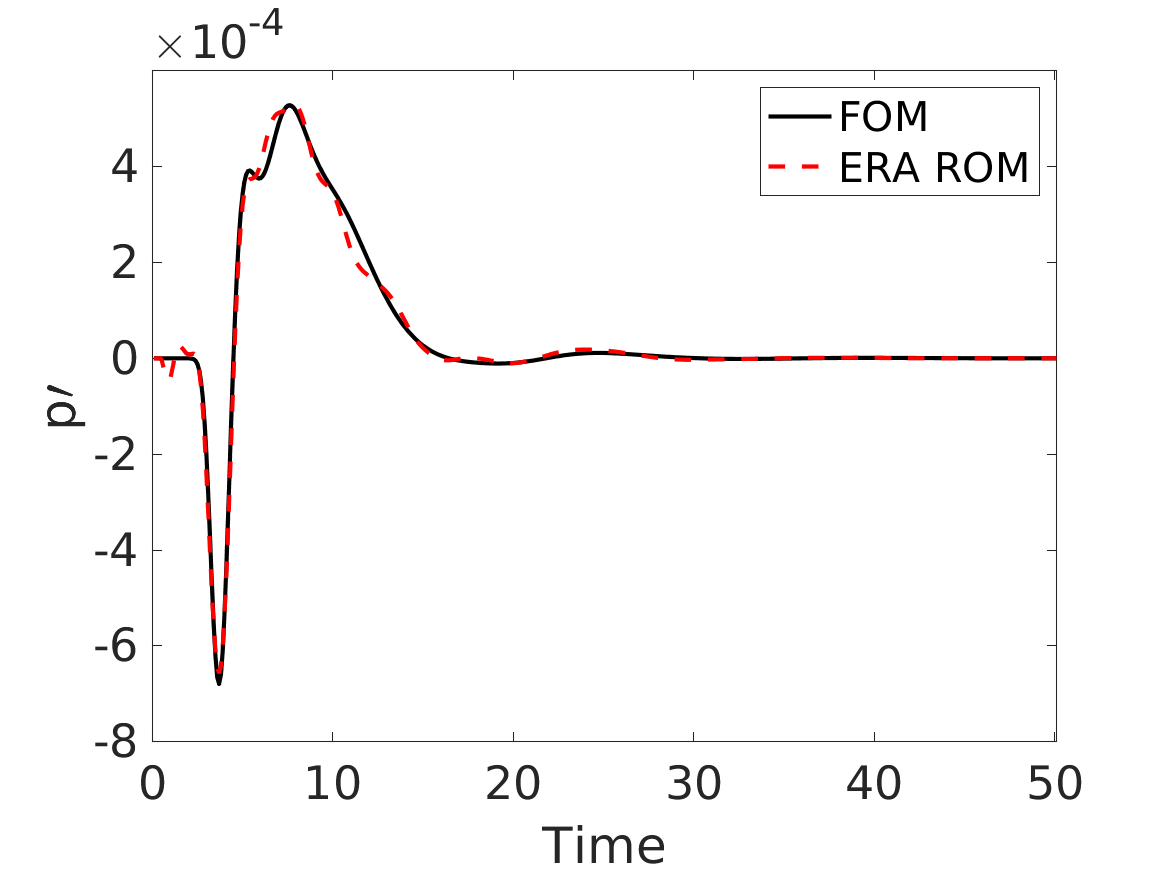}
    \rput(0.5,0.1){\psscalebox{0.5}{\color{black} \textbf{a)}}}
    \vspace{0.1cm}
  \end{minipage}
  \centering
  \begin{minipage}[a]{0.49\textwidth}
    \includegraphics[trim=4 -0.1 4 4, clip, width=\textwidth]{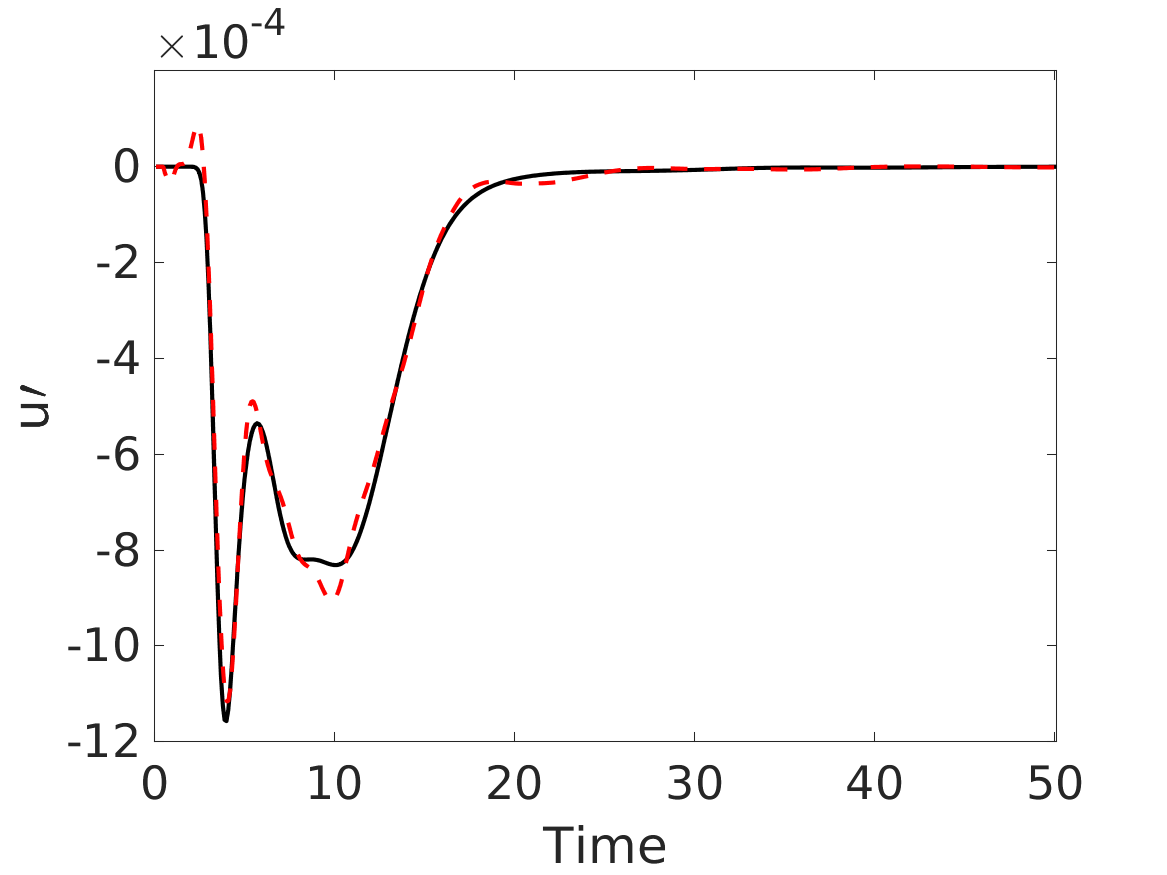}
    \rput(0.5,0.1){\psscalebox{0.5}{\color{black} \textbf{b)}}}
    \vspace{0.1cm}
  \end{minipage}
  \centering
  \begin{minipage}[a]{0.49\textwidth}
    \includegraphics[trim=4 -0.1 4 4, clip, width=\textwidth]{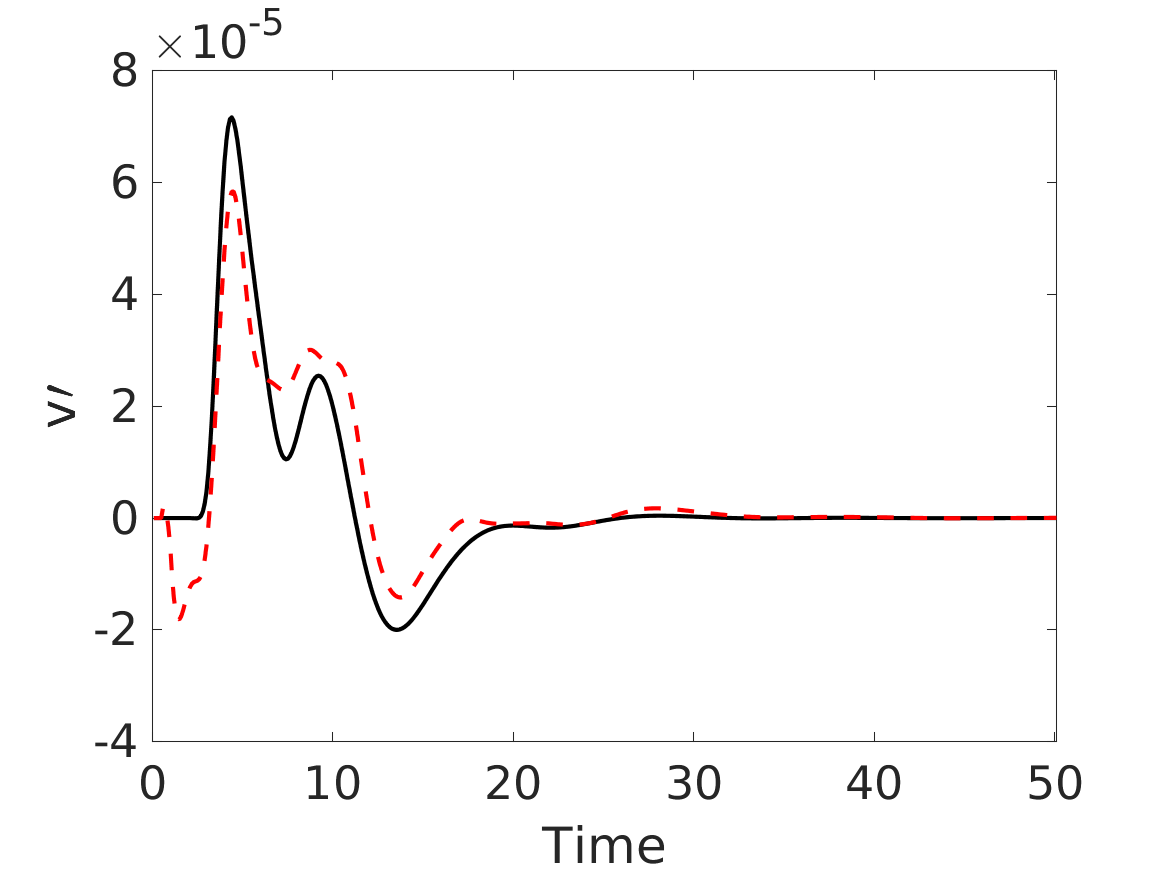}
    \rput(0.5,0.1){\psscalebox{0.5}{\color{black} \textbf{c)}}}
  \end{minipage}
  \centering
  \begin{minipage}[a]{0.49\textwidth}
    \includegraphics[trim=4 -0.1 4 4, clip, width=\textwidth]{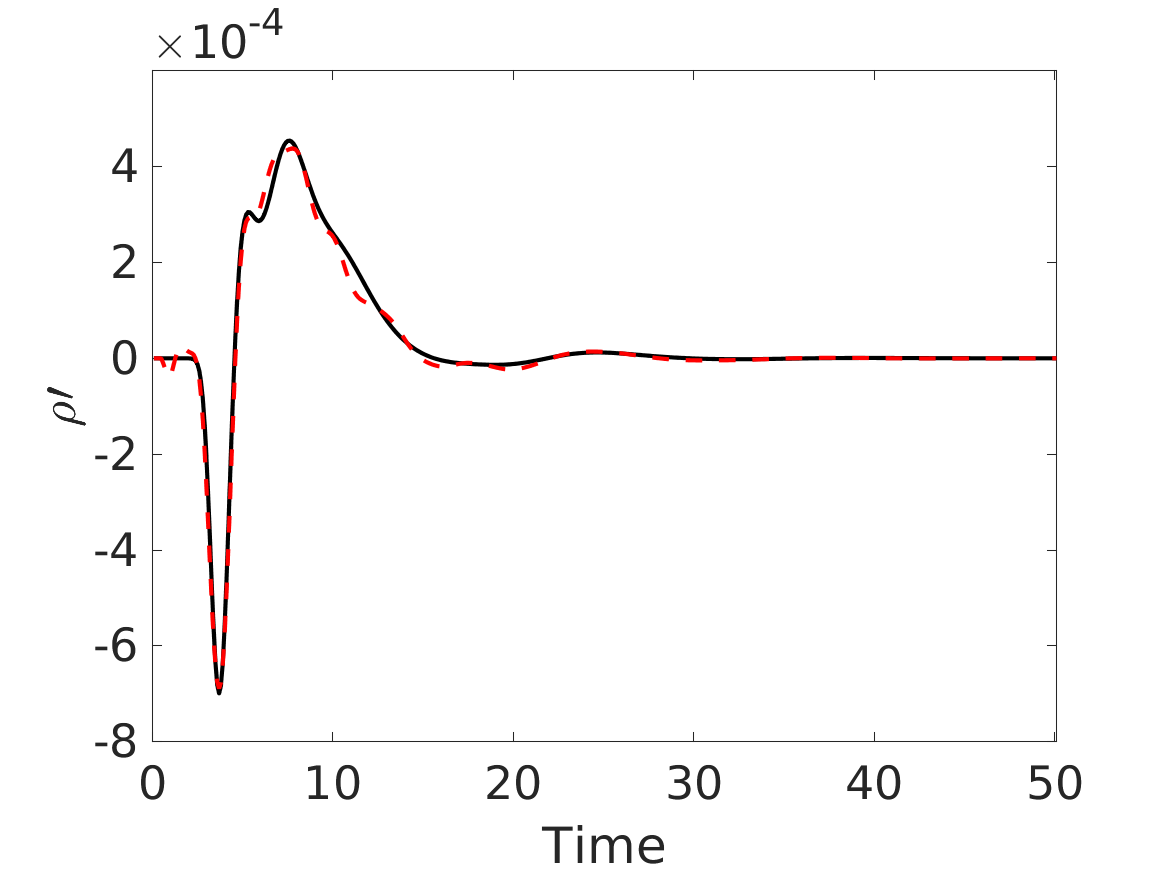}
    \rput(0.5,0.1){\psscalebox{0.5}{\color{black} \textbf{d)}}}
  \end{minipage}
  \centering
  \caption{Comparison of pressure (a), u-velocity (b), v-velocity (c), and density (d) computed by the linearized FOM and the ERA ROM at $(x,y) = (0.3510,-0.1096)$. The freestream is perturbed with the non-periodic square wave.} 
   \label{step_probe_linear}
\end{figure}

%\subsubsection{ROM Predictions for the Non-periodic Square Wave Input}
The performance of ROMs trained with the linearized FOM snapshots is tested next in response to the non-periodic square wave input in equation~\ref{ug_square}. The ROM outlined in Table~\ref{t:tanintlin} is used here for prediction. As shown in Figure~\ref{step_probe_linear}, for a probe located at $(x,y) = (0.3510,-0.1096)$, the ROM solution demonstrates a good agreement with the linearized FOM. Also note that in terms of numerical oscillations, the performance of the ROM trained by the linearized solver is more accurate than the ROM trained by the nonlinear solver in Figure~\ref{step_probe}. This highlights the fact that the Gaussian input response is a true representative of the sequence of Markov parameters in the linearized system, while in the nonlinear system, it only approximately represents the Markov parameters. The better accuracy of the ROM based on the linearized solver in response to the non-periodic input is further demonstrated in terms of the relative error in \nameref{appB}.

\subsection{ROMs based on the High-resolution Linearized Euler Solver}
\label{gappy_results}
\subsubsection{Actuator Selection using Low-fidelity Gappy POD}
\begin{figure}[h!]
  \centering
  \begin{minipage}[a]{0.49\textwidth}
    \includegraphics[trim=4 4 4 4, clip, width=\textwidth]{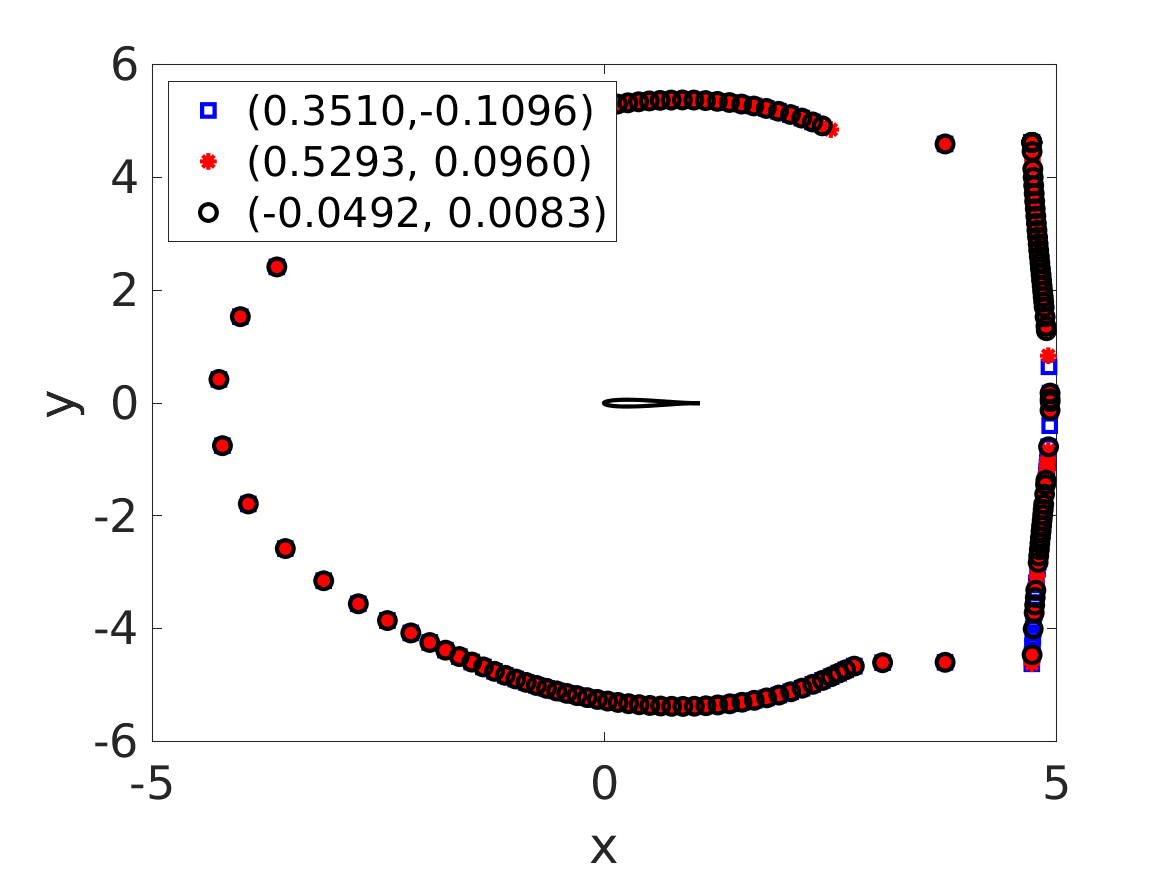}
    \rput(0.5,0.1){\psscalebox{0.5}{\color{black} \textbf{a)}}}
  \end{minipage}
  \centering
  \begin{minipage}[a]{0.49\textwidth}
    \includegraphics[trim=4 4 4 4, clip, width=\textwidth]{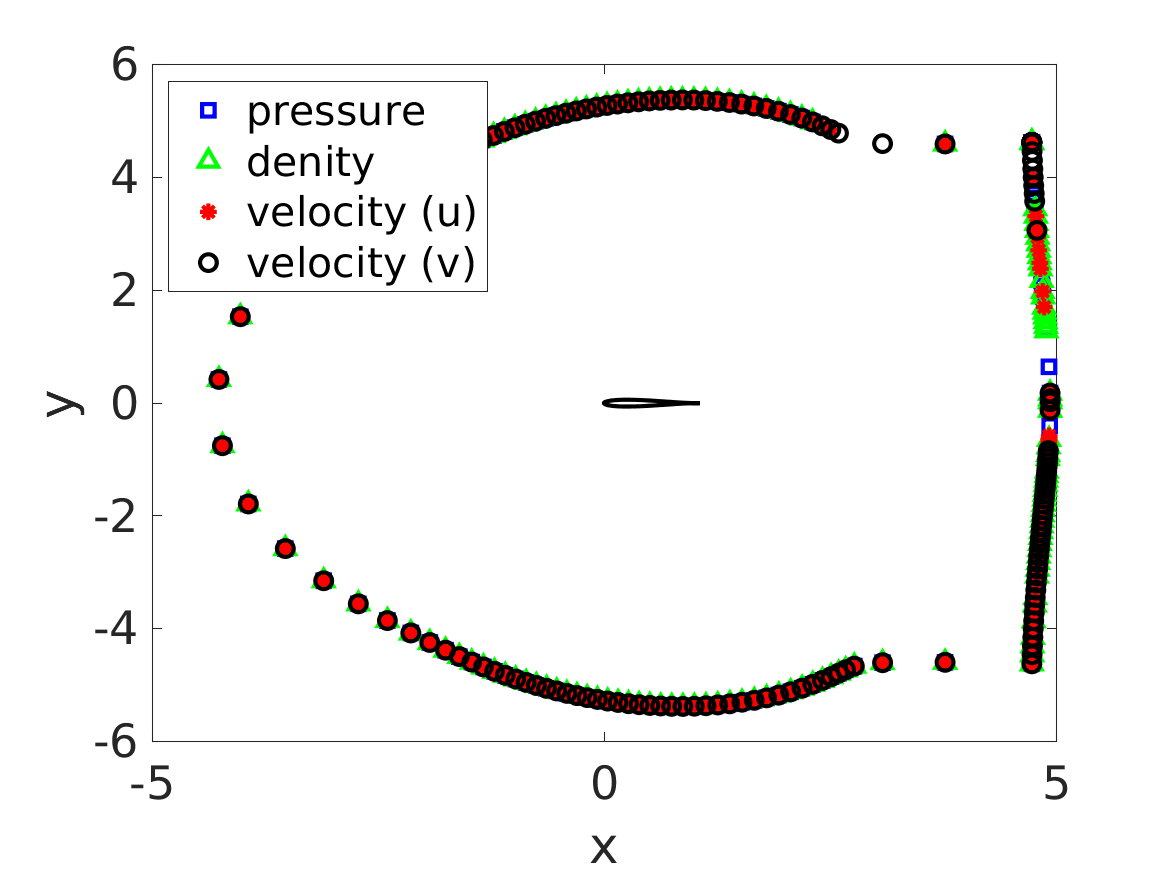}
    \rput(0.5,0.1){\psscalebox{0.5}{\color{black} \textbf{b)}}}
  \end{minipage}
  \centering
  \caption{Location of the critical input channels selected by QR factorization with column pivoting. The airfoil is located at the origin. (a) is the location of the pivots when gappy POD is trained with the pressure Markov sequence at $(x, y)=(0.3510,-0.1096)$, $(x, y)=(0.5293,0.0960)$, and $(x, y)=(-0.0492,0.0083)$. (b) is the pivot locations when gappy POD is trained with pressure, density, and velocity Markov sequences at $(x, y)=(0.3510,-0.1096)$.}
   \label{qr_pivots}
\end{figure}

The low-fidelity gappy POD framework described in section~\ref{gappy} is used here to identify the critical input channels along the far-field boundary to make offline ROM computations tractable as the grid resolution increases. The training data for gappy POD is obtained using the Gaussian input response of the linearized solver over a coarser grid with $101 \times 51$ cells.
10 POD modes for u component of velocity, and 11 modes for the rest of the variables are retained that capture $99\%$ of the energy for each variable in the gappy POD process to compute the critical input channel locations by QR factorization with column pivoting based on the Markov sequence measured at a point located below the airfoil with coordinates $(x, y)=(0.3510,-0.1096)$. Following the recommendations by Peherstorfer et al. \cite{Peherstorfer2020}, we use over-sampling to maximize the reconstruction accuracy to ensure a rich dataset will be collected for training the high-resolution ERA ROMs. $150$ input channels (actuators) are selected by the QR factorization with column pivoting applied to the POD basis. Figure~\ref{qr_pivots} (a) shows the location of the selected input channels (i.e., the QR factorization pivots) when gappy POD is separately trained with the pressure Markov sequences at $(x, y)=(0.3510,-0.1096)$, $(x, y)=(0.5293,0.0960)$, and $(x, y)=(-0.0492,0.0083)$. The three sets of input channels share 125 channel locations. That is, only 25 input channels are selected in each set that are different from the other two sets. Figure~\ref{qr_pivots} (b) shows the input channels obtained by the Markov sequences of all four variables at $(x, y)=(0.3510,-0.1096)$. Similarly, 125 out of the 150 selected channels are at the same location for the different variables.

\begin{figure}[h!]
  \centering
  \begin{minipage}[a]{0.49\textwidth}
    \includegraphics[trim=4 -0.1 4 4, clip, width=\textwidth]{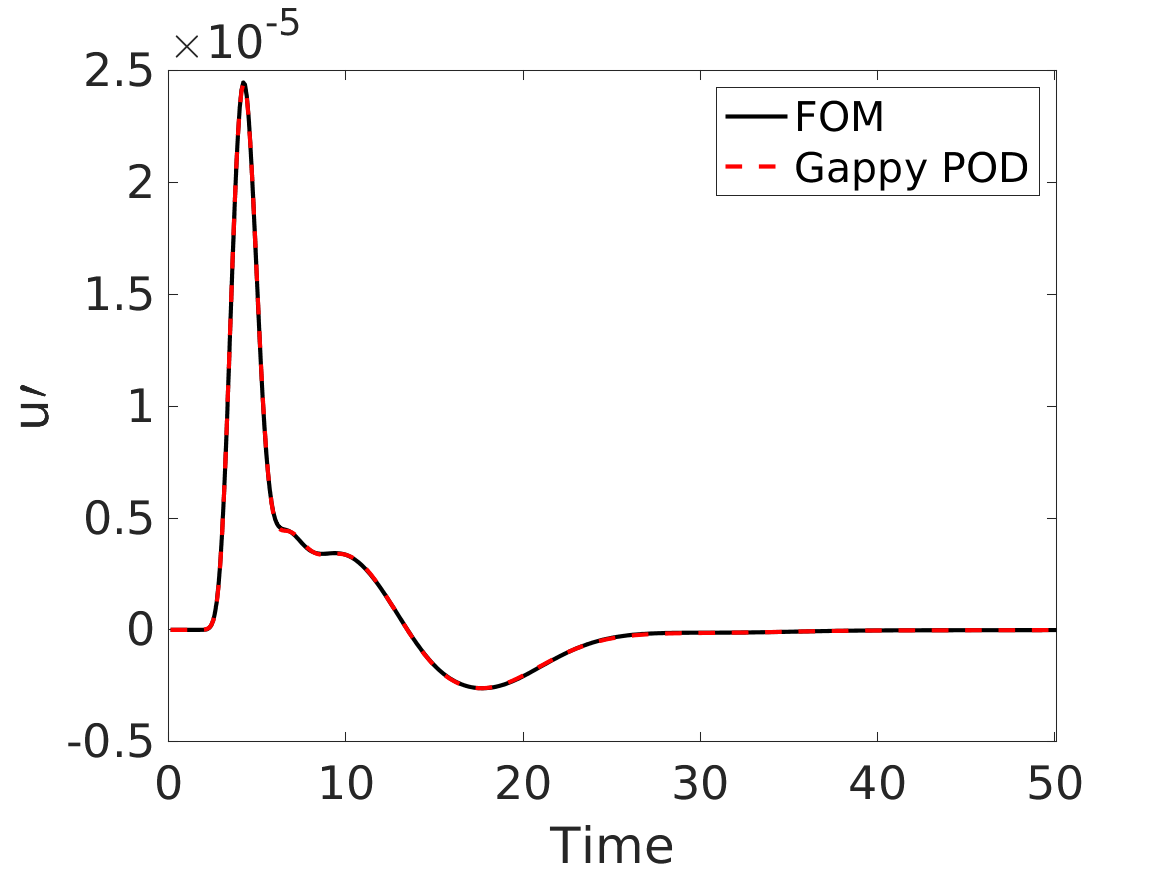}
    \rput(0.5,0.1){\psscalebox{0.5}{\color{black} \textbf{a)}}}
    \vspace{0.1cm}
  \end{minipage}
  \centering
  \begin{minipage}[a]{0.49\textwidth}
    \includegraphics[trim=4 -0.1 4 4, clip, width=\textwidth]{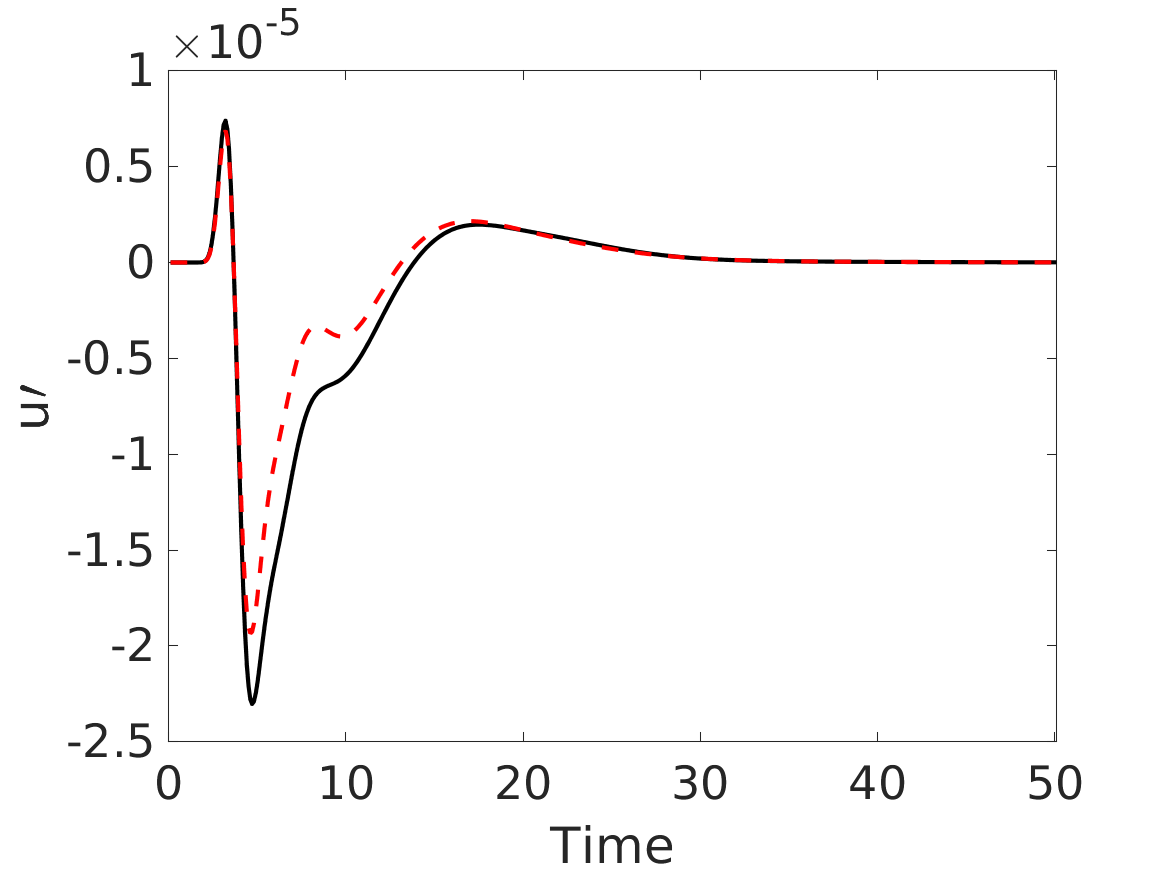}
    \rput(0.5,0.1){\psscalebox{0.5}{\color{black} \textbf{b)}}}
    \vspace{0.1cm}
  \end{minipage}
  \centering
  \caption{Markov sequence reconstructions for the u component of velocity by the gappy POD approach at a) $(x, y) = (0.3510, -0.1096)$ and b) $(x, y) = (0.5293, 0.0960)$. The active input channel is located at $(x, y) = (-2.1701, 4.0420)$.}
   \label{gappy_reconst}
\end{figure}

Figure~\ref{gappy_reconst} compares the FOM solution with the reconstruction by the low-fidelity gappy POD approach, when using as the training data the Markov sequences of velocity u corresponding to each of the far-field points (input channels) measured by a probe located at $(x, y) = (0.3510, -0.1096)$.
The observable measures the Markov sequences for the subset of channels selected by QR factorization. Figure~\ref{gappy_reconst} demonstrates the dynamic response when the input channel located at $(x, y) = (-2.1701, 4.0420)$ is perturbed by the Gaussian input.
Note that the Markov sequence measured by the velocity probe at $(x, y) = (0.3510, -0.1096)$ was included in the training data, while the Markov sequence measured by the probe at $(x, y) = (0.5293, 0.0960)$ was not used during training the gappy POD method.

\subsubsection{ROM Training}
\begin{figure}[h!]
  \centering
  \begin{minipage}[a]{0.43\textwidth}
    \includegraphics[trim=4 4 4 4, clip, width=\textwidth]{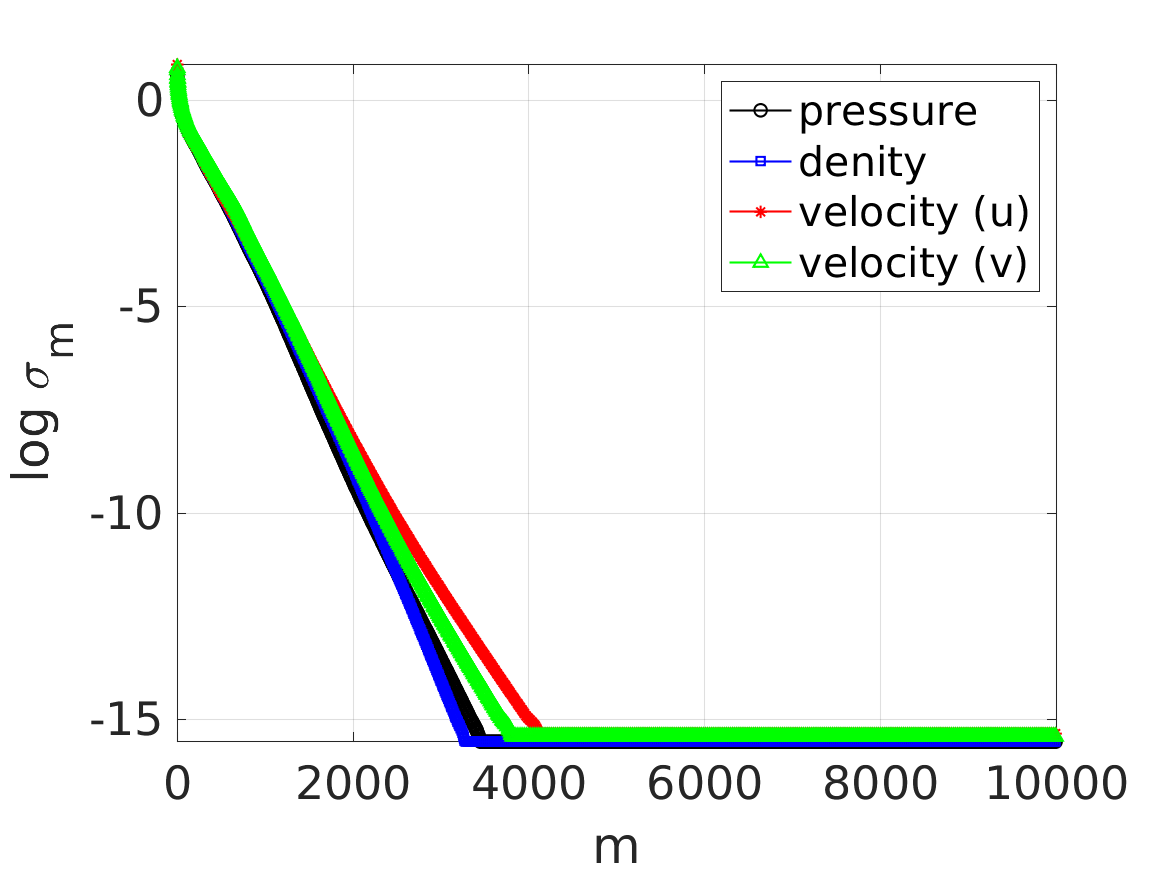}
    \rput(0.5,0.1){\psscalebox{0.5}{\color{black} \textbf{a)}}}
  \end{minipage}
  \centering
  \begin{minipage}[a]{0.43\textwidth}
    \includegraphics[trim=4 4 4 4, clip, width=\textwidth]{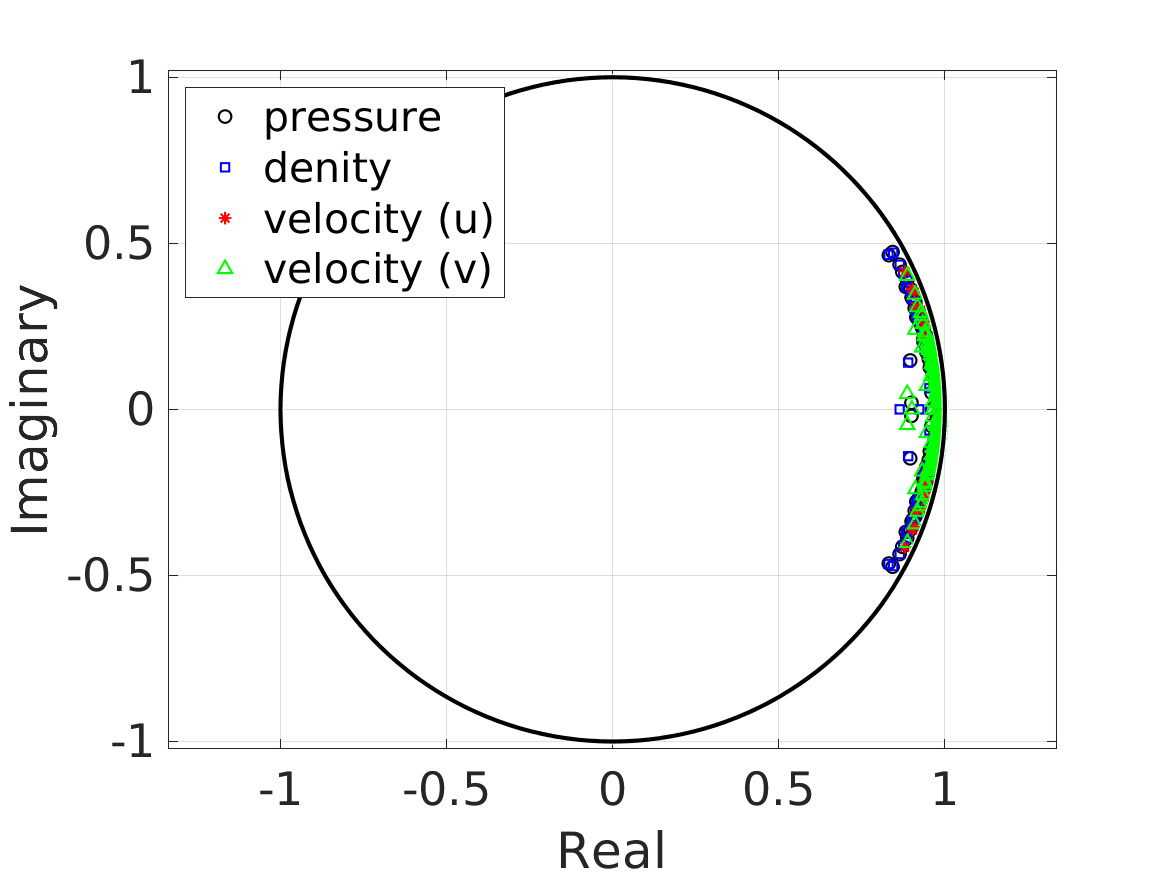}
    \rput(0.5,0.1){\psscalebox{0.5}{\color{black} \textbf{b)}}}
  \end{minipage}
  \centering
  \caption{a) Singular values of the Hankel matrix. b) Eigenvalues of the ERA ROMs. The ERA ROMs are constructed with the Markov sequence collected for the channels identified by the gappy POD method.}
   \label{sv_rom_eigs_gappy}
\end{figure}

Since the distribution of the channels obtained by training gappy POD with the velocity probe measurements covers almost every cluster of channels corresponding to the different probe locations in Figure~\ref{gappy_reconst} (a), this set of channels is used to compute the Gaussian input response of the linearized FOM with the $401 \times 101$ grid. The Markov sequences constructed for these 150 channels are then used to construct the Hankel matrix. Figure~\ref{sv_rom_eigs_gappy} (a) shows the decay of the Hankel singular values, which is slower than the case with snapshots computed by the linearized solver over the coarser grid, and faster than the nonlinear solver over the coarser grid. The ERA ROMs are then constructed to predict the dynamics for the high-resolution case. The eigenvalues of the ERA ROMs are shown in Figure~\ref{sv_rom_eigs_gappy}. All four ROMs are stable with eigenvalues residing inside the unit circle. 
The number of retained balancing and tangential modes are shown in Table~\ref{t:tanintlingappy}. The retained tangential modes capture $80\%$ of the energy in the training data. Note that the number of input channels is reduced from 604 in the original high-resolution setup to 150 using the actuator selection approach and further reduced using tangential interpolation. The retained Hankel singular vectors capture $80\%$ of the input-output energy. Also note that computing ROM matrices for the original high-resolution problem without the gappy POD approach was beyond our computational resources.
\begin{table}[h!]
 \begin{center}
  \caption{The number of the retained tangential modes for the ROM of each variable out of a total of 40,000 left and 150 right tangential modes and the dimension of the balanced ROMs.}
  \label{t:tanintlingappy}
  \begin{tabular}{lllll}\hline
        & Variable & Left modes & Right modes & ROM dimension \\\hline
        & Pressure & 211 & 49 & 115 \\\hline
        & Velocity (u) & 209 & 67 & 90  \\\hline
        & Velocity (v) & 221 & 57 & 120  \\\hline
        & Density & 215 & 49 & 116  \\\hline
  \end{tabular}
 \end{center}
\end{table}

\subsubsection{ROM Prediction}
\begin{figure}[h!]
  \centering
  \begin{minipage}[a]{0.49\textwidth}
    \includegraphics[trim=4 0.1cm 4 4, clip, width=\textwidth]{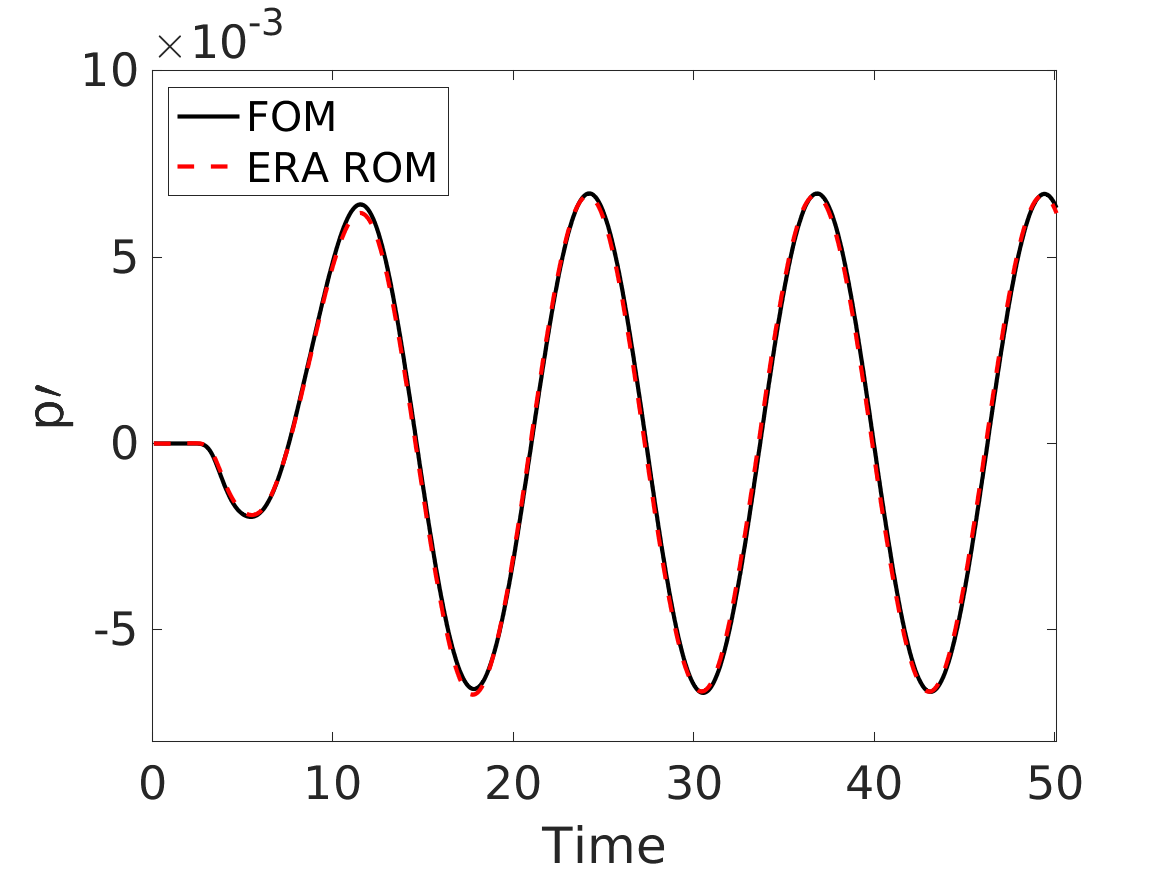}
    \rput(0.5,0.1){\psscalebox{0.5}{\color{black} \textbf{a)}}}
    \vspace{0.1cm}
  \end{minipage}
  \centering
  \begin{minipage}[a]{0.49\textwidth}
    \includegraphics[trim=4 0.1cm 4 4, clip, width=\textwidth]{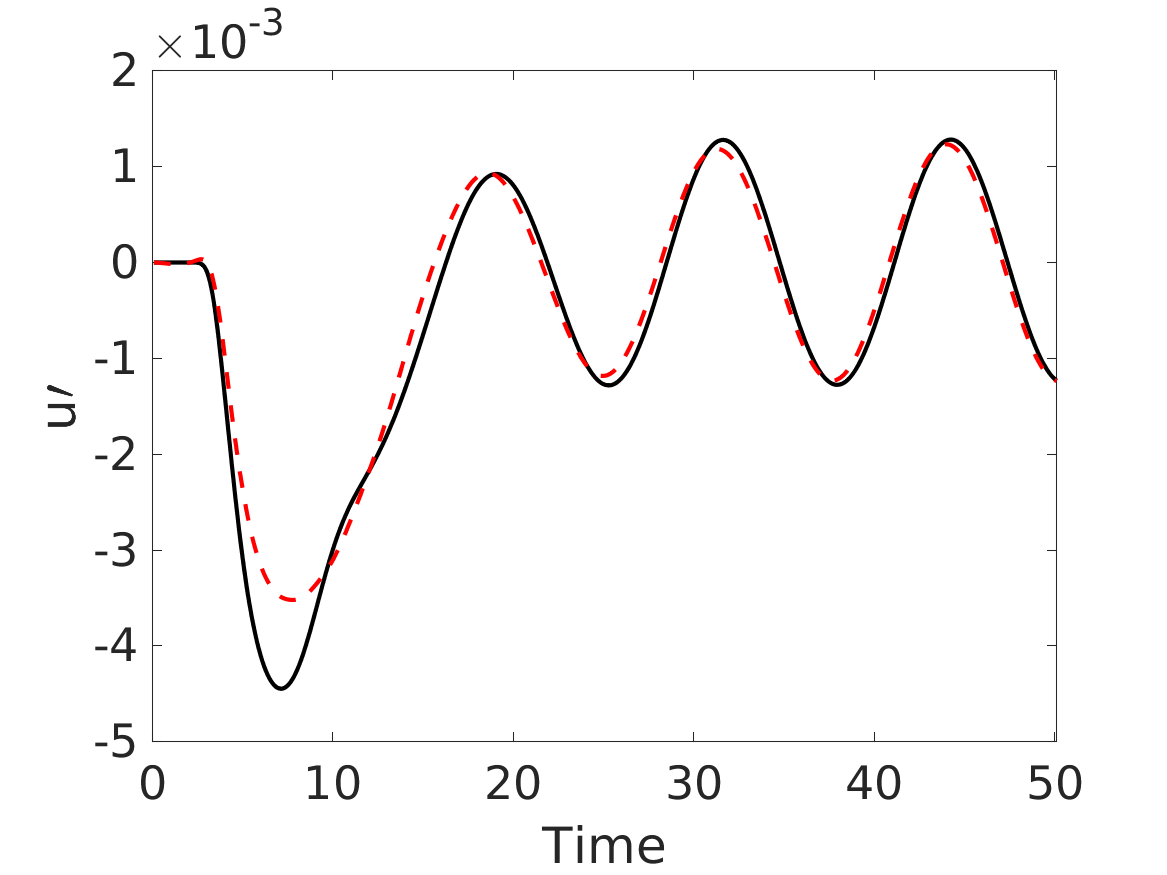}
    \rput(0.5,0.1){\psscalebox{0.5}{\color{black} \textbf{b)}}}
    \vspace{0.1cm}
  \end{minipage}
  \centering
  \begin{minipage}[a]{0.49\textwidth}
    \includegraphics[trim=4 0.1cm 4 4, clip, width=\textwidth]{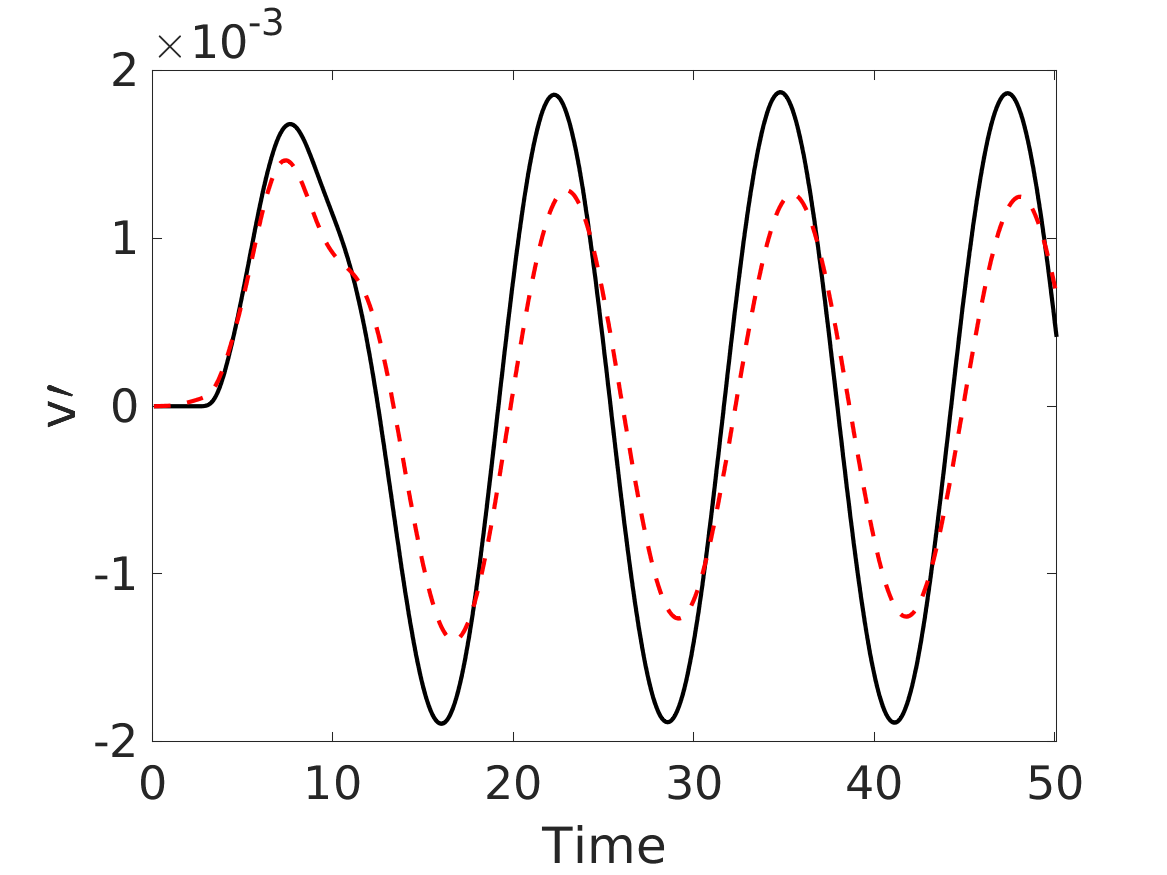}
    \rput(0.5,0.1){\psscalebox{0.5}{\color{black} \textbf{c)}}}
  \end{minipage}
  \centering
  \begin{minipage}[a]{0.49\textwidth}
    \includegraphics[trim=4 0.1cm 4 4, clip, width=\textwidth]{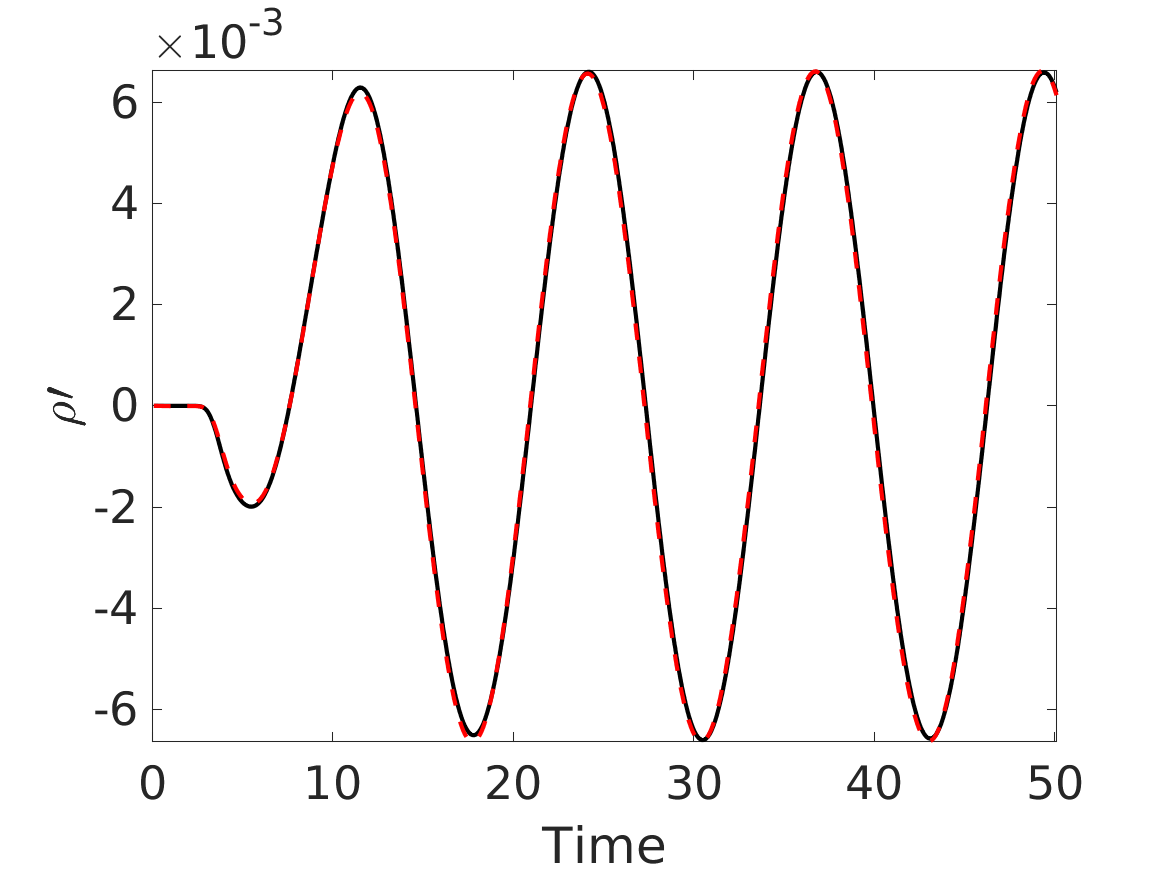}
    \rput(0.5,0.1){\psscalebox{0.5}{\color{black} \textbf{d)}}}
  \end{minipage}
  \centering
  \caption{Comparison of the pressure (a), u-velocity (b), v-velocity (c), and density (d) computed using the $401 \times 101$ grid by the linearized FOM and the ERA ROM at $(x,y) = (0.4628,-0.3129)$. The ROMs are trained using the input channels identified by gappy POD. The freestream is perturbed with the sinusoidal input.}
   \label{sine_probe_linear_gappy}
\end{figure}

%\subsection{ROM Predictions for the Sinusoidal input}
Figure~\ref{sine_probe_linear_gappy} presents the prediction of the ERA ROMs for probes located at $(x,y) = (0.4628,-0.3129)$ when the input channels are perturbed by the sinusoidal input in equation~\ref{ugtest}. Note that the FOM response is obtained by perturbing 604 channels, which comprises of the entire far-field boundary cells in the $401 \times 101$ grid, while the ROM response is obtained by perturbing all of the ROM input channels, that is, the 150 input channels chosen by the gappy POD method. Clearly, the intensity of the acoustic response in the two cases cannot be the same, therefore, the amplitude of the sinusoidal input is scaled in the ROM to match the FOM response in the first cycle. Note that we do not need to repeat this adjustment as we change the type of the input signal. In other words, the same scaling factor can be used to maximize prediction accuracy also in response to other types of inputs. 

\begin{figure}[h!]
  \centering
  \begin{minipage}[a]{0.49\textwidth}
    \includegraphics[trim=4 4 4 -0.05, clip, width=\textwidth]{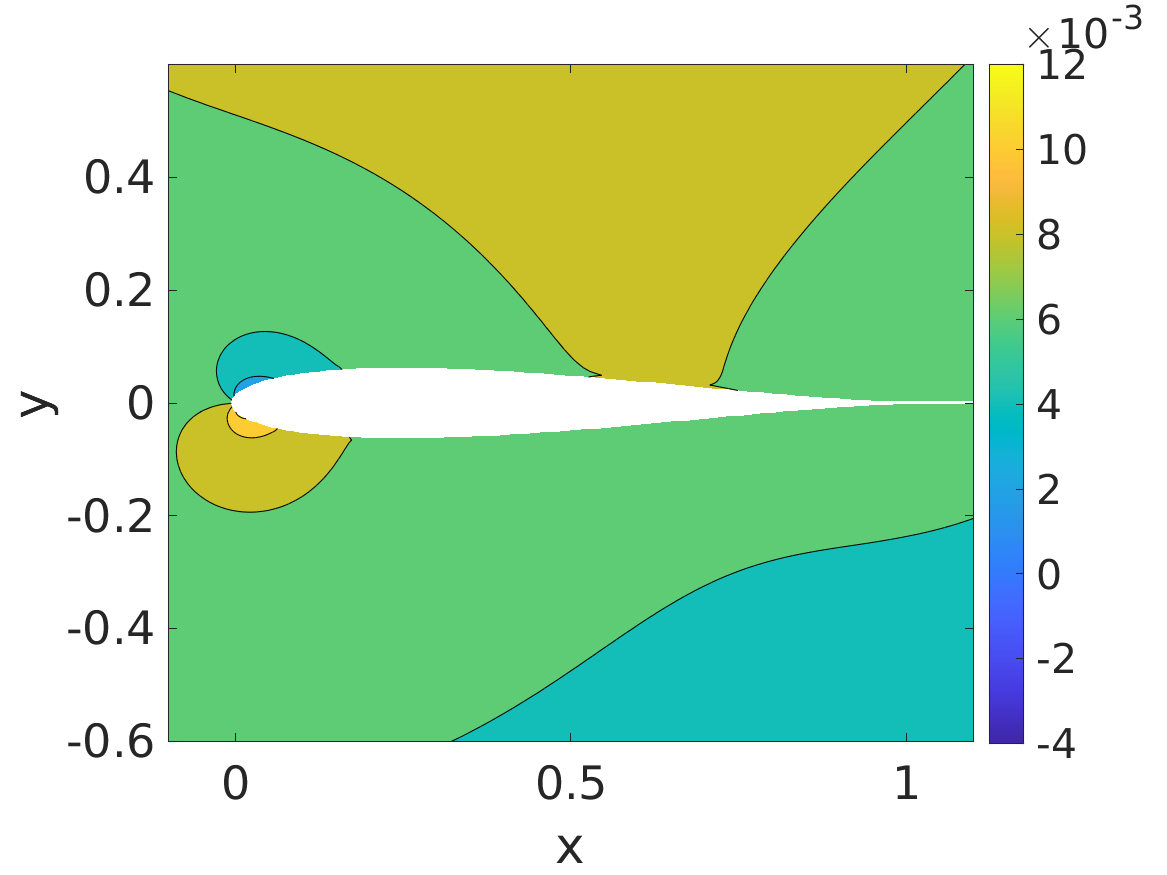}
    \rput(0.5,0.1){\psscalebox{0.5}{\color{black} \textbf{a)}}}
  \end{minipage}
  \centering
  \begin{minipage}[a]{0.49\textwidth}
    \includegraphics[trim=4 4 4 -0.05, clip, width=\textwidth]{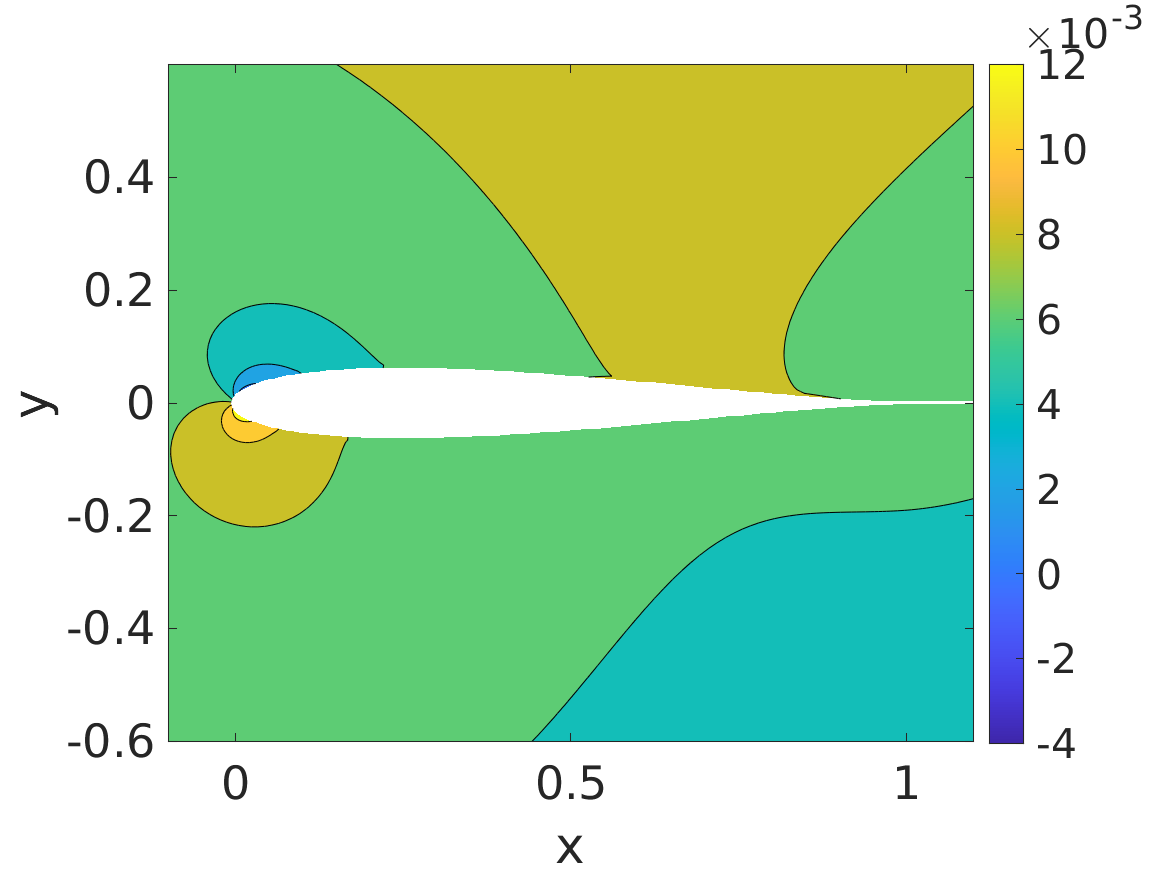}
    \rput(0.5,0.1){\psscalebox{0.5}{\color{black} \textbf{b)}}}
  \end{minipage}
  \centering
  \caption{Pressure contours computed by the linearized FOM (a) and ERA ROM (b) at $t=50$. The ERA ROM is trained using the input channels identified by gappy POD. The freestream velocity is perturbed by the sinusoidal input.}
   \label{pcontour_linear_gappy}
\end{figure}

The flow field is visualized in Figure~\ref{pcontour_linear_gappy} that compares the pressure perturbation contour computed by the ERA ROM with limited actuators against the FOM for a snapshot at $t=50$. ROM prediction matches the FOM despite a reduction in the input channels from 604 in the FOM to only 150 channels in the ROM.

\begin{figure}[h!]
  \centering
  \begin{minipage}[a]{0.49\textwidth}
    \includegraphics[trim=4 -0.1 4 4, clip, width=\textwidth]{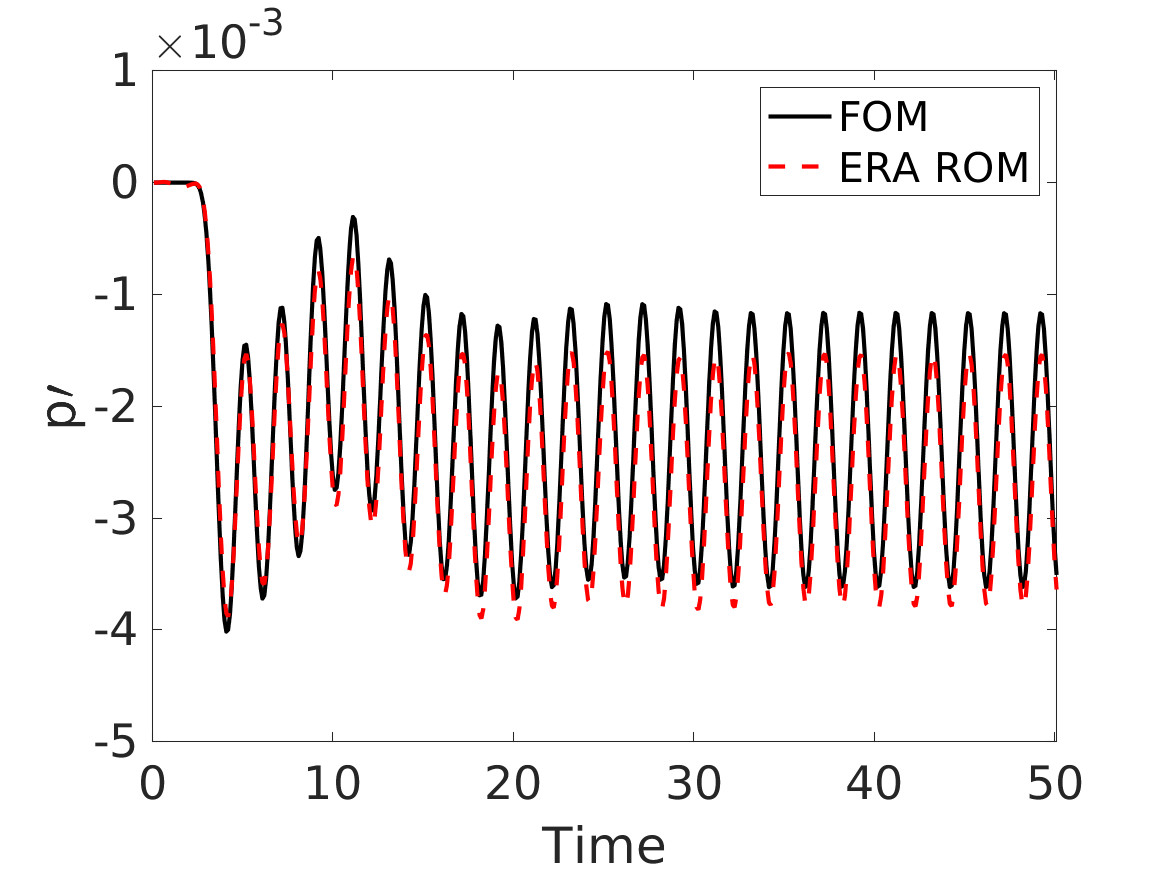}
    \rput(0.5,0.1){\psscalebox{0.5}{\color{black} \textbf{a)}}}
    \vspace{0.1cm}
  \end{minipage}
  \centering
  \begin{minipage}[a]{0.49\textwidth}
    \includegraphics[trim=4 -0.1 4 4, clip, width=\textwidth]{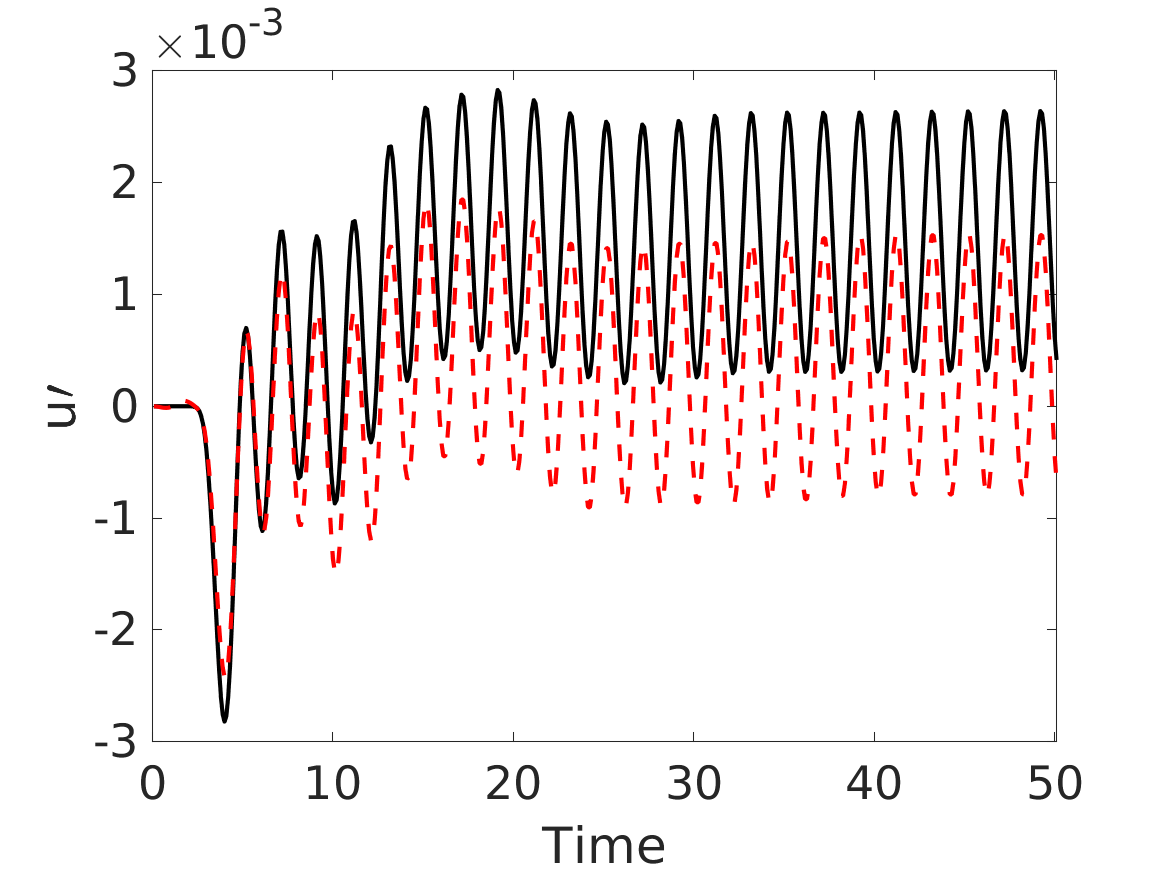}
    \rput(0.5,0.1){\psscalebox{0.5}{\color{black} \textbf{b)}}}
    \vspace{0.1cm}
  \end{minipage}
  \centering
  \begin{minipage}[a]{0.49\textwidth}
    \includegraphics[trim=4 -0.1 4 4, clip, width=\textwidth]{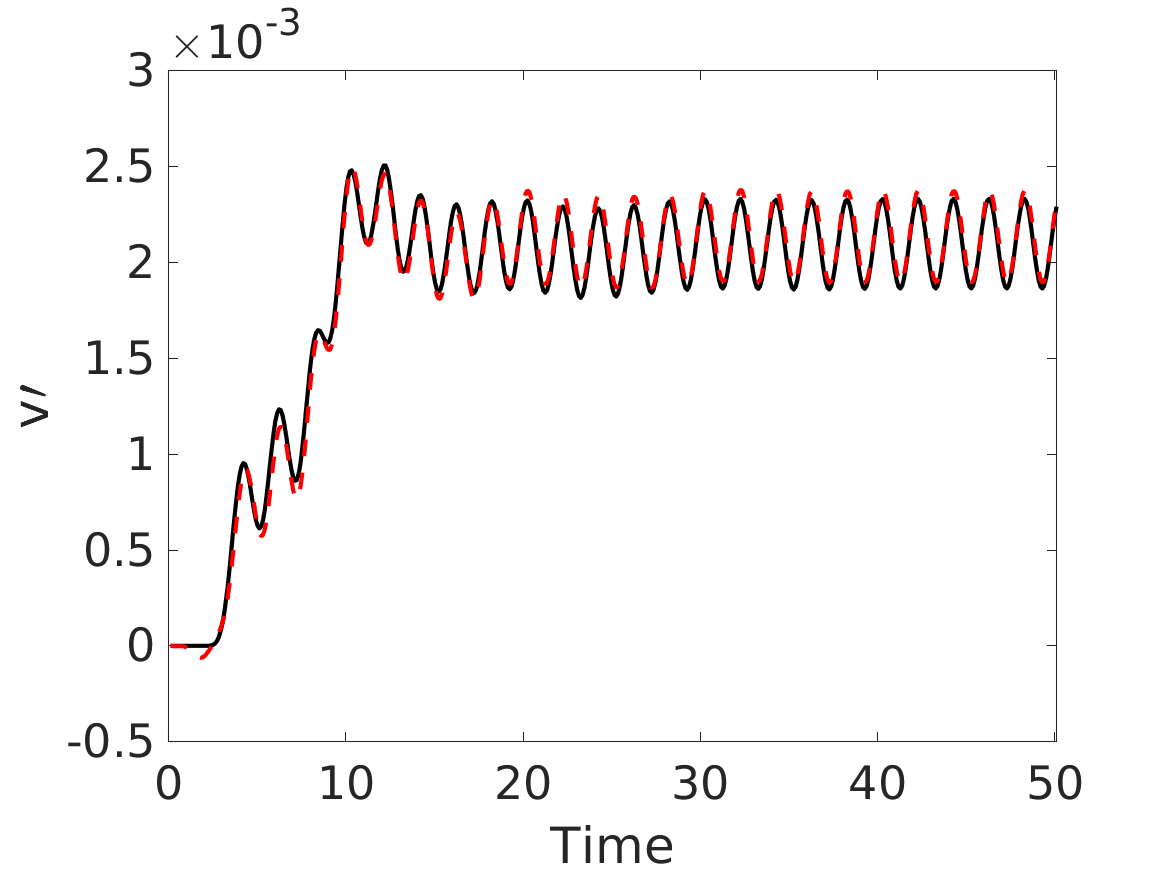}
    \rput(0.5,0.1){\psscalebox{0.5}{\color{black} \textbf{c)}}}
  \end{minipage}
  \centering
  \begin{minipage}[a]{0.49\textwidth}
    \includegraphics[trim=4 -0.1 4 4, clip, width=\textwidth]{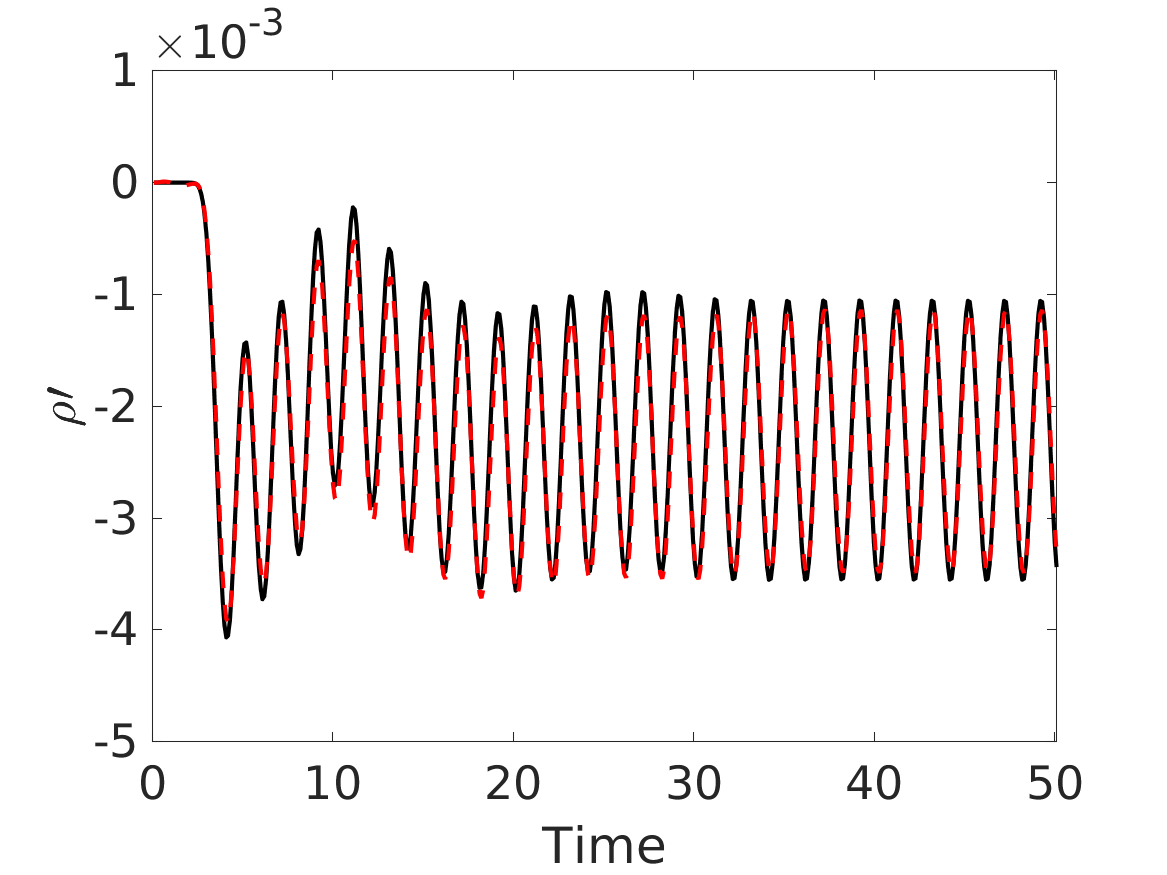}
    \rput(0.5,0.1){\psscalebox{0.5}{\color{black} \textbf{d)}}}
  \end{minipage}
  \centering
  \caption{Comparison of pressure (a), u-velocity (b), v-velocity (c), and density (d) computed by the linearized FOM and the ERA ROM at $(x, y) = (0.2920, 0.3221)$. The ROMs are trained using the input channels identified by gappy POD. The freestream is perturbed with the triangular wave.} 
   \label{tri_probe_linear_gappy}
\end{figure}

%\subsection{ROM Predictions for the Triangular Wave Input}
Figure~\ref{tri_probe_linear_gappy} shows ROM predictions measured by a probe located at $(x, y) = (0.2920, 0.3221)$, when the far-field boundary cells are perturbed by the triangular wave input. Despite the sharp gradients in the triangular wave input, the ERA ROM predictions of the acoustic response with respect to the unseen input signal closely follow the FOM solution.

\begin{figure}[h!]
  \centering
  \begin{minipage}[a]{0.49\textwidth}
    \includegraphics[trim=4 -0.1 4 4, clip, width=\textwidth]{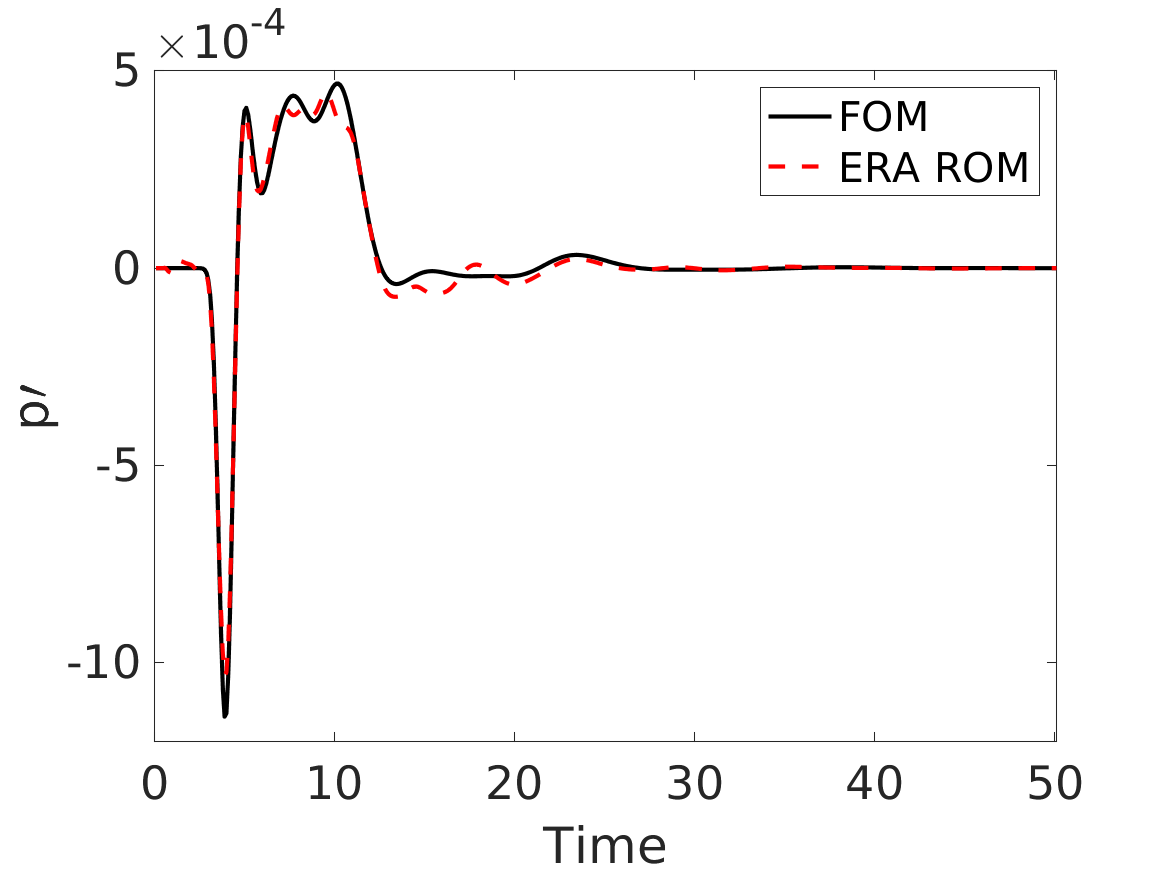}
    \rput(0.5,0.1){\psscalebox{0.5}{\color{black} \textbf{a)}}}
    \vspace{0.1cm}
  \end{minipage}
  \centering
  \begin{minipage}[a]{0.49\textwidth}
    \includegraphics[trim=4 -0.1 4 4, clip, width=\textwidth]{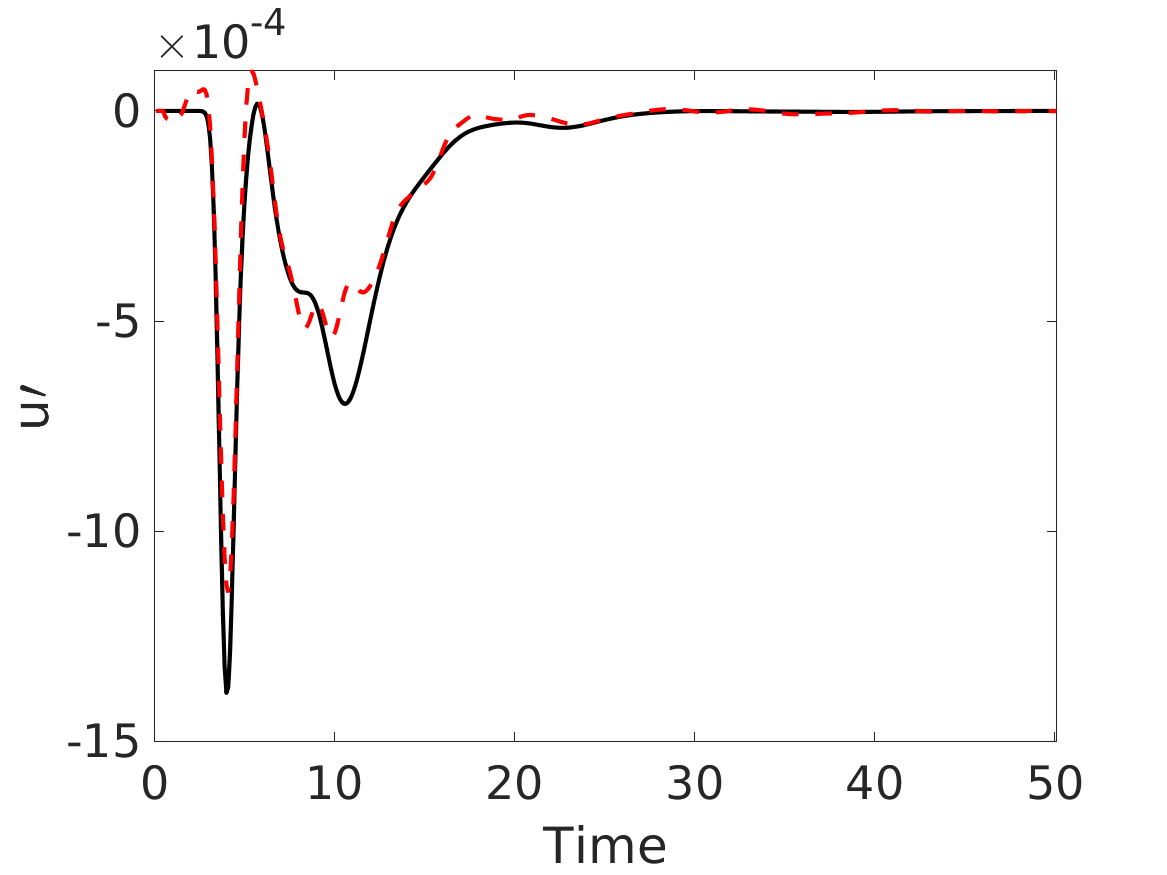}
    \rput(0.5,0.1){\psscalebox{0.5}{\color{black} \textbf{b)}}}
    \vspace{0.1cm}
  \end{minipage}
  \centering
  \begin{minipage}[a]{0.49\textwidth}
    \includegraphics[trim=4 -0.1 4 4, clip, width=\textwidth]{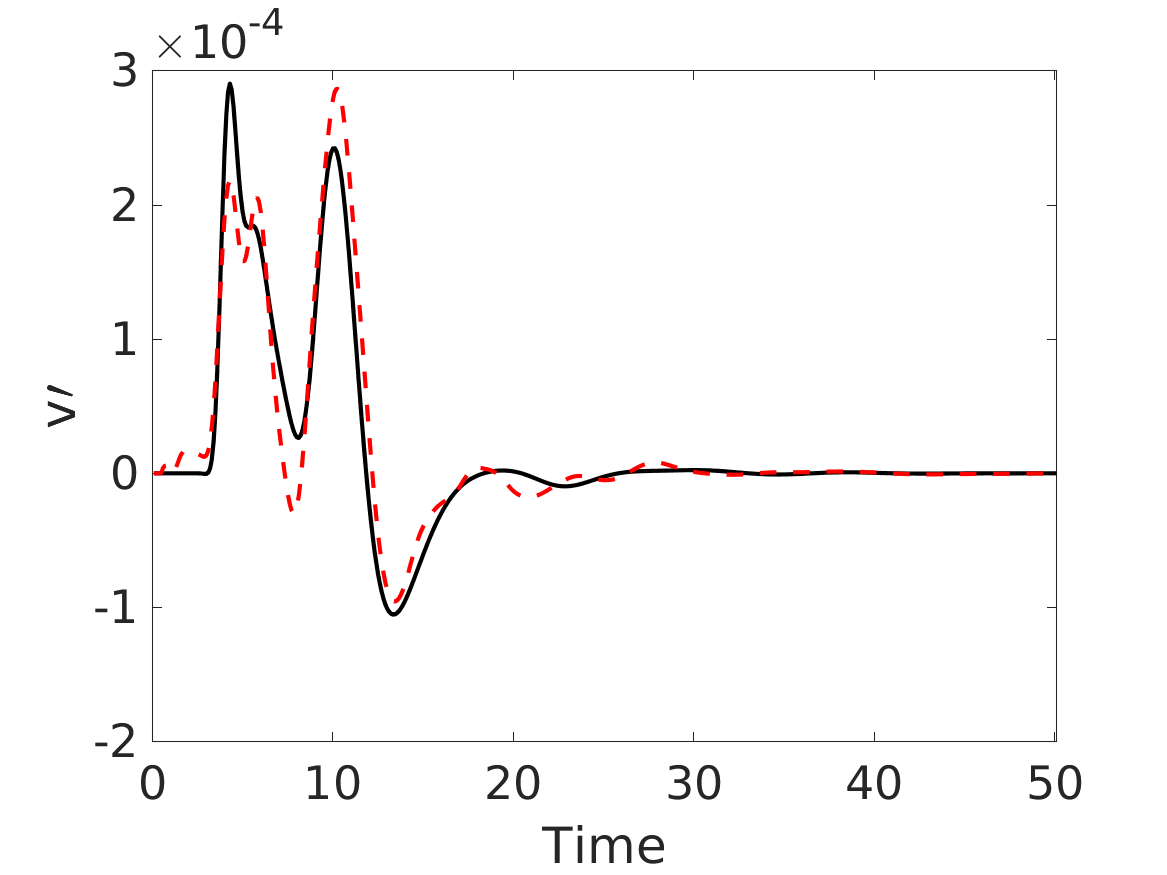}
    \rput(0.5,0.1){\psscalebox{0.5}{\color{black} \textbf{c)}}}
  \end{minipage}
  \centering
  \begin{minipage}[a]{0.49\textwidth}
    \includegraphics[trim=4 -0.1 4 4, clip, width=\textwidth]{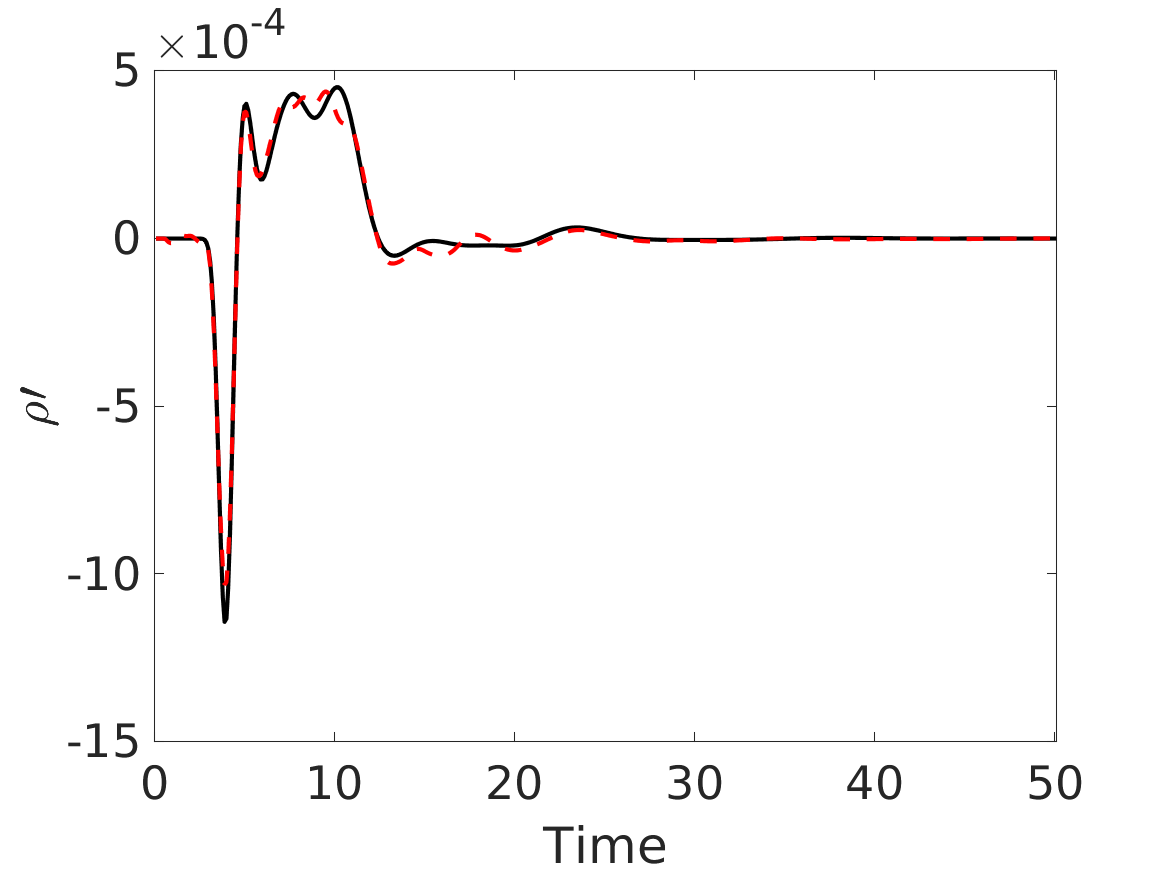}
    \rput(0.5,0.1){\psscalebox{0.5}{\color{black} \textbf{d)}}}
  \end{minipage}
  \centering
  \caption{Comparison of pressure (a), u-velocity (b), v-velocity (c), and density (d) computed by the linearized FOM and the ERA ROM at $(x,y) = (0.4628,-0.3129)$. The ROMs are trained using the input channels identified by gappy POD. The freestream is perturbed with the non-periodic square wave.} 
   \label{step_probe_linear_gappy}
\end{figure}

%\subsection{ROM Predictions for the Non-periodic Square Wave Input}
To evaluate the predictive capability of the high-resolution ROMs in response to transient non-periodic inputs, Figure~\ref{step_probe_linear_gappy} demonstrates the response of the ROM measured by a probe located below the airfoil at $(x,y) = (0.4628,-0.3129)$, when the entire far-field boundary is perturbed with the non-periodic square input signal. Comparison of the ROM predictions against the linearized FOM shows a very good agreement in the acoustic response prediction for the airfoil. We should reiterate here that the ROM response to both the triangular wave and the square wave are computed using the scaling factor adjusted based on the sinusoidal input. Hence, as we mentioned earlier, the scaling factor can be adjusted based on the first few time steps of any input signal, and it does not need further adjustment as we change the input type.

Note that the ROMs constructed here based on the prescribed set of actuators are not accurate close to the far-field boundary. This is not surprising, as only a limited number of far-field boundary cells are perturbed to generate these ROMs. Therefore, accuracy is clearly affected at the vicinity of the input channels that are not excited through the ROM training process, as the fluctuations have not reached the states in that area to capture their dynamic response. Nonetheless, it is the airfoil acoustic response that we are interested in here, which becomes dominant closer to the airfoil surface and away from the far-field boundary. The ERA ROM predictions have shown to be accurate in this area. For this reason, a global relative error metric such as the one in equation~\ref{rel_e} is not a suitable measure to demonstrate the predictive performance of the ROMs created here for the solutions in the finer grid. 

By using the gappy POD approach to choose the most effective actuators and building the ERA ROMs from the Markov sequence corresponding to a limited number of input channels, a speed up factor of 257.97 is achieved against the linearized FOM. However, the main speed up here is achieved in the offline stage of model reduction, where the Gaussian input response snapshots are generated and collected in response to separate activation of each input channel, tangential interpolation is applied to the sequence of Markov parameters, the Hankel matrix is constructed and finally, the ROM matrices are computed. Knowing that computing the Gaussian input response for each channel with the $401 \times 101$ grid takes 46.66 minutes, reducing the number of input channels by a factor of 4 decreases the computation time of the training data by 353 hours. This is in addition to the computational savings as a result of a reduction in the size of the Hankel matrix, the cost of computing the SVD of the Hankel matrix and computation cost of the ROM matrices, which were otherwise unaffordable to compute in this application. Unfortunately, since constructing the ROMs from the Markov sequences of the original 604 input channels is beyond our available computational resources, we are not able to provide an offline speed up factor here.

%probes: front of the airfoil (195,10), bottom of the airfoil (140,30), top of the airfoil (250,30)

\section{Conclusions}
\label{sec:conclusion}
 Recursive processes such as parametric studies for certification and design optimization require very efficient simulation tools. Towards this end we propose a data-driven model reduction approach based on balanced truncation for rapid aerodynamic and aeroacoustic response predictions. This approach uses a system identification method called the eigensystem realization algorithm (ERA) to create a minimum realization  that balances the system Gramians when trained with a sufficiently sampled Markov sequence. Transformation of the system to balance the Gramians results in a new coordinate system in which the modes that are preserved are equally highly controllable and observable. Therefore, components such as acoustic waves that are dynamically observable are  preserved despite the fact they do not carry a significant amount of energy.

We created ERA ROMs for prediction of the acoustic response of a NACA0021 airfoil. The training data for the ROMs is generated by the linearized and nonlinear Euler solvers, where the gust fluctuations are modeled by periodic freestream perturbations. Since the entire far-field boundary needs to be perturbed to replicate the true physics of the problem, the system is characterized with a large number of input channels (604 for a computational grid with $401 \times 101$ cells). On the other hand, we are interested in full field predictions  to investigate the validity of the model reduction approach. As a result, the multi input-multi output system gives rise to a large Hankel matrix and an expensive SVD to compute the ROM matrices. We used the tangential interpolation method to reduce the number of inputs and outputs by projecting the training data onto the leading left and right tangential directions, yet incurring a high cost of constructing ROM matrices. Therefore, we pursued a different route to enable data-driven balancing transformation of this system.

We first demonstrated the predictive performance of the ERA ROMs for the solution in a coarser computational grid with $101 \times 51$ cells in two scenarios: 1. when the training snapshots (the Gaussian input response) are computed by solving the nonlinear Euler equations, and 2. when the sequence of Markov parameters for training the ERA ROMs is computed by solving the linearized Euler equations.
The created ROMs were able to replicate the predictions by the linearized and nonlinear FOMs, despite the ERA ROMs were trained with a Gaussian input and tested with unseen input signals. 

Next, we defined an actuator selection problem to enable model reduction for prediction of the solution in the finer $401 \times 101$ grid. Specifically, we used the Markov sequence computed by the snapshots from the coarser $101 \times 51$ grid in a gappy POD framework to identify the actuators (input channels) that affect the dynamics the most. Then we queried the high-resolution ($401 \times 101$) FOM for the Markov sequence of only the channels selected by the QR factorization pivots in the gappy POD method. This approach reduced the number of channels by a factor of 4 and the wall-clock time for training snapshots generation by hundreds of hours. We then constructed the ERA ROMs based on the reduced set of Markov sequences. The actuator selection framework along with the tangential interpolation approach enabled application of data-driven balanced truncation to the high-resolution problem. The results demonstrate that the ERA ROMs are capable of predicting the aeroacoustic response in truly predictive scenarios with orders of magnitude smaller computational time compared to the FOM. While there is much to be demonstrated on more complex problems, we remark that aeroacoustic propagation may be an ideal application for balanced truncation because of the linearity of wave propagation dynamics.

\section{Appendix A}
\label{appA}
In this section we provide more details about the construction of the linearized FOM matrices. As shown in section~\ref{sec:hfm}, the Jacobian matrix is evaluated at the equilibrium point as $\mathbf{J}=\left[- \left(\pd{(\mathbf{f}+\mathbf{g})}{\mathbf{q}}\right)_C + \left(\pd{(\mathbf{f} + \mathbf{g})}{\mathbf{q}}\right)_L - \left(\pd{(\mathbf{f} + \mathbf{g})}{\mathbf{q}}\right)_R \right]_{\mathbf{q} = \Bar{\mathbf{q}}}$. Here, the flux Jacobians are computed using the Roe flux formulation as,
\be
\left(\pd{\mathbf{f}}{\mathbf{q}}\right)_{i+1/2,j} = \frac{1}{2} \left[\left(\pd{\mathbf{f}}{\mathbf{q}}\right)_{i,j} + \left(\pd{\mathbf{f}}{\mathbf{q}}\right)_{i+1,j}\right] - \left(\mathbf{R} \left|\boldsymbol{\Lambda}\right| \mathbf{L}\right)_{i+1/2,j},
\ee
where, we drop $\mathbf{q} = \Bar{\mathbf{q}}$ for simplicity and assume all of the terms are evaluated at the equilibrium point (i.e., the steady-state solution). Similarly, the derivatives of the fluxes in the y direction are obtained as,
\be
\left(\pd{\mathbf{g}}{\mathbf{q}}\right)_{i,j+1/2} = \frac{1}{2} \left[\left(\pd{\mathbf{g}}{\mathbf{q}}\right)_{i,j} + \left(\pd{\mathbf{g}}{\mathbf{q}}\right)_{i,j+1}\right] - \left(\mathbf{R} \left|\boldsymbol{\Lambda}\right| \mathbf{L}\right)_{i,j+1/2}.
\ee
Here, the term with the subscript C contains all of the flux derivative terms evaluated at the current cell $(i,j)$ and the corresponding dissipation terms. Similarly, the terms with the subscripts L and R contain all the flux derivatives evaluated at the left and right neighboring cells, respectively, along with their corresponding Roe dissipation terms. The right and left eigenvectors are the columns and rows of the $\mathbf{R}$ and $\mathbf{L}$ matrices, respectively,
\be
\mathbf{R} =
\begin{bmatrix}
    1 & 1 & 1 & 0 \\
    u_{ave} - c_{ave} n_x & u_{ave} & u_{ave} + c_{ave} n_x & n_y \\
    v_{ave} - c_{ave} n_y & v_{ave} & v_{ave} + c_{ave} n_y & -n_x \\
    \frac{c_{ave}^2}{(\gamma - 1)} + \nu - c_{ave} \tilde{u} & \nu & \frac{c_{ave}^2}{(\gamma - 1)} + \nu + c_{ave} \tilde{u} & u_{ave} n_y - v_{ave} n_x 
\end{bmatrix},
\ee
\be
\mathbf{L} = 
\begin{bmatrix}
    \frac{(\gamma - 1) \nu + c_{ave} \tilde{u}}{2 c_{ave}^2} & \frac{(1 - \gamma) u_{ave} - c_{ave} n_x}{2 c_{ave}^2} & \frac{(1 - \gamma) v_{ave} - c_{ave} n_y}{2 c_{ave}^2} & \frac{(\gamma - 1)}{2 c_{ave}^2} \\
    1 - \frac{(\gamma - 1) \nu}{c_{ave}^2} & \frac{(\gamma - 1) u_{ave}}{ c_{ave}^2} & \frac{(\gamma - 1) v_{ave}}{ c_{ave}^2} & \frac{(1 - \gamma)}{ c_{ave}^2} \\
    \frac{(\gamma - 1) \nu - c_{ave} \tilde{u}}{2 c_{ave}^2} & \frac{(1 - \gamma) u_{ave} + c_{ave} n_x}{2 c_{ave}^2} & \frac{(1 - \gamma) v_{ave} + c_{ave} n_y}{2 c_{ave}^2} & \frac{(\gamma - 1)}{2 c_{ave}^2} \\
    \frac{v_{ave} - \tilde{u} n_y}{n_x} & n_y & \frac{n_y^2 - 1}{n_x} & 0
\end{bmatrix},
\ee
and $\mathbf{\Lambda}$ is the diagonal matrix of eigenvalues $\lambda_1 = \tilde{u} - c_{ave}$, $\lambda_{2,4} = \tilde{u}$, and $\lambda_3 = \tilde{u} + c_{ave}$, where, $\nu = \frac{u_{ave}^2 + v_{ave}^2}{2}$, $\tilde{u} = u_{ave} n_x + v_{ave} n_y$, $n_x$ and $n_y$ are the outward unit vectors normal to the cell faces in the x and y directions. Quantities with the subscript $ave$ are the Roe-averaged quantities, where, $u_{ave} = \frac{\sqrt{\rho_L} u_L + \sqrt{\rho_R} u_R}{\sqrt{\rho_L} + \sqrt{\rho_R}}$, $v_{ave} = \frac{\sqrt{\rho_L} v_L + \sqrt{\rho_R} v_R}{\sqrt{\rho_L} + \sqrt{\rho_R}}$, $h_{ave} = \frac{\sqrt{\rho_L} h_L + \sqrt{\rho_R} h_R}{\sqrt{\rho_L} + \sqrt{\rho_R}}$, and $c_{ave} = \sqrt{(\gamma - 1) (h_{ave} - \nu)}$. Computing the boundary fluxes in equation~\ref{b_mat} follows the same procedure.

To compute the derivative of the flux vectors with respect to the conservative variables $\mathbf{q}= \left [
q_1 \ \ q_2 \ \ q_3 \ \ q_4
\right ]^T$, the flux terms are first written in terms of the conservative variables as,
\be
\mathbf{f} = 
\begin{bmatrix}
    f_1 \\
    f_2 \\
    f_3 \\
    f_4
\end{bmatrix} =
\begin{bmatrix}
    q_2 \\
    \frac{q_2^2}{q_1} + p \\
    q_2 \frac{q_3}{q_1} \\
    q_4 \frac{q_2}{q_1} + \frac{q_2}{q_1} p
\end{bmatrix}, \qquad
\mathbf{g} = 
\begin{bmatrix}
    g_1 \\
    g_2 \\
    g_3 \\
    g_4
\end{bmatrix} =
\begin{bmatrix}
    q_3 \\
    q_2 \frac{q_3}{q_1} \\
    \frac{q_3^2}{q_1} + p \\
    q_4 \frac{q_3}{q_1} + \frac{q_3}{q_1} p
\end{bmatrix},
\ee
and pressure is obtained as $p = \left(\gamma - 1 \right) \left(q_4 - \frac{q_2^2 + q_3^2}{2q_1}\right)$. It is straight forward now to derive the Jacobian terms $\pd{\mathbf{f}}{\mathbf{q}}$ and $\pd{\mathbf{g}}{\mathbf{q}}$,
\be
\begin{aligned}
    & \pd{(f_1 + g_1)}{q_1} = 0, 
      \ \ \pd{(f_1 + g_1)}{q_2} = n_x, 
      \ \ \pd{(f_1 + g_1)}{q_3} = n_y, 
     \ \ \pd{(f_1 + g_1)}{q_4} = 0, \\
    & \pd{(f_2 + g_2)}{q_1} =  \left[\frac{(\gamma - 3)}{2} u^2 + \frac{(\gamma - 1)}{2} v^2 \right] n_x - u v n_y, 
     \ \ \pd{(f_2 + g_2)}{q_2} = (3 - \gamma) u n_x + v n_y, \\
    & \pd{(f_2 + g_2)}{q_3} = (1 - \gamma) v n_x + u n_y, 
     \ \ \pd{(f_2 + g_2)}{q_4} = (\gamma - 1) n_x, \\ 
    & \pd{(f_3 + g_3)}{q_1} = - uv n_x + \left[\frac{(\gamma - 3)}{2} v^2 + \frac{\gamma - 1}{2} u^2 \right] n_y, \\
    & \pd{(f_3 + g_3)}{q_2} = v n_x - (\gamma - 1) u n_y, 
     \ \ \pd{(f_3 + g_3)}{q_3} = u n_x + (3 - \gamma) v n_y, 
     \ \ \pd{(f_3 + g_3)}{q_4} = (\gamma - 1) n_y, \\
    & \pd{(f_4 + g_4)}{q_1} = \left[- \gamma \frac{u e}{\rho} + (\gamma - 1) u^3 + (\gamma - 1) uv^2 \right] n_x + \left[- \gamma \frac{ve}{\rho} + (\gamma - 1) u^2 v + (\gamma - 1) v^3 \right] n_y, \\
    & \pd{(f_4 + g_4)}{q_2} = \left[\gamma \frac{e}{\rho} - \frac{3}{2} (\gamma - 1) u^2 - (\gamma - 1) \frac{v^2}{2} \right] n_x + (1 - \gamma) uv n_y, \\
    & \pd{(f_4 + g_4)}{q_3} = (1 - \gamma) uv n_x + \left[\gamma \frac{e}{\rho} - (\gamma - 1) \frac{u^2}{2} - \frac{3}{2} (\gamma - 1) v^2 \right] n_y, \\
    & \pd{(f_4 + g_4)}{q_4} = \gamma u n_x + \gamma v n_y.
\end{aligned}
\ee

\section{Appendix B}\label{appB}
\begin{figure}[h!]
  \centering
  \begin{minipage}[a]{0.49\textwidth}
    \includegraphics[trim=4 -0.1 4 4, clip, width=\textwidth]{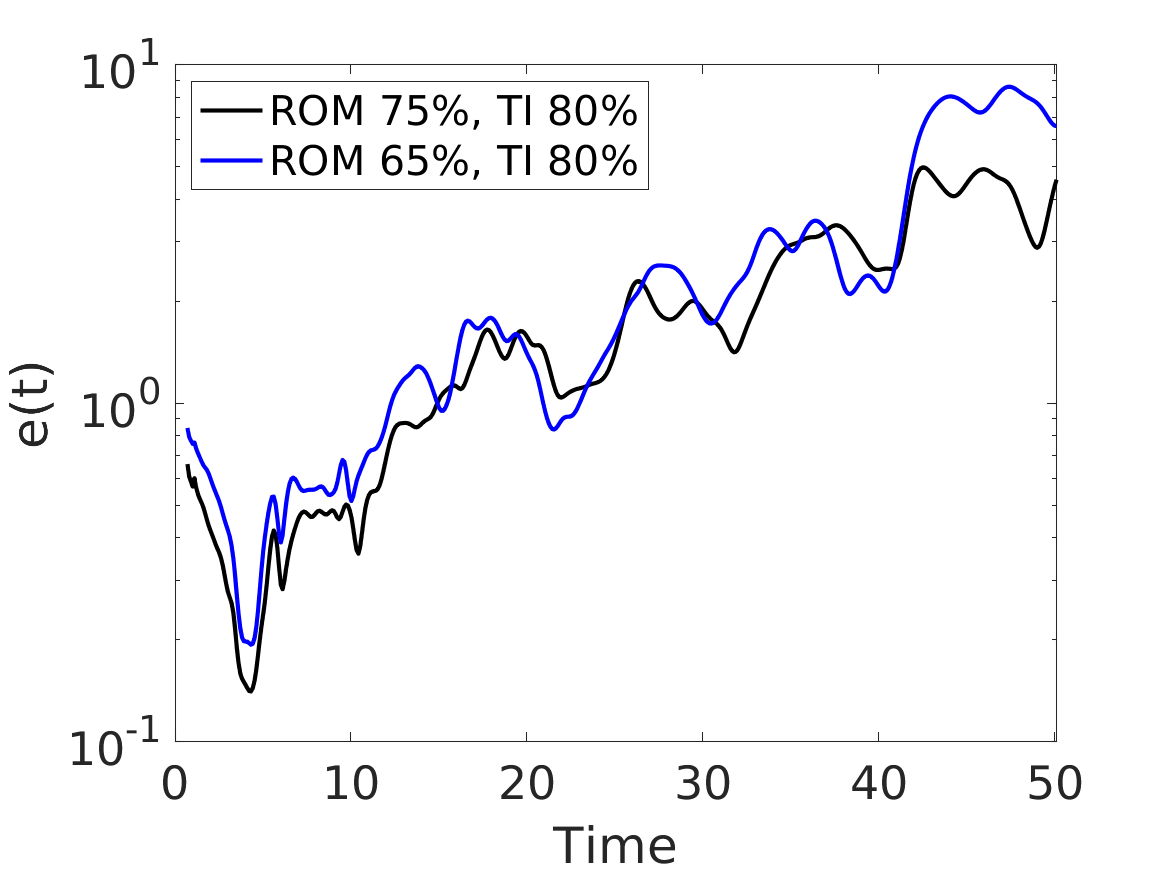}
    \rput(0.5,0.1){\psscalebox{0.5}{\color{black} \textbf{a)}}}
    \vspace{0.1cm}
  \end{minipage}
  \centering
  \begin{minipage}[a]{0.49\textwidth}
    \includegraphics[trim=4 -0.1 4 4, clip, width=\textwidth]{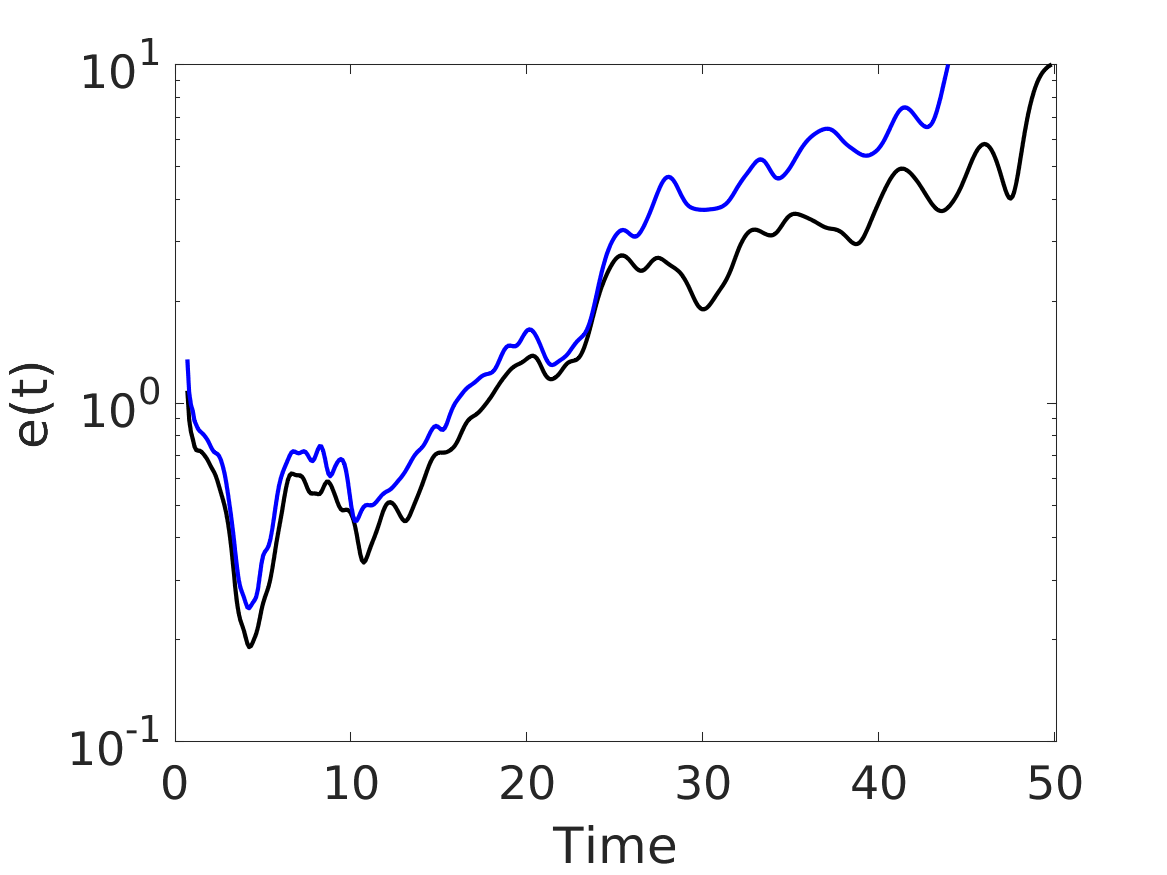}
    \rput(0.5,0.1){\psscalebox{0.5}{\color{black} \textbf{b)}}}
    \vspace{0.1cm}
  \end{minipage}
  \centering
  \begin{minipage}[a]{0.49\textwidth}
    \includegraphics[trim=4 -0.1 4 4, clip, width=\textwidth]{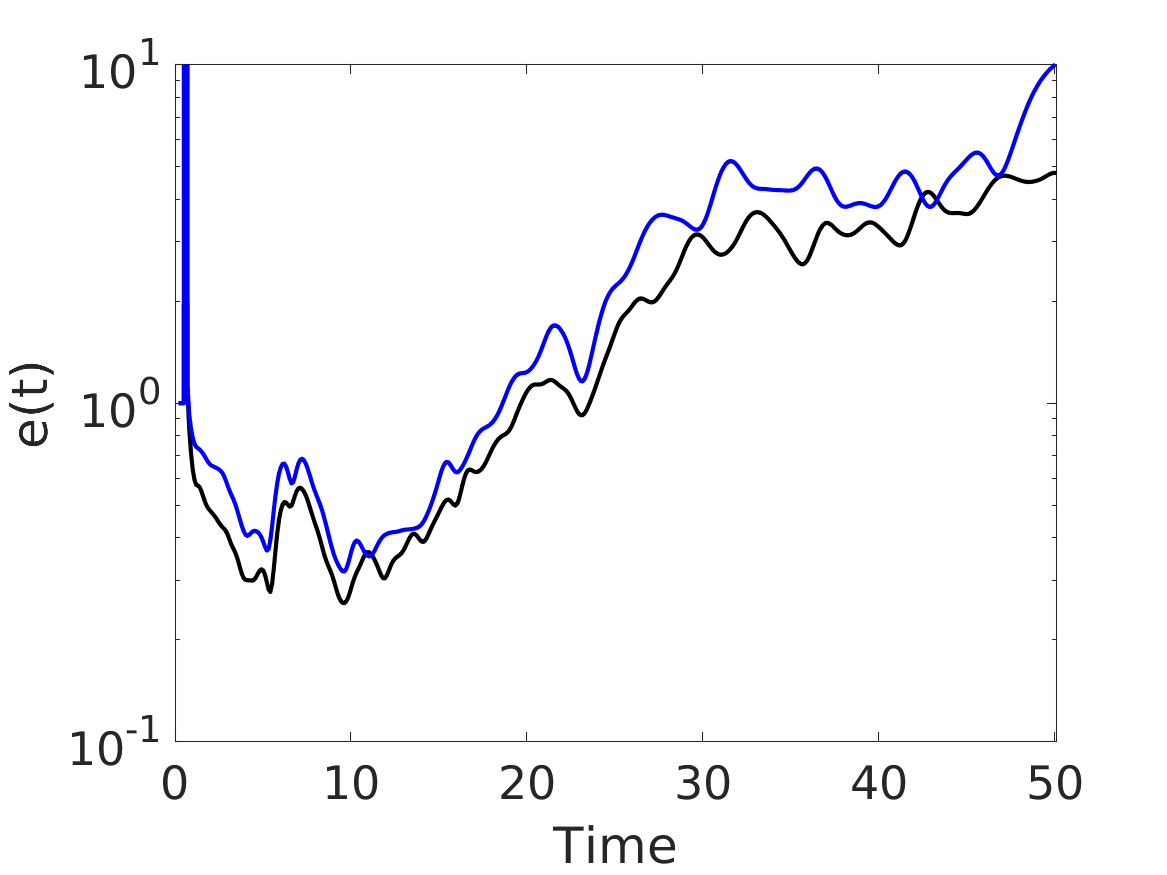}
    \rput(0.5,0.1){\psscalebox{0.5}{\color{black} \textbf{c)}}}
  \end{minipage}
  \centering
  \begin{minipage}[a]{0.49\textwidth}
    \includegraphics[trim=4 -0.1 4 4, clip, width=\textwidth]{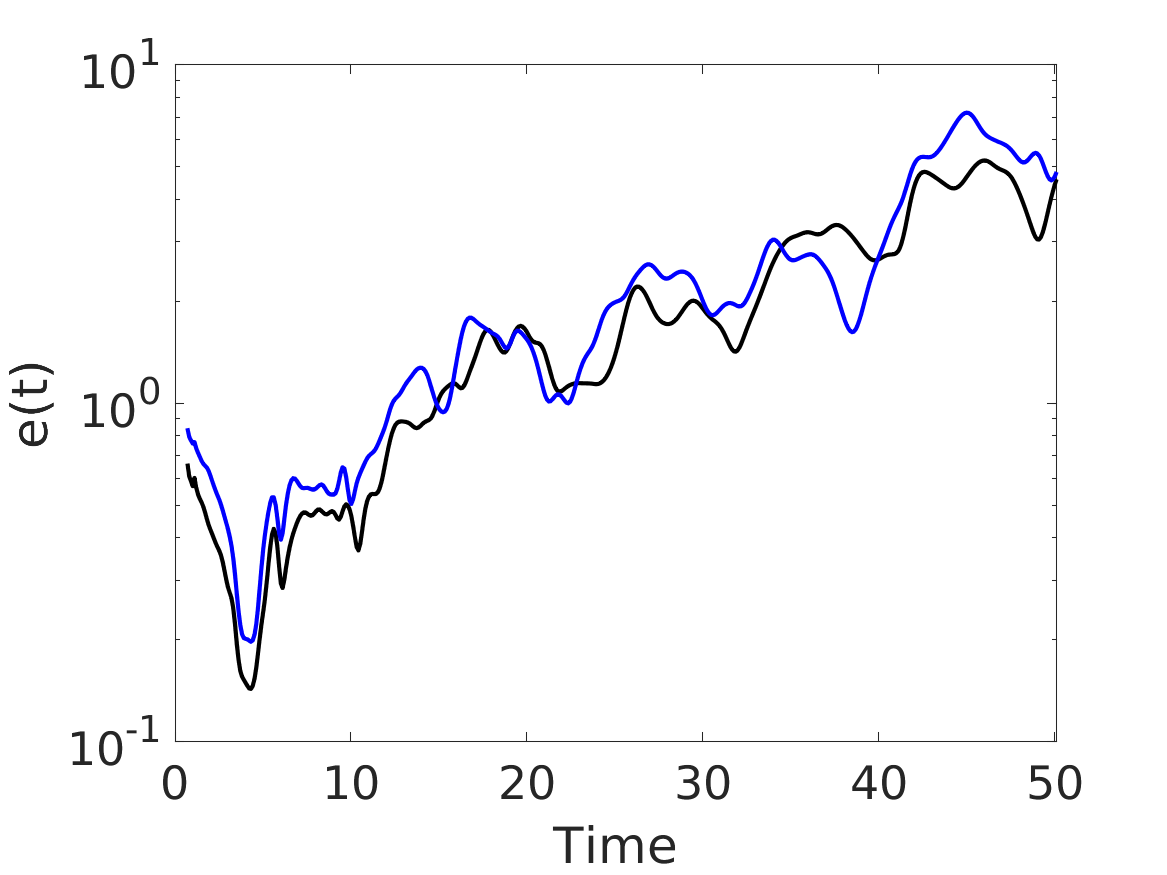}
    \rput(0.5,0.1){\psscalebox{0.5}{\color{black} \textbf{d)}}}
  \end{minipage}
  \centering
  \caption{Relative errors computed for pressure (a), u-velocity (b), v-velocity (c), and density (d). The freestream is perturbed with the non-periodic square wave, while ROMs are trained with the snapshots of the Gaussian input response generated by the nonlinear FOM.} 
   \label{step_error}
\end{figure}

The evolution of the relative error in equation~\ref{rel_e} is demonstrated in Figure~\ref{step_error} for the ROM response to the non-periodic square wave input. ROMs are constructed with Hankel singular vectors that capture $65\%$ and $75\%$ of the input-output energy. The retained tangential modes capture $80\%$ of the energy in the training snapshots.
The performance of the two ROMs is almost identical for the non-periodic input, showing that the smaller ROM (with $65\%$ energy content) is also capturing the main dynamical signatures in the acoustic response.

\begin{figure}[h!]
  \centering
  \begin{minipage}[a]{0.55\textwidth}
    \includegraphics[trim=4 -0.1 4 4, clip, width=\textwidth]{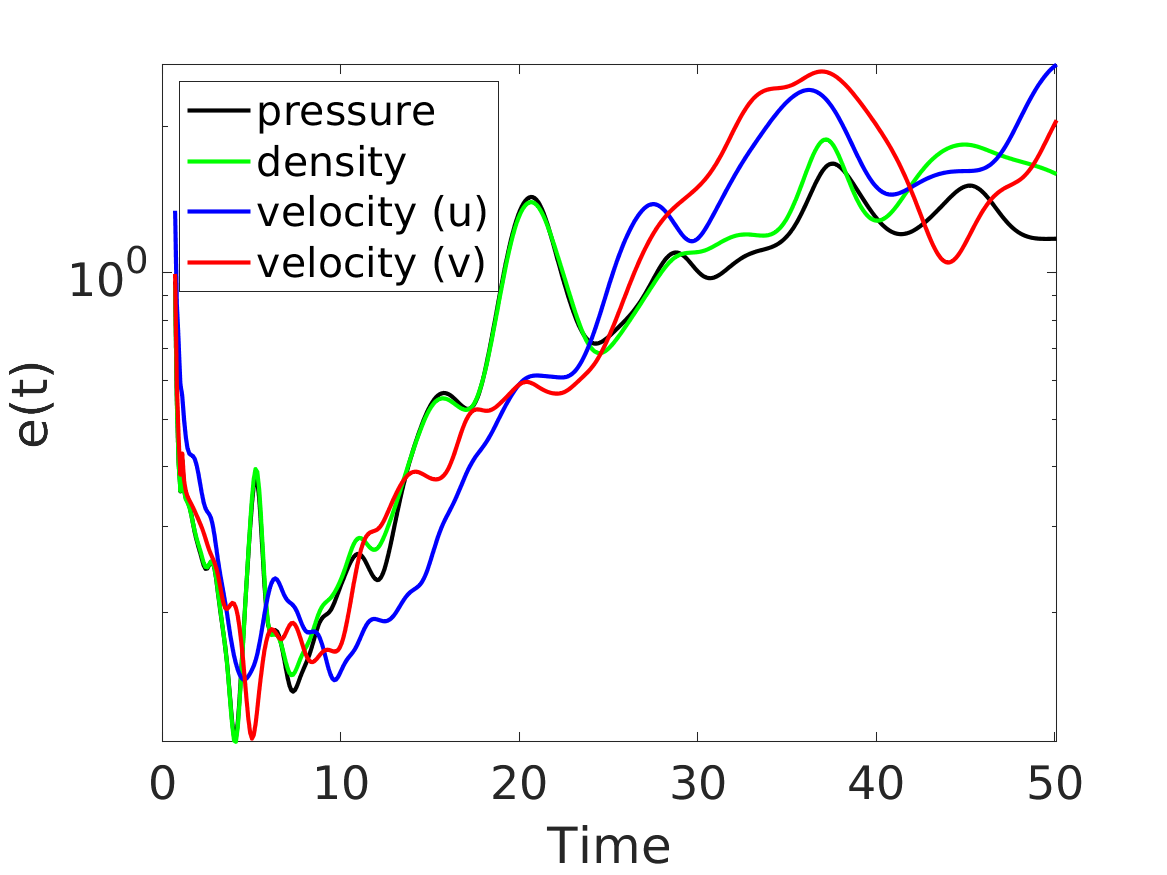}
  \end{minipage}
  \centering
  \caption{Relative errors computed for ERA ROMs of different variables.
  The freestream is perturbed with the non-periodic square wave, while ROMs are trained with the snapshots of the Gaussian input response generated by the linearized FOM.} 
   \label{step_error_linear}
\end{figure}

The same error metric is evaluated in Figure~\ref{step_error_linear} for the ROM trained with the snapshots generated by the linearized FOM. ERA ROMs in this figure are constructed based on the balancing modes that capture $80\%$ of the input-output energy. As also seen in the probe measurements in Figure~\ref{step_probe_linear}, ROM predictions agree well with the FOM and numerical errors are not as prevalent as in the ROMs trained with the nonlinear solver. The better performance of the balanced ROMs based on the linearized solver is not surprising as balanced truncation and ERA are originally developed for linear systems. In addition, the faster decay of the Hankel singular values in the ROM trained with the linearized FOM snapshots allows for a more effective dimensionality reduction when compared against the ROM trained with the nonlinear FOM snapshots. In other words, the former is able to more accurately represent its FOM with about the same number of modes as the latter.

\section*{Funding Sources}
Funded by Air Force under grant FA9550-17-1-0195.

\section*{Acknowledgments}
The authors gratefully acknowledge the Air Force for supporting this research through the Center of Excellence under grant FA9550-17-1-0195.

\bibliography{ref}

\end{document}